\documentclass{jpp}
\usepackage{graphicx}
\usepackage{hyperref}
\usepackage[utf8]{inputenc}
\usepackage[T1]{fontenc}
\usepackage{amsmath}
\usepackage{amssymb}
\usepackage{xcolor}
\usepackage{graphicx}
\usepackage{caption}
\usepackage{subcaption}

\shorttitle{Trapped-particle orbits and modes in quasi-symmetric stellarators and tokamaks}
\shortauthor{E. Rodríguez, R.J.J. Mackenbach}

\title{Trapped-particle precession and modes in quasi-symmetric stellarators and tokamaks: \\ 
a near-axis perspective}

\author{E. Rodríguez,\aff{1} R.J.J. Mackenbach\aff{1,2}}

\affiliation{\aff{1}Max Planck Institute for Plasma Physics, 17491 Greifswald, Germany
\aff{2}Eindhoven University of Technology, 5612 AZ Eindhoven, The Netherlands}

\begin{document}

\maketitle

\begin{abstract}
This paper presents the calculation of the bounce-averaged drift of trapped particles in a near-axis framework for axisymmetric and quasisymmetric magnetic fields that possess up-down and stellarator symmetry respectively. This analytic consideration provides important insight on the dependence of the bounce-averaged drift on the geometry and stability properties of the field. In particular, we show that, although the maximum-$\mathcal{J}$ property is unattainable in quasisymmetric stellarators, one may approach it through increased plasma $\beta$ and triangular shaping, { albeit going through a reduced precession scenario with potentially higher particle losses}. The description of trapped particles allows us to calculate the available energy of trapped electrons analytically in two asymptotic regimes, providing insight into the behaviour of this measure of turbulence. It is shown that the available energy is intimately related to MHD-stability, providing a potential synergy between this measure of gyrokinetic turbulence and MHD-stability.
\end{abstract}

\section{Introduction}
It has long been known that the bounce-averaged drift that a trapped particle experiences is central to both linear and nonlinear stability of gyrokinetic trapped-particle modes \citep{kadomtsev1967plasma,rosenbluth1968,helander2013collisionless,helander2017available}. This motion of trapped particles can serve as an energy source or sink for various instabilities, and thus their study is central to understanding their behaviour in any plasma-field scenario. 
% A clear example of their importance may be seen in the efforts to achieve so-called maximum-$\mathcal{J}$ fields, in which the density-gradient driven trapped-particle mode is fully stabilised \citep{proll2012resilience,proll2022turbulence}.  
\par
The behaviour of trapped particles depends crucially on the class of fields considered. In an effort to study stellarators, so-called omnigeneous fields \citep{hall1975,bernardin1986,cary1997,landreman2012,helander2014theory} are of particular interest. In such fields, composed of nested flux surfaces on which field lines live, trapped particles have, by definition, a vanishing averaged drift in the direction normal to flux surfaces (i.e. radially). This restricts the dynamics of trapped particles to an average drift \textit{within} flux surfaces, often referred to as \textit{precession}, and denoted by $\omega_\alpha$. Many authors have investigated the behaviour of this quantity \citep{white1984hamiltonian,roach1995trapped}, and analytic expressions exist for large-aspect ratio tokamaks with circular cross-sections and small non-axisymmetric perturbations \citep{connor1983effect,hegna2015effect}. For general omnigeneous stellarators (and even more so, upon relaxing omnigeneity), expressions for $\omega_\alpha$ rarely allow for analytical calculation {\citep{velasco2023robust}}. This ends up {impeding} the dissection of the underlying physics and effects.
\par
This paper carries out such calculations in a more general scenario. To make the problem tractable, we consider two main simplifications. First, we specialise to quasi-symmetric  \citep{boozer1983transport,nuhren1988,rodriguez2020necessary} and axisymmetric fields with stellarator symmetry \citep{dewar1998} and up-down symmetry respectively, two special sub-classes of the wider class of omnigenous systems. The central feature of both of these classes is the symmetry of their magnetic field magnitude, $|\mathbf{B}|$. This reduces the complexity of the particle dynamics significantly, especially in the region close to the magnetic axis (the centremost field line of the field, around which flux surfaces accrue). This leads to the second simplifying consideration in this paper: the asymptotic description near the axis. In this near-axis approach, the field magnitude may be directly parametrised, and the framework developed by \cite{garrenboozer1991a,landreman2019constructing} may be employed directly. Both these simplifying considerations enable an explicit description of the trapped particle motion, whose construction and interpretation we present in Sections~\ref{sec:Jpar} and \ref{sec:w_alpha}.
\par
Once the particle precession is known, we next investigate its role on trapped-particle mode stability. We do so by studying the available energy (\AE{}) of trapped electrons \citep{helander2020available}; that is, a measure of the available thermal energy that may be liberated by appropriate rearrangments of the electron distribution function. We compute \AE{} analytically in Section~\ref{sec:AE}, explicitly showing its dependence on various important parameters. This enables a direct comparison to other physically relevant properties such as MHD stability and flux surface shaping. 
\par

\section{Asymptotic expression for the second adiabatic invariant}  \label{sec:Jpar}
The description of the trapped particle precession requires the evaluation of the bounce-averaged drift around flux-surfaces, denoted as $\omega_\alpha$. To calculate such quantity, then, we must begin by appropriately defining flux surfaces and a notion of direction over them. To this end, we first introduce the Clebsch representation \citep{d2012flux} of the magnetic field; namely, $\mathbf{B}=\nabla\psi\times\nabla\alpha$, where $\psi$ is the magnetic toroidal flux divided by $2\pi$ and $\alpha$ is an angular potential defined as $\alpha = \theta - \iota \varphi$. Here $\theta$ and $\varphi$ are straight-field line \citep{d2012flux} poloidal and toroidal angular coordinates respectively, and $\iota$ is the rotational transform. The flux surfaces are assumed to be nested and correspond to constant pressure surfaces, following a magnetic field that is in equilibrium, $\mathbf{j}\times\mathbf{B}=\nabla p$. The angle $\alpha$ can be interpreted as a field line label within flux surfaces (following $\mathbf{B}\cdot\nabla\alpha=0$). Thus, we define the precession $\omega_\alpha$ as the rate at which trapped particles change field line within a flux surface, formally,
\begin{equation}
    \omega_\alpha=\overline{\mathbf{v}_D\cdot\nabla\alpha}. \label{eqn:wa_def}
\end{equation}
This is the common bounce-averaged binormal drift \citep{kadomtsev1967plasma, white1984hamiltonian,helander2014theory}. Here, the overline operator denotes a bounce-averaging: that is, a time average over the back-and-forth motion of the trapped particle along the field line { (thus assuming a `thin-orbit') }, 
\begin{equation}
    \overline{\dots}=\frac{\oint \dots \frac{\mathrm{d}\ell}{v_\parallel}}{\oint \frac{\mathrm{d}\ell}{v_\parallel}},
\end{equation}
where $v_\parallel$ is the parallel velocity, the arc-length along a magnetic field-line is parametrised by $\ell$, and the domain of integration is taken to be a simply connected region that satisfies $v_\parallel(\ell) \geq 0$ (which is typically referred to as a bounce-well). This is an integral at constant $\psi$ and $\alpha$, but also particle energy $H$ and first moment $\mu$. 
\par
It is convenient to write the bounce-average drift in Eq.~(\ref{eqn:wa_def}) in terms of derivatives of a single scalar quantity, $\mathcal{J}_\parallel$ \citep{helander2014theory}. This scalar quantity is the so-called second adiabatic invariant, 
\begin{equation}
    \mathcal{J}_\parallel = \int m v_\parallel \: \mathrm{d} \ell,
\end{equation}
and is an approximately conserved quantity of trapped particles. Importantly, this quantity serves as the Hamiltonian of the trapped-averaged dynamics of trapped-particles, meaning, as can be shown explicitly \citep{helander2014theory, calvo2017effect}, that 
\begin{equation}
    \omega_\alpha \stackrel{\cdot}{=} - \frac{1}{q} \frac{ \left( \partial \mathcal{J}_\parallel / \partial \psi \right)_{\mu,H,\alpha}}{\left( \partial \mathcal{J}_\parallel / \partial H \right)_{\mu,\psi,\alpha}} = \frac{\Delta \alpha}{\tau_b}.
    \label{eq:def_omegaalpha_dJ}
\end{equation}
Here $q$ is the charge of the particle (which we shall take to be $q=-1$ for electrons), and $\Delta \alpha$ and $\tau_b$ have been defined as the total $\alpha$-excursion and elapsed time in a particle bounce, respectively. 
\par
Because $\mathcal{J}_\parallel$ encodes the dynamics of trapped particles in a single scalar expression (and more generally, also allows one to calculate the radial drift), we shall explicitly calculate the asymptotic expression for $\mathcal{J}_\parallel$ as a first step towards finding $\omega_\alpha$.
\par
\subsection{Expanding the second adiabatic invariant}
Let us write $\mathcal{J}_\parallel$ explicitly as a function of the field line following coordinate $\ell$ (where we have taken the particle mass $m=1$),
\begin{equation}
\mathcal{J}_\parallel = \sqrt{2 H} \int \sqrt{1 - \lambda \hat{B}(\ell)} \: \mathrm{d} \ell. \label{eqn:Jpar_ell}
\end{equation}
Here we have introduced the pitch-angle $\lambda = \mu B_0 / H$ (using for the first adiabatic invariant $\mu=(2H-v_\parallel^2)/B$), which distinguishes between {different trapped particles (deeper or more shallowly trapped)}, and we have normalized the magnetic field by some reference field strength $\hat{B} = B/B_0$. The integral is taken between bounce points, i.e., between points at which $\hat{B}=1/\lambda$. We are assuming there is no electric field within each flux surface, and as such no electrostatic potential appears in Eq.~(\ref{eqn:Jpar_ell}). 
\par
It will prove convenient to express the field-line following coordinate $\ell$ in terms of Boozer angles \citep{boozer1981plasma}. We define for that purpose the helical angle,
\begin{equation}
    \chi\stackrel{\cdot}{=}\theta-N\varphi,
\end{equation}
where $N$ is an integer that defines a helical angle, foreseeing the application to quasi-symmetric devices with a helical symmetrry. Using this angular parameterisation,  the Boozer-coordinate Jacobian $\mathcal{J}=B_\alpha(\psi)/B^2$, where $B_\alpha=G+\iota I$ (in the standard Boozer notation \citep{boozer1981plasma,helander2014theory}), and defining $\bar{\iota}=\iota-N$, the second adiabatic invariant in these coordinates may now be written as
\begin{equation}
    \mathcal{J}_\parallel= \frac{\sqrt{2H}}{\overline{\iota}} \frac{B_\alpha}{B_0} \int \frac{1}{\hat{B}} \sqrt{1 - \lambda \hat{B}} \: \mathrm{d} \chi. \label{eqn:invariant-in-boozerchi}
\end{equation}
It is crucial to note that $\mathcal{J}_\parallel$ depends directly on the magnetic field magnitude along a field line, with minimal involvement of other geometric elements. This simplifies the calculation of $\mathcal{J}_\parallel$ ostensibly when considering quasi- and axisymmetric configurations, and makes them rather analogous. 
\par
In order to construct a representative $\hat{B}$, we resort to the inverse-coordinate near-axis framework \citep{garrenboozer1991a,landreman2019constructing}, in which the asymptotic form of $\hat{B}$ near the axis is rather simple. We will employ essential results from the near-axis theory of quasisymmetric equilibrium fields as needed without re-deriving them, and refer to the literature for details \citep{garrenboozer1991a, landreman2019constructing}. That way, and to second order in the distance from the magnetic axis, $r=\sqrt{2\psi/B_0}$, we write $\hat{B}$, following Eq.~(1) in \cite{garrenboozer1991a} or Eq.~(2.15) in \cite{landreman2019constructing}, and noting that this behaviour goes beyond the particular form of equilibrium assumed \citep{rodriguez2021weak}, as
\begin{equation}
    \hat{B}=\bar{B}+ \delta B(\varphi,\chi),
    \label{eq: magnetic field grouping}
\end{equation}
where we have separated,
\begin{equation}
    \overline{B} = 1 - r \eta \cos \chi,
    \label{eq:first-order-field}
\end{equation}
and the second order,
\begin{equation}    
    \delta B= r^2\left(B_{20}+B_{22}^C\cos2\chi+B_{22}^S\sin2\chi \right).
    \label{eq:second-order-field}
\end{equation} 
In the ideal quasi- or axisymmetric limit, all parameters ($\eta$, $B_{20}$, $B_{22}^C$ and $B_{22}^S$) are constants instead of functions of the toroidal angle $\varphi$, and for stellarator symmetry $B_{22}^S=0$. From here on we shall assume that this condition is satisfied exactly. Note that in practice this condition is only achieved approximately: the near-axis expansion generally fails to find exactly quasi-symmetric solutions for equilibria at second order in $r$ (unless exactly axisymmteric fields are considered). This is commonly referred to as the Garren-Boozer oveerdetermination problem \citep{garrenboozer1991b}, and results from a clash between the symmetry and the equilibrium \citep{rodriguez2021anis,rodriguez2021weak}. In that sense the idealised framework is only an approximation, but it will prove useful, and, as we shall see, practical in approximating quasisymmetric configurations. The constants that define $|\mathbf{B}|$ can then be interpreted as parameters that describe different configurations. In fact, with these parameters together with a magnetic axis shape, the field and flux surfaces may be constructed explicitly \citep{landreman2019constructing}. Thus, expressing the dependence of the second adiabatic invariant on these parameters will provide a direct link to the configuration and its distinguishing features. { A note of caution: although we are considering the asymptotic limit in the distance to the magnetic axis, we cannot approach it closer than the gyroradius related `potato-orbit' size \citep[Ch.~7]{helander2005collisional} without violating the `thin orbit limit' of our $\mathcal{J}_\parallel$ calculation.\footnote{The back-of-the-envelope calculation is as follows. Simplify things by considering the tokamak notation $\eta\sim B_0/G_0\sim1/R$ and $q=1/\iota$ for the safety factor. The adiabatic invariant calculation is correct when, along the bounce-orbit, the guiding centre approximately follows the field-line. That is, the argument $v_\parallel\mathrm{d}\ell$ should not change significantly over the distance the particle drifts in a bounce time. As $v_\parallel\sim v_t\sqrt{\epsilon/R}$, the argument changes $\mathrm{d}(v_\parallel\mathrm{d}\ell)/\mathrm{d}r\sim\mathcal{J}_\parallel/r$. As the transit time is $\omega_t^{-1}\sim qR/v_t$ and $\mathbf{v}_d\cdot\nabla r\sim v_t\rho/R$, the error $\Delta\mathcal{J}_\parallel/\mathcal{J}_\parallel\sim\sqrt{q^2\rho^2R/r^3}$. Thus, $r\gg r_p=(q^2\rho^2R)^{1/3}$, where $r_p$ is the range where potato orbits come into play \citep{helander2005collisional}.}}
\par
The way that we have grouped the terms in Eq.~\eqref{eq: magnetic field grouping} might be unexpected, given that we have mixed asymptotically unequal terms: the constant on-axis field ($r=0$) with the first order variation. However, since we are interested in trapped particles, variation in $|\mathbf{B}|$ along the field line is necessary, otherwise trapped particles would not exist. Therefore, in our perturbative description of the problem we must take $\Bar{B}$ to constitute the leading order magnetic field magnitude. It would then appear natural to write, 
\begin{equation}
        \mathcal{J}_\parallel=\sqrt{2H}\frac{B_\alpha}{\bar{\iota}B_0}\int_{\chi_b^-}^{\chi_b^+}\frac{\sqrt{1-\lambda(\bar{B}+ \delta B)}}{\bar{B}+\delta B}\mathrm{d}\chi. \label{eqn:J_epsilon}
\end{equation} 
where for every $\lambda$, the bounce-points $\chi_b^\pm$ (left and right) of the integral are given by,
\begin{equation}
    \frac{1}{\lambda}=\bar{B}(\chi_b^\pm)+\delta B(\chi_b^\pm), \label{eqn:lam_chi_b}
\end{equation}
and attempt to expand it in powers of $\delta B$ (i.e. expand around smallness of $\delta B$). There are however two important sources of conflict in doing so. First, and formally, evaluating $\sqrt{1-\lambda \Bar{B}}$ at the bounce points $\chi_b^\pm$ can lead to imaginary contributions near these points without additional careful consideration of the bounce points. Secondly, physically, there are also issues related to the behaviour of classes such as barely trapped particles, which under an infinitesimal perturbation may undergo a finite (non-infinitesimal) change. This translates into diverging expressions in the perturbative construction.
\par 
To deal with these issues consistently, we start by defining a correction field $\delta B_b (\lambda)$, defined to be the perturbed field $\delta B(\chi)$ evaluated, for a given particle class $\lambda$, at the bouncing points, see Eq.~(\ref{eqn:lam_chi_b}). We shall assume, for simplicity, that the device is stellarator symmetric about the bottom of the magnetic well (i.e. we ignore the $B^S_{22}$ component), so that $\delta B_b (\lambda)$ is unique (i.e., it does not have a left and right values). Introducing this correction, let us rewrite $\mathcal{J}_\parallel$ for a stellarator symmetric field in the following form,
\begin{equation}
        \mathcal{I}(\epsilon)=\sqrt{2H}\frac{B_\alpha}{\bar{\iota}B_0}\int_{-\chi_b}^{\chi_b}\frac{\sqrt{1-\lambda[(\bar{B}+\delta B_b)+ \epsilon(\delta B-\delta B_b)]}}{\bar{B}+\epsilon\delta B}\mathrm{d}\chi,
\end{equation}
so that $\mathcal{J}_\parallel=\lim_{\epsilon \rightarrow 1^-} \mathcal{I}(\epsilon)$. Expressing the integral in this form ensures that the integrand upholds positive definiteness for all $\epsilon \in [0,1)$, evading the issue of the integrand becoming imaginary. This furthermore circumvents the need to expand the bounce points of the integral. Expressed in this form, the integral may now be Taylor expanded in $\epsilon$,
\begin{equation}
    \mathcal{I} \approx \underbrace{\mathcal{I}(0)}_{\stackrel{\cdot}{=}\mathcal{J}_\parallel^{(0)}} + \epsilon\underbrace{\left.\frac{\partial \mathcal{I}}{\partial \epsilon}\right|_{\epsilon=0}}_{\stackrel{\cdot}{=}\mathcal{J}_\parallel^{(1)}} \label{eqn:I_J_par}
\end{equation}
where we shall take $\epsilon \rightarrow 1$ in the final answer so that $\mathcal{J}_\parallel\approx\mathcal{J}_\parallel^{(0)}+\mathcal{J}_\parallel^{(1)}$. If each of these terms is evaluated to the right order, we shall retrieve a consistent expression for the adiabatic invariant $\mathcal{J}_\parallel$. 

\subsection{Leading order expression}
Let us start by investigating the leading order term of the expansion in $\epsilon$. Setting the expansion parameter $\epsilon$ to zero results in the following integral,
\begin{equation}
    \mathcal{J}_\parallel^{(0)} = \sqrt{2H}\frac{B_\alpha}{\bar{\iota}B_0}\int\frac{\sqrt{1-\lambda(1 +\delta B_b - r \eta \cos \chi )}}{1 - r \eta \cos \chi}\mathrm{d}\chi.
    \label{eq: expression for Jparr(1)}
\end{equation}
This integral is highly reminiscent of that occurring in the magnetic field of a large-aspect ratio tokamak with circular cross-section, which has been considered by many authors before in various asymptotic regimes \citep{connor1983effect,roach1995trapped,helander2005collisional,hegna2015effect}, and as such the derivation closely mirrors these calculations. One may refer to Appendix~\ref{app: second adiabatic invariant details} for a complete derivation. The main step required is to re-express the integral in terms of a trapping parameter $k$, which we define as
\begin{equation}
    k^2 = \sin^2 \left( \frac{\chi_b}{2} \right), \label{eqn:def-k2}
\end{equation}
where $\chi=0$ is defined to be the magnetic well minimum. The most deeply trapped particles reside here, and thus have $k^2 = 0$. The most shallowly trapped particles reside at the tops of the well, namely $\chi_b=\pi$, and thus correspond to $k^2 = 1$. The two trapped particle classes are connected monotonically, in the sense that $\lambda$ and $k$ maintain the order of trapped classes. With this definition, the integral may be expressed in terms of complete elliptic integrals of the first and second kind (also known as Legendre’s Integrals, see e.g. \citet{DLMF}, Sec.~19), which we define as
\begin{subequations}
\begin{align}
    K(k) & \stackrel{\cdot}{=} \int_0^{\pi/2} \frac{ \mathrm{d}\zeta}{\sqrt{1 - k^2 \sin^2 \zeta} },   \label{eq:expansion-new-coordinates} \\
    E(k) & \stackrel{\cdot}{=} \int_0^{\pi/2} \sqrt{1 - k^2 \sin^2 \zeta} \: \mathrm{d}\zeta
    \label{eq:expansion-new-coordinates-b}.
\end{align}
\end{subequations}
With these definitions, one can express $\mathcal{J}_\parallel^{(0)}$ in closed form expanding $\mathcal{I}(0)$ around the smallness of $r$. The result is, to order $O(r^{5/2})$,
\begin{equation}
    \mathcal{J}_\parallel^{(0)} = 4\sqrt{H r\eta }\frac{B_{\alpha0}}{\bar{\iota}_0B_0} \left[ I_1(k) +r\eta\left(I_1(k)\left(\frac{1}{2}-k^2\right) + I_2(k)\right) + O(r^2)\right],
    \label{eq:J(1)}
\end{equation}
where the functions $I_1$ and $I_2$ describe the behaviour of different trapped-particle classes via $k$,
\begin{subeqnarray}
    I_1(k) &&= 2\left[ (k^2 - 1) K(k) + E(k) \right],   \label{eq:trapping function I1}\\
    I_2(k) &&=  \frac{2}{3} \left[(2 k^2-1) E(k)-(k^2-1) K(k)\right].
    \label{eq:trapping function I2}
\end{subeqnarray}
\par
One can readily interpret the leading form of $\mathcal{J}_\parallel^{(0)}$ by referring back to its basic form in terms of a parallel velocity and bounce distance, $\mathcal{J}_\parallel\sim v_\parallel \ell$. The scaling with $\sqrt{H r\eta }$ is directly related to the reduced parallel velocity that trapped particles must have in order for them to be trapped.\footnote{From energy conservation, one can readily show that
$
    mv_\parallel^2/2 \approx H r \eta \left(  2 k^2 - 1 \right) \hat{B},
    % &= H \left(1 - \lambda \hat{B} \right) \\
    % &= H \left(1 - \frac{1}{1+r\eta (2k^2-1)} \hat{B} \right) \\ 
    % &
$
and hence that $v_\parallel\sim\sqrt{H r\eta}$.} 
The factor $B_{\alpha0}/\bar{\iota}_0B_0$ represents the changes in the ``connection length'' along the magnetic field-line. In terms of geometric quantities, and using Amp\'{e}re's law, one can estimate this length-scale to be $B_{\alpha0}/\bar{\iota}_0B_0\sim R_0/(\iota_0-N)$,
where $R_0$ is the major radius of the device and $N$ represents the helicity of the $|\mathbf{B}|$ symmetry determined by the shape of the axis \citep{landreman2019constructing,rodriguez2022phases}. As the major radius increases, the total distance travelled by a trapped particle grows, and so does $\mathcal{J}_\parallel$. Similarly, decreasing $\bar{\iota}$ increases the distance between bounce points, as field lines become further misaligned with respect to $|\mathbf{B}|$ contours. Finally, we observe that $I_1$, which describes the trapped-particle class dependence of $\mathcal{J}_\parallel$ to leading order, monotonically increases with $k$, as so does the bounce distance. Under such a perspective, it is then clear that it must vanish for the deeply trapped particles (i.e. $I_1(k=0)=0$).
\par
To provide a full description of $\mathcal{J}_\parallel$ to next order, it is important to note that although the expression we found is correct to order $O(r^{5/2})$, it does not correspond to the value of $\mathcal{J}_\parallel$ to that order. The expression is incomplete, as it is missing the contribution from $\mathcal{J}_\parallel^{(1)}$.

\subsection{The first order correction}
To evaluate the second order term we follow Eq.~(\ref{eqn:I_J_par}), for which we need the following integral,
\begin{equation}
        \mathcal{J}_\parallel^{(1)}=
        -\sqrt{2H}\frac{B_\alpha}{\bar{\iota}B_0}\int_{-\chi_b}^{\chi_b} \left( \frac{\lambda}{2} \frac{\delta B-\delta B_b}{\sqrt{1-\lambda(\bar{B}+\delta B_b)}} + \frac{\delta B \sqrt{1 - \lambda (\bar{B} + \delta B_b)}}{\bar{B}^2}  \right)\mathrm{d}\chi.
    \label{eq:J-epsilon-expansion}
\end{equation}
The second term in the integrand is a factor $r$ smaller than the first one (which may be verified by writing expressions explicitly in terms of $k$), and hence, for our current expansion, only the first term needs to be taken into account. This term requires rewriting to involve the integration variable $\chi$ explicitly. Using the near-axis form of $|\mathbf{B}|$ for a quasi- and stellarator symmetric field, Eq.~(\ref{eq:second-order-field}), 
\begin{subequations}
    \begin{gather}
        \delta B-\delta B_b= r^2 B_{22}^C \left( \cos 2\chi - \cos 2 \chi_b  \right).
    \end{gather}
\end{subequations}
From this point the procedure is analogous to the steps followed in the leading order case; the details are given, once again, in Appendix~\ref{app: second adiabatic invariant details}. We simply denote the central result here, which is the expression for $\mathcal{J}_\parallel^{(1)}$ expanded to leading order in $r$,
\begin{equation}
    \mathcal{J}_\parallel^{(1)} \approx - 4 \sqrt{H r \eta} \frac{B_\alpha}{\bar{\iota}B_0 } \frac{r B_{22}^C}{\eta} \mathcal{I}_{22}^C(k),
\end{equation}
where the function $\mathcal{I}_{22}^C(k)$ is defined to be,
\begin{equation}
    \mathcal{I}_{22}^C(k)\stackrel{\cdot}{=} I_2(k)-(2k^2-1)I_1(k).
\end{equation}
Combining this result with Eq.~(\ref{eq:J(1)}), we may complete the asymptotic form of the second adiabatic invariant to order $O(r^{5/2})$,
\begin{equation}
    \mathcal{J}_\parallel\approx4\sqrt{H r\eta }\frac{B_{\alpha 0}}{\Bar{\iota}_0 B_0}\left\{I_1(k) +r\eta\left[\left(\frac{1}{2}-k^2\right)I_1(k)+I_2(k)-\frac{B_{22}^C}{\eta^2}\mathcal{I}_{22}^C(k)\right]\right\}. \label{eqn:2nd_ad_inv_full}
\end{equation}
% We remind ourselves that this expression is valid in the exact quasisymmetric limit with stellarator symmetry\footnote{It would in principle be straightforward to relax this condition. However, it would require treatment of left and right parts of the well-integrals separately. Thus, for brievity, we do not present said calculations, and instead focus on the relevant stellarator-symmetric case.}, e.g. a tokamak with up-down symmetry. As we shall see in the next section, when these conditions are met the above equation can accurately predict the trapped-particle drift, and may furthermore be used to understand how pressure gradients and shaping of flux surfaces affect this drift.

\section{Trapped particle precession} \label{sec:w_alpha}
With the second adiabatic invariant constructed, we are in a position to evaluate the bounce-average precession. We remind ourselves that we considered the exact quasisymmetric limit and stellarator symmetry\footnote{It would in principle be straightforward to relax this condition. However, it would require treatment of left and right parts of the well-integrals separately. Thus, for brevity, we do not present said calculations, and instead focus on the relevant stellarator-symmetric case.} (e.g. a tokamak with up-down symmetry) when constructing $\mathcal{J}_\parallel$. Because of the idealised omnigeneous nature of the field, we need not worry about the field-line dependence (i.e., $\alpha$ dependence), as the behaviour is `identical' in all field lines as far as the precession is concerned. This is apparent in Eq.~(\ref{eqn:2nd_ad_inv_full}). The formalism presented could however be extended to incorporate a description of said $\alpha$ dependence upon controlled deviations from omnigeneity. We present in Appendix~\ref{sec:app-radial-drifts} an explicit estimation of the radial average drift in configurations that only achieve quasisymmetry approximately, providing a previously lacking physically meaningful measure of deviations from quasisymmetry within the near-axis framework, which could aid as an optimisation target \citep{landreman2022map,rodriguez2022measures,rodriguez2022phases}.
\par
Let us nevertheless return to the calculation of precession in our idealised scenario, Eq. \eqref{eq:def_omegaalpha_dJ}. We have almost all that is needed to compute $\omega_\alpha$. The only remaining step is taking partial derivatives of Eq.~(\ref{eqn:2nd_ad_inv_full}) with respect to $\psi$ and the particle energy $H$. Acknowledging the dependence of the trapped particle label $k$ on both these variables, the result of this calculation may be written as 
\begin{equation}
    \omega_{\alpha}=\omega_{\alpha,0}+\omega_{\alpha,1}+O( r),
    \label{eq:trapped-particle-precession-full}
\end{equation}
where the leading term scales like $1/r$ and $\omega_{\alpha,1}\sim O(r^0)$ (details of this derivation may be found in Appendix~\ref{sec:appPrecess}).
\par
The leading order term $\omega_{\alpha,0}$ is 
\begin{equation}
    \omega_{\alpha,0}=2 \frac{H \eta}{r B_0}\left(\frac{E(k)}{K(k)}-\frac{1}{2}\right)\stackrel{\cdot}{=}\frac{H \eta}{r B_0}G(k), \label{eqn: precession leading order}
\end{equation}
which is precisely of the form found by \citet{connor1983effect} for a large-aspect ratio tokamak, without magnetic shear or a pressure gradient. Elements of pressure and shaping are involved in the present approach (although not explicitly) through the next order correction, $\omega_{\alpha,1}$, 
\begin{equation}
    \omega_{\alpha,1}= \frac{H \eta}{r B_0}\left[ r \eta \mathcal{G}(k) + \frac{r B_{20}}{\eta}\mathcal{G}_{20}(k) + \frac{r B_{22}^C}{\eta}\mathcal{G}_{22}^C(k)\right],
    \label{eq: second order precession}
\end{equation}
where the functions $\mathcal{G}$ may be expressed in terms of elliptic integrals,
\begin{subequations}
    \begin{align}
        \mathcal{G}(k)&\stackrel{\cdot}{=}-4 \left(\frac{E(k)}{K(k)}\right)^2 + 2 (3 - 2 k^2) \frac{E(k)}{K(k)} - (1 - 2 k^2), \label{eqn:G(k)} \\
        \mathcal{G}_{20}(k)&\stackrel{\cdot}{=}-2,  \label{eqn:G22(k)} \\
        \mathcal{G}_{22}^C(k)&\stackrel{\cdot}{=}-4\Bigg[\left(\frac{E(k)}{K(k)}\right)^2 - 2 k^2 \frac{E(k)}{K(k)} + \left(k^2-\frac{1}{2}\right) \Bigg].
    \end{align} \label{eqn:G_funcs_2}
\end{subequations}
% The natural representation of this correction involves the prefactors to these magnetic field harmonics explicitly. Our next step will be to analyse the terms and their $k$-dependence, individually.

\subsection{Leading order precession: a tokamak-like behaviour} 
Let us start by analysing the leading order precession of trapped particles focusing on $\omega_{\alpha,0}$, Eq.~\eqref{eqn: precession leading order}. The expression includes physics in two ways: the overall scaling factors in front, and the $k$ dependence through $G(k)$, which describes the behaviour of different classes of trapped particles. We first investigate the former.
\par
Precession is proportional to $\eta$, parameter defined in Eq.~\eqref{eq:first-order-field} as a measure of the leading order variation of $|\mathbf{B}|$ over flux surfaces. This variation is however intimately linked to the near-axis elliptical shaping of cross-sections (see \cite{garrenboozer1991b,landreman2019constructing,rodriguez2023mhd}). In fact, for up-down symmetric cross-sections, the elongation along the horizontal axis (i.e., ratio of horizontal to vertical axes of the cross-section) is $\mathcal{E}=(\eta/\kappa)^2$, where $\kappa$ is the curvature of the magnetic axis at the point where the cross-section is being assessed. Thus, for a fixed elongation, $\eta\sim\kappa$. In the special case of an axisymmetric field, this means that $\omega_\alpha \sim\sqrt{\mathcal{E}}/R_0$, where $R_0$ is the major radius. Going back to $\omega_{\alpha,0}$ then, for a fixed cross-sectional shape, increasing the major radius reduces precession, consequence of the field becoming more straight and the gradients in $|\mathbf{B}|$ (and with it the drift) becoming smaller. In quasisymmetric fields, the local curvature of the axis defines a `major radius', which leads to strongly curved magnetic axis shapes having increased precession. Quasi-helically symmetric fields require more strongly shaped magnetic axes (for a fixed average major radius), and thus will tend to have a stronger precession. This provides a qualitative separation between quasi-axisymmetric (QA) and quasi-helically symmetric (QH) stellarators (\cite{rodriguez2022phases}, see some values of $\eta$ in Table~\ref{tab:etaDesigns}).\footnote{In the language of \cite{rodriguez2022phases}, the qualitative difference between the QA and QH configurations holds for cases that live deep in their corresponding quasisymmetric phases. Close to phase transitions, i.e. axis shapes close to having vanishing curvature points (and hence stronger flux surface shaping), these differences fade, and in particular so will any `distinct' $\eta$-behaviour.} 
\begin{table}
    \centering
    \begin{tabular}{ccccccccc}
        & ARIESCS & ESTELL & GAR & HSX & NCSX & QHS48 & Precise QA & Precise QH \\\hline
        $\eta R_{00}$ & 0.59 & 0.79 & 0.73 & 1.54 & 0.55 & 1.15 & 0.68 & 1.51 \\
        \hline
    \end{tabular}
    \caption{\textbf{Some $\eta$ values for QS-optimised configurations.} This table presents the values of $\eta$ for many quasisymmetric designs, normalised to have an axis whose average major radius is $R_{00}=1$. This is a form of comparing them on the same basis. The largest values of $\eta$ correspond to quasi-helically symmetric configurations (namely HSX, QHS48 and precise QH), although there exist significant variations within these classes.}
    \label{tab:etaDesigns}
\end{table}
\par
We finally take note of the divergent nature of $\omega_\alpha$ with the radial coordinate. The $1/r$ behaviour may initially appear worrying, but it can be easily understood in the following terms. From the form of the poloidal drift, we estimate $ \mathbf{v}_D \cdot \nabla \theta \sim v_{\nabla B} |\nabla \theta|$, where $v_{\nabla B}$ denotes the gradient drift driven by the radial variation of $B$. As $\nabla \theta \sim 1/r$, and the gradient drift does not scale with $r$ to leading order, the result is the $1/ r$ dependence (the result of a finite drift velocity over an ever shrinking surface).
\par
\begin{figure}\centering
\begin{subfigure}{.5\textwidth}
  \centering
  \includegraphics[width=0.6\textwidth]{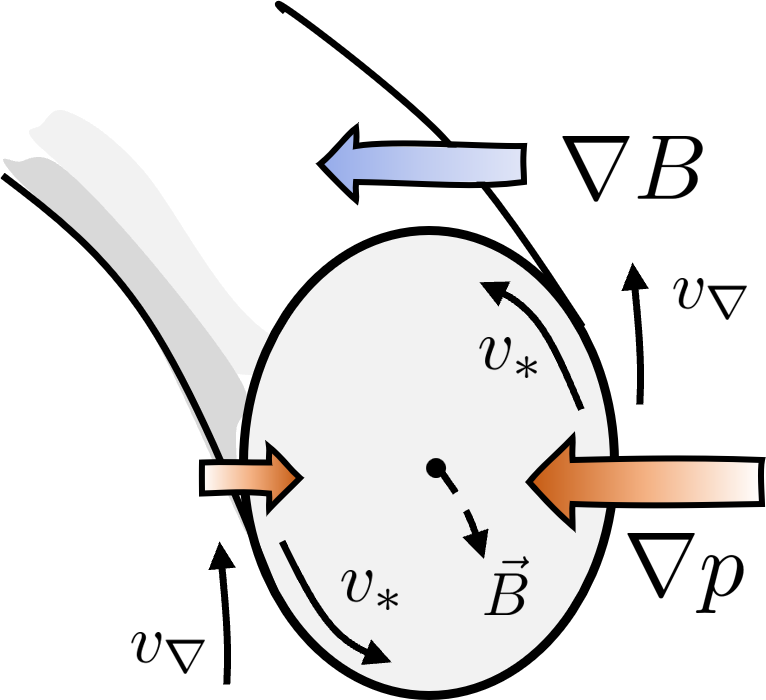}
  \caption{Schematic diamagnetic and gradient drifts}
  \label{fig:electron_drift_diagram}
\end{subfigure}%
\begin{subfigure}{.5\textwidth}
  \centering
  \includegraphics[width=0.8\textwidth]{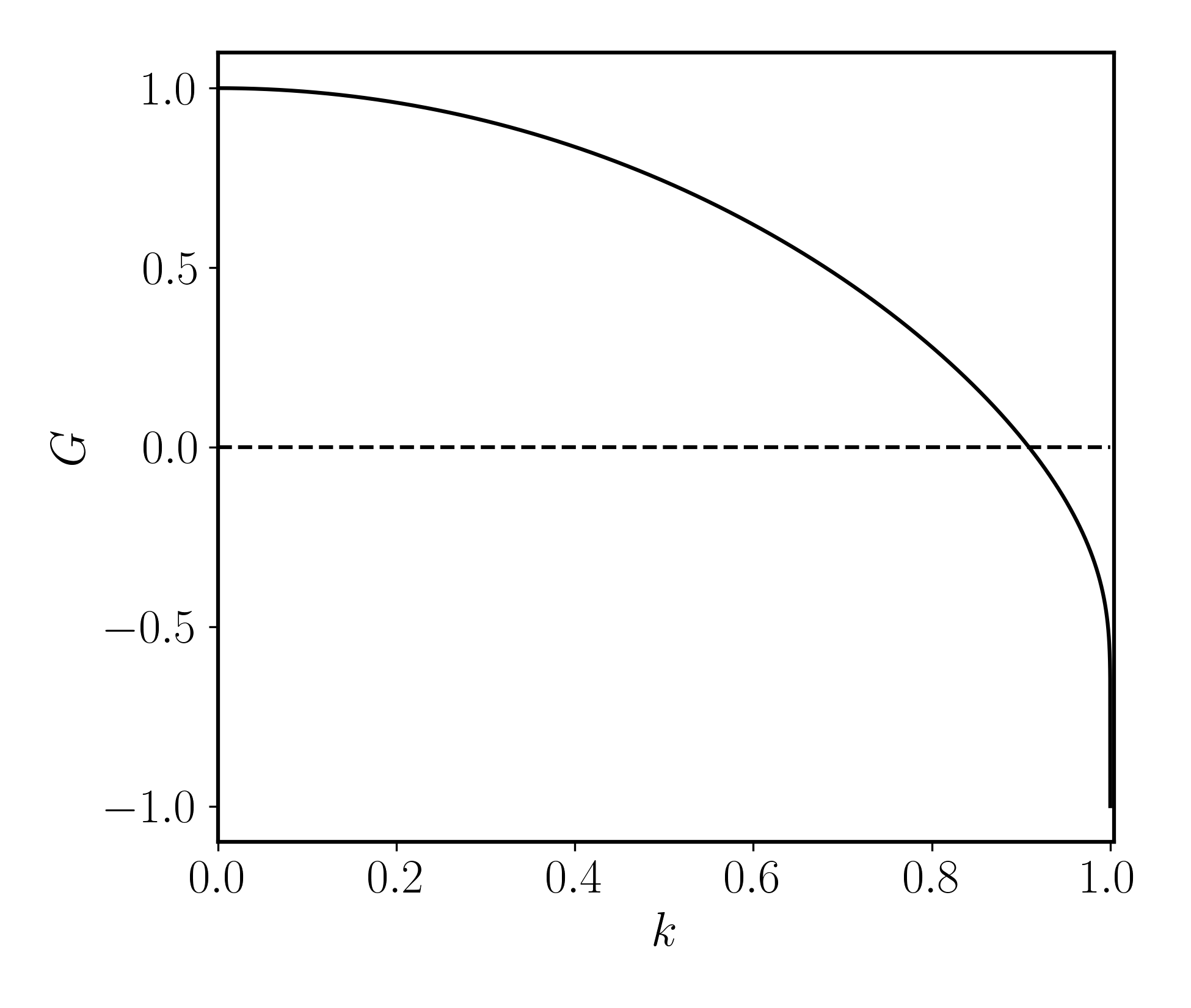}
  \caption{Plot showcasing $G(k)$}
  \label{fig:circ_tok_wa_1_e}
\end{subfigure}
\caption{\textbf{Precession of trapped particles near the magnetic axis.} Left: the diagram shows the main drifts in a magnetic configuration with a dominant $\nabla B$ direction, such as in a tokamak. The electron particle $\nabla B$ drift, $v_\nabla$, and the diamagnetic drifts, $v_*$, are shown (the latter proportional to $-\mathbf{B}\times\nabla p$). Resonance between the two occurs on the outboard side (where the deeply trapped particles live), defining the bad curvature region. Right: the plot shows the function $G(k)$ as a function of the trapped-particle class $k$, and thus the behaviour of precession as a function of trapped particle class near the magnetic-axis.}
\label{fig: leading order precession and schematic}
\end{figure}

We next shift our attention to the dependency on the trapped class dependence of Eq.~(\ref{eqn: precession leading order}). A plot of $G$ as a function of $k$ is presented in Fig.~\ref{fig:circ_tok_wa_1_e}. The plot shows that for the vast majority of trapped particles, the drift of the electrons is positive. Physically, positive values imply that the trapped particles precess in the direction of the diamagnetic drift (see Fig.~\ref{fig:electron_drift_diagram}), an important feature which will become relevant for the discussion on trapped particle modes later. This behaviour only changes for the barely trapped particle classes which end up spending a significant fraction of their bounce-time near the maximum of $|\mathbf{B}|$, where there is `good curvature'. The transition point occurs at $k_0$, where $G(k_0)=0$, roughly at $k_0\sim0.9$. Such a class of particles is, to leading order, stationary. The existence of these groups of trapped particles precessing in opposite directions proves the impossibility of making quasisymmetric configurations exactly maximum-$\mathcal{J}$. This simply follows from the definition of maximum-$\mathcal{J}$ as the property of a field that guarantees $\partial_\psi\mathcal{J}_\parallel<0$ for all trapped particles, which in terms of the electron precession is equivalent to $\omega_\alpha(k)<0$ for all $k$. Of course, this is not to say that the behaviour of a quasisymmetric field cannot become closer to maximum-$\mathcal{J}$, as higher order shaping and equilibrium parameters modify the leading order behaviour above.  However, one may not achieve it exactly everywhere, and especially, close to the axis. In practice, one may only get around this issue at a finite radius if the higher order contributions are strong enough.
\par
\par 
Comparing this result against previous investigation, we find, as we already pointed, the behaviour to be identical to that shown in the work by \cite{connor1983effect} for a large-aspect-ratio circular tokamak. That this correspondence is found in the more general quasisymmetric case should not come as a surprise, given the existing isomorphism between axisymmetric and quasisymmetric fields \citep{boozer1983transport}. We have, however, gained generality beyond circular-shaped cross-sections as $\eta$ allows for non-unity ellipticity. We also note that the asymptotic considerations here and in the literature are in many respects different. Many previous investigations have focused on employing radially local solutions to the MHD-equation (see e.g. \citet{MercierLuc1974,miller1998noncircular,hegna2000local}) to discuss precession, which weakens the coupling between $|\mathbf{B}|$ and the geometry of the field that exists in the near-axis treatment. We take that additional coupling in the near-axis consideration to form part of a more globally consistent description of the field. This results in higher order effects showing quantitative differences (though qualitative trends are retained), as we shall now explore.

\subsection{The higher order effects on precession}
% The term that describes the effects of a radial growth of the average magnetic field, $B_{20}$, is simply

% The final component of the first order correction measures the effect of the second-order cosine term of the magnetic field,

\begin{figure}
    \centering
    \includegraphics[width=\textwidth]{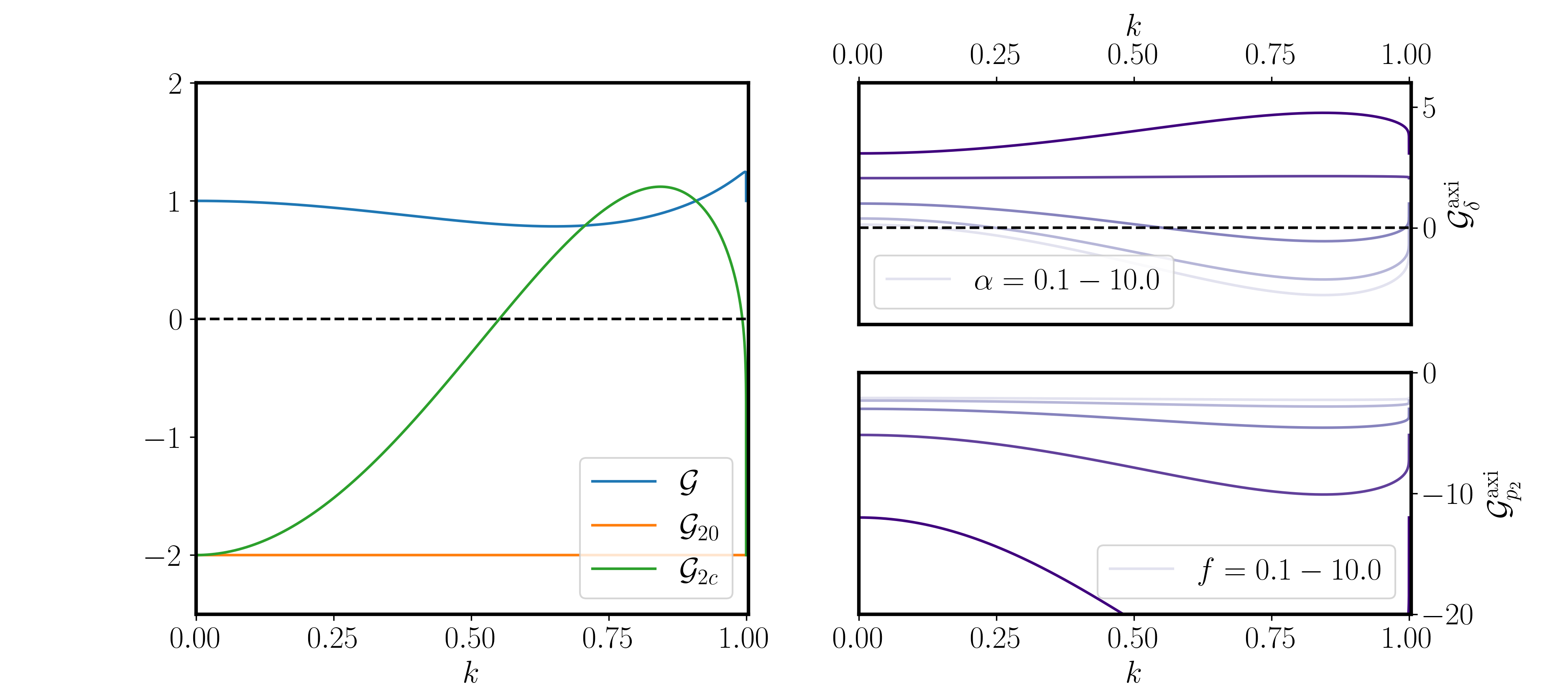}
    \caption{\textbf{Effect of second order quantities on precession.} (Left) The plot shows the dependence on the trapped particle class $k$ of the three components of $\omega_{\alpha,1}$. (Top right) Dependence of precession on triangularity of an axisymmetric tokamak as a function of $k$, for a number of values of $\alpha=\mathcal{E}^2$, where $\mathcal{E}$ is the elongation in the major radius direction. (Bottom right) Dependence of precession on pressure gradient in an axisymmetric tokamak as a function of $k$, for a number of values of $f=B_0^2(1+\alpha)^2/R_0^2I_2^2(\alpha+3)$. } 
    \label{fig:w_a1_gen}
\end{figure}

% Through the relation of the magnetic field $\delta B$ to other physical aspects of the field, the functions $\mathcal{G}$, $\mathcal{G}_{20}$, and $\mathcal{G}_{22}^C$ contain information on the effects of elements such as pressure gradient or triangularity. We remark in passing that one could also include shear in the current expression, although it formally corresponds to a higher order correction, as is shown in Appendix~\ref{sec:appPrecess}. 
Let us now focus on the effect of second order elements on precession. In the form presented in Eq. \eqref{eq: second order precession}, the order $r$ correction on the leading order precession consists of three different terms: an ``intrinsic'' term (purely a consequence of consistency of the field with its elliptical shape), a term that is proportional to $B_{20}$ (and thus the radial increase of the average $B$), and another proportional to $B_{22}^C$. The behaviour of each one of these terms with $k$ is illustrated in the left plot of Figure~\ref{fig:w_a1_gen}.
\par
From these three contributions, that coming from $B_{20}$ (often called the \emph{magnetic well} term) is the simplest: a positive $B_{20}$ pushes particles against the diamagnetic drift. That is, deeply trapped particles decrease their precession rate, while barely trapped ones precess even faster. This behaviour is a direct consequence of the influence of $B_{20}$ on the gradient $\nabla B$. The magnetic well term reinforces the outwards magnetic field gradient everywhere, affecting all particles equally and in the direction opposed to the diamagnetic drift. More precisely, the drift $v_{\nabla B}\sim\mathbf{B}\times\nabla(1/B)$ is driven by the gradient of $1/B$, and thus it is the change in the gradient of $1/B$ that most directly affects precession. As $\partial_\psi(1/B) \sim \eta^2/2-B_{20}$, this explains not only the $B_{20}$ contribution, but also the ``intrinsic'' $\mathcal{G}$ one.
\par
The direct effect of $B_{20}$ relates precession naturally to MHD stability. MHD stability of interchange modes is improved by the enhancement of the so-called magnetic well of the field \citep{greene1997}, which corresponds in the near-axis limit to increasing $B_{20}$ (the radial derivative of the average $B$) \citep{landreman2020magnetic,kim2021,rodriguez2023mhd}. Thus, there is a natural synergy between improving MHD stability and making particles precess in the direction opposite to the diamagnetic drift. This behaviour, obvious from the $B_{20}$ dependence of $\omega_{\alpha,1}$, aligns with the general observation made in Sec.~3.7 of \cite{helander2014theory} relating the `averaged' behaviour of precession over a flux surface and all particle classes to the magnetic well. However, the problem of MHD stability is more subtle, as precession is also affected by the variation of the magnetic field $B_{22}^C$ explicitly, and MHD stability is further influenced by pressure (as becomes clear when considering the Mercier criterion \citep{mercier1962,mercier1974,greene1962,bauer2012,freidberg2014} to assess it). We shall revisit this relation on more solid grounds later.
\par
Setting $B_{20}$ aside for now, the $B_{22}^C$ contribution to precession introduces a richer $k$ dependence than considered so far at this order. This is so because different trapped particles experience different modifications of the magnetic field through the $\chi$ dependence of $|\mathbf{B}|\sim B_{22}^C\cos 2\chi$. Naturally, both the most deeply and shallowly trapped particles, who live at $\chi=0,\pi$ respectively, feel the same effect, marked by $\mathcal{G}_{22}^C(0)=\mathcal{G}_{22}^C(1)$. In between these classes, particles perform an average of the gradient over their bounce, leading to a maximum at $k \sim 0.8$. 
\par 
The components $B_{22}^C$ and $B_{20}$ have proven especially convenient to describe the higher order effects on drifts. However, they do not provide a good sense for what the field looks like in terms of its geometry or its equilibrium. A physical discussion requires us to make this connection, which we shall do within the near-axis framework. To that end, we introduce the pressure gradient supported by the field as $p_2=(B_0~\mathrm{d}p/\mathrm{d}\psi|_{\psi=0})/2$, as well as the triangularity of cross-sections (normalised by $r$)\footnote{In fact, $\delta$ is also normalised by a factor of $r$. This is so because upon approaching the axis the triangular shaping disappears in favour of elliptical cross-sections, and thus triangularity clearly changes with $r$. Because the leading order is proportional to $r$ we normalise it out.} as $\delta$. These two parameters can substitute $B_{22}^C$ and $B_{20}$ to write the precession explicitly in terms of $p_2$ and $\delta$.
\par
The details of this linear mapping between parameters were presented in \cite{rodriguez2023mhd}. We include in this paper only the key elements necessary to make that connection. This is particularly important to make sense of what $\delta$ truly represents. For an up-down symmetric cross-section, we define triangularity as the relative displacement of the vertical turning points of a cross section respect to its centre-point along the symmetry line normalised to its width \citep[Appendix~C]{rodriguez2023mhd}. And $\delta$, as its asymptotic form in $r$, normalised by $r$. For simplicity, we define this triangularity in the near-axis, Frenet-Serret frame. This makes the regular notion of triangularity in the lab-frame (i.e., the triangularity of the cross-section at a constant cylindrical angle) generally different by an offset when the magnetic axis is not perpendicular to a constant cylindrical angle plane (see some details in Appendix~\ref{app: relation to geometry}). However, by changing $\delta$ we are changing triangularity in the lab frame by the same amount, although $\delta$ is generally not the triangularity there. The axisymmetric case is an exception in which this { correspondence} is exact.
We also pick the sign of triangularity to be defined with respect to the direction of curvature of the axis; i.e., positive triangularity, $\delta>0$, indicates a shift of the turning points in the direction of the curvature.\footnote{One must be careful with how the sign of $\eta$ is defined in this paper and in the definition of $\delta$. For the derivation of the precession in this paper we assumed $\eta>0$, but used the unusual convention of writing $B=B_0(1-r\eta\cos\chi)$, with a minus sign. For the triangularity to have the meaning above, one must include a $\mathrm{sign}(\eta)$ to the definition presented in \cite{rodriguez2023mhd}, which is defined assuming $\eta>0$ in the opposite convention to that considered here.} 
\par
In the general quasisymmetric scenario, following this definition of $\delta$, there appears to be a multitude of `triangularities'. Each cross-section around the torus is generally different (but for an $N$-fold symmetry), but it is sufficient to describe the triangularity of any of its cross-sections to describe the field uniquely (given some choice of lower order parameters and a pressure gradient).\footnote{This statement is almost always true. There are some special cases in which, however, the triangularity is not the most appropriate geometric feature to explicitly involve in the parametrisation of the field, and instead the Shafranov shift \citep{shafranov1963, wessonTok,rodriguez2023mhd} should be chosen. We do not consider this special case.} Once such a cross-section has been chosen, then one should interpret $\delta$ as a measure of its triangularity in the Frenet-Serret frame, and any discussion about the effect of modifying triangularity should be interpreted as the effect of changing the triangularity of that very cross-section. We shall conveniently choose an up-down symmetric cross-section to represent the quasisymmetric stellarator, and when it comes to analysing real configurations, we shall take the most vertically elongated one (often referred to as the bean-shaped cross-section \citep{rodriguez2023mhd}). We do so by analogy with the axisymmetric scenario. As this cross-section is changed, the remainder of the field must also change as a consequence of satisfying both equilibrium and quasisymmetry simultaneously. 
\par
In short, the pressure gradient and the triangularity as defined above provide sufficient information and a minimal second order parametrisation for both axisymmetric and quasisymmetric configurations.

\subsection{Relation to geometric and equilibrium parameters}
Let us then proceed to write and analyse the precession of trapped particles $\omega_{\alpha,1}$ in terms of triangularity, $\delta$, and pressure gradient, $p_2$. The details of how to do so are presented in Appendix~\ref{app: relation to geometry} and rely heavily on \cite{rodriguez2023mhd}. The result reads,
\begin{equation}
    \omega_{\alpha,1}=\frac{H \eta}{B_0 r}\left[ r\eta \tilde{\mathcal{G}}(k) - \frac{r\mu_0p_2}{|\eta| B_0^2}\mathcal{G}_{p_2}(k) + \frac{r \delta}{2}\mathcal{G}_{\delta}(k)\right].
\end{equation}
The function $\mathcal{G}_{p_2}$ encodes the effect of the pressure gradient and $\mathcal{G}_\delta$ that of the triangularity, and their full explicit forms may be found in Appendix~\ref{app: relation to geometry}. The function $\tilde{\mathcal{G}}$ is a complicated functions of lower order quantities that we do not present explicitly and shall ignore for the analysis in this paper. For a discussion on the form and meaning of the other contributions, we specialise now to a generally shaped, up-down symmetric tokamak configuration.
\par
In the axisymmetric limit the functions become,
\begin{subequations}
    \begin{equation}
        \mathcal{G}_{p_2}^\mathrm{axi}= -2\left[1+ \left(\frac{B_0}{R_0 I_2}\right)^2\frac{(1+\bar{\alpha})^2}{\bar{\alpha}+3}\left(1 +\frac{\mathcal{G}_{22}^C(k)}{4}\right)\right], \label{eqn:Gp_axi}
    \end{equation}
    \begin{equation}
    \mathcal{G}_{\delta}^\mathrm{axi}= -6 \left(\frac{1-\bar{\alpha}}{3+\bar{\alpha}} \right)-\frac{3-\bar{\alpha}}{3+\bar{\alpha}}\mathcal{G}_{22}^C(k),
    \end{equation} \label{eqn:G_delta_G_p2}
\end{subequations}
using the definitions in Eqs.~(\ref{eqn:G_funcs_2}). Here the parameter $\bar{\alpha}=(\eta R_0)^4=\mathcal{E}^2$ is the square of the elongation of the flux surfaces along the major radial direction. To arrive at such an expression we used the relation $\bar{\iota}_0=2\sqrt{\bar{\alpha}}G_0I_2/B_0^2(1+\bar{\alpha})$, which holds true for a tokamak, whose rotational transform must be fully  driven by a toroidal plasma current. 
\par
Let us analyse the behaviour of the finite pressure term first. It is clear from Eq.~(\ref{eqn:Gp_axi}) that $\mathcal{G}_{p_2}^\mathrm{axi}<-2$ for all $k$ and possible combinations of parameters, as $\mathcal{G}_{22}^C\geq-2$, and therefore $1+\mathcal{G}_{22}^C/4\geq1/2$. This negative sign of $\mathcal{G}_{p_2}^\mathrm{axi}$ indicates that the usual peaked pressure profile (i.e., $p_2<0$ for the assumed $\mathrm{sgn}(\psi)=+1$) leads to an increase in precession in the direction opposite to the diamagnetic frequency; a direct consequence of the magnetic well term discussed in the previous section. A finite $\beta$ (at fixed triangularity) assists in making the behaviour of trapped particles more maximum-$\mathcal{J}$. This is a well-known effect \cite{rosenbluth1971,connor1983effect}, referred to as the diamagnetic effect of the toroidal field. In fact, we note that in the circular tokamak limit, the expression becomes almost identical to the result of \citet{connor1983effect}, see the details in Appendix~\ref{app: relation to geometry}. { Although maximum-$\mathcal{J}$ is being approached by increasing $\beta$ and this is generally regarded as a positive effect, at an intermediate stage particle precession reaches $\omega_\alpha\sim0$ for many trapped particles. This slow precession scenario is generally associated to enhanced fast particle losses (especially of deeply trapped particles) whenever deviations from QS exist \citep{nemov2008poloidal,velasco2021model,bader2021modeling}.}
% Alongside this effect, and considering the role of pressure on the ideal-MHD limit, finite plasma-$\beta$ decreases MHD stability.
\par
Different trapped particles are affected differently by pressure as a consequence of the evolving Shafranov shift \citep{shafranov1963, wessonTok,rodriguez2023mhd}, which perturbs the field in a non-symmetric way. This underlying role of the Shafranov shift is clear from the contribution of the factor $f=B_0^2(1+\bar{\alpha})^2/R_0^2I_2^2(\bar{\alpha}+3)$, which shows an amplified effect of pressure for low currents (i.e., or low rotational transform), larger horizontal elongation or smaller major radius. All of these are known to increase the Shafranov-shift effect, and will enhance the counter-precession of particles with respect to the diamagnetic drift (see Fig.~\ref{fig:w_a1_gen}).
\par
Because to leading order deeply trapped particles co-rotate with the diamagnetic drift, we may estimate when the plasma $\beta$ is sufficient to reverse their direction. We interpret the resulting as the critical plasma $\beta$ for which the field becomes barely maximum-$\mathcal{J}$. Formally, this involves equating two different asymptotic orders, which goes against the very nature of the asymptotic treatment. One may nevertheless interpret this as an estimate of the precession at a `finite' radius.\footnote{One needs to be careful here, as increasing the pressure gradient will also increase the Shafranov shift, and with it the second order magnetic field shaping. This will start to incur on our asymptotic ordering. Anyhow, equating the different orders still results in a convenient measure.} This shows that one may try to increase the maximum-$\mathcal{J}$ behaviour of a QS by enpowering some of the higher order contributions (pressure and other shaping). In practice the effective radius in which the leading order is dominant may be small enough that we may refer to the field as maximum-$\mathcal{J}$. As shown in the examples of Figure~\ref{fig:numerical-verification}, though, this does not seem to be the natural tendency of QS fields, and certainly is not asymptotically. 
\par

Focusing on the behaviour of $k=0$ (such deeply trapped particles are typically the least maximum-$\mathcal{J}$), $\omega_{\bar{\alpha},0}=H\eta/rB_0$ from Eq.~(\ref{eqn: precession leading order}), and for the pressure we have $\omega_{\alpha,1}=-(H/B_0)\mu_0|p_2|(2+f)/B_0^2$. Equating the two, we find
\begin{equation}
    \frac{\mu_0 |p_2|}{B_0^2} \sim \frac{\eta}{r(2+f)} \sim \frac{\sqrt{\mathcal{E}}}{R_0 r(2+f)}.
\end{equation}
For a parabolic pressure profile, $a^2 p_2 = -p_0$, where $a$ is the minor radius and $p_0$ the pressure on axis, one finds that the critical plasma $\beta$ on axis is, 
\begin{equation}
    \beta_\mathrm{crit} \stackrel{\cdot}{=} 2a^2\frac{\mu_0 p_2}{B_0^2} \sim \frac{a^2}{R_0 r} \frac{2\sqrt{\mathcal{E}}}{2 + f}. \label{eqn:crit_beta}
\end{equation}
We thus see that the most susceptible fields are those with a large-aspect ratio, vertical elongation (small $\mathcal{E}$) and lower current (large $f$). As expected, these finite $\beta$ effects become more pronounced as we move away from the magnetic axis. { For further illustration, consider the scenario of a circular-shaped tokamak with a representative safety factor of $q=2$ and aspect ratio $\sim3$, for which at the edge $\beta_\mathrm{crit}\sim(a/R_0)/(1+q^2/2)\sim 11\%$. Reversing the precession of deeply trapped particles is thus predicted to require a significant plasma $\beta$.}
\par
\begin{figure}
    \centering
    \includegraphics[width=0.8\textwidth]{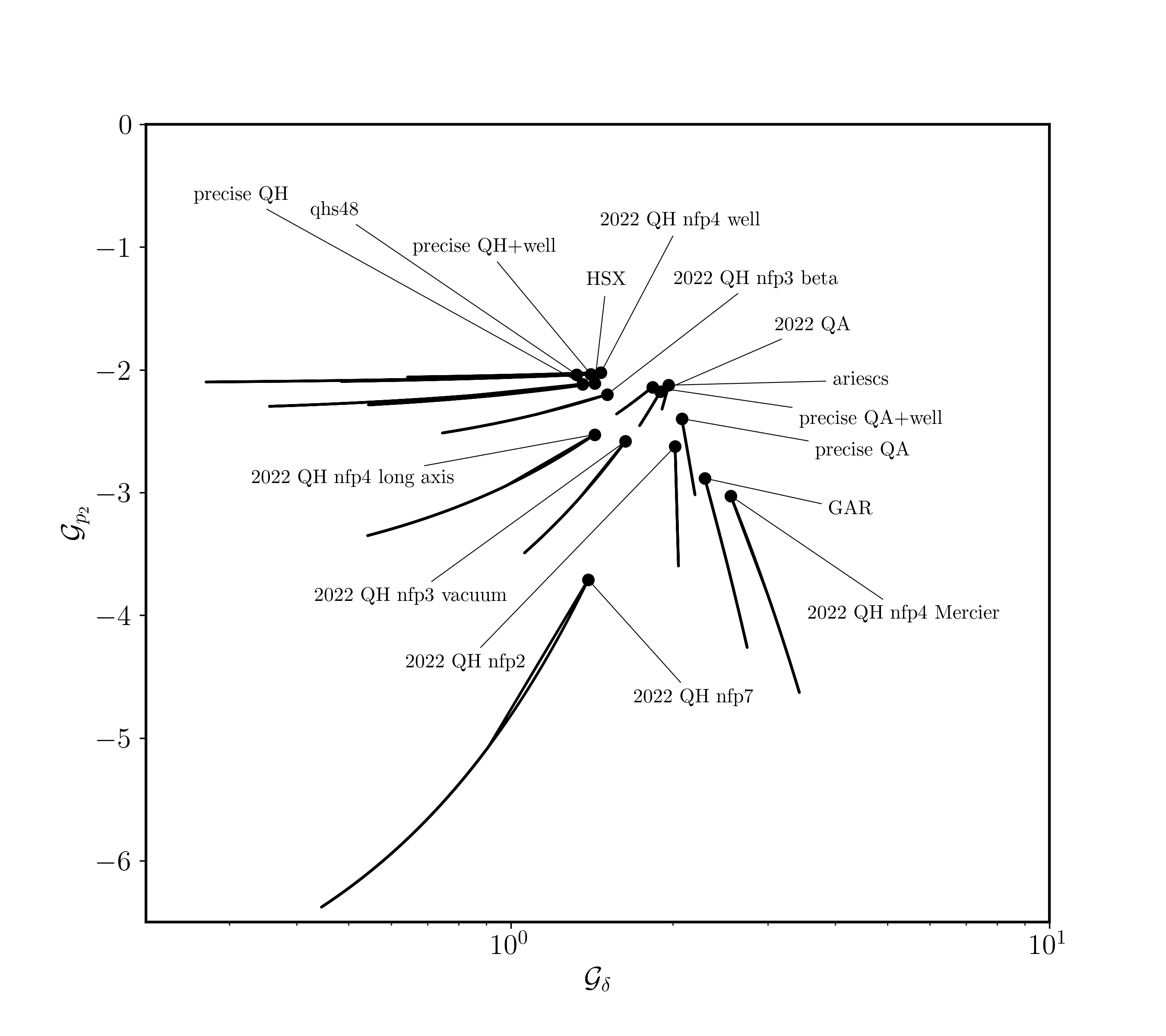}
    \caption{\textbf{Effect of triangularity and pressure gradient in the precession of trapped electrons in some QS configurations.} The plot shows the values of $\mathcal{G}_{\delta}$ and $\mathcal{G}_{p_2}$ for a number of quasisymmetric configurations in the ideal quasisymmetric limit represented by their most vertically elongated up-down symmetric cross-section (in some configurations this occurs at $\varphi=\pi/N$ rather than $\varphi=0$). The scatter plot corresponds to the values of both $k=0,~1$ for different configurations, while each segment represents the other $k$ values. { The near-axis models can be found in the acknowledged repositories; some more details, such as their QS quality, can be found in Appendix~\ref{sec:appCritEst}. } }
    \label{fig:qs_config_wa_1}
\end{figure}
The effects of triangularity are markedly more involved than those of pressure, which prevents us from writing a parameter-insensitive bound like we did for the effect of pressure (see Figure~\ref{fig:w_a1_gen}). Depending on the value of elongation, $\bar{\alpha}=\mathcal{E}^2$, the precession due to triangularity will tend to be in one direction or the other, a behaviour that also changes depending on the particle class considered. There exists, though, a critical value of $\bar{\alpha}$ beyond which $\mathcal{G}_\delta^{\mathrm{axi}}>0$ for all $k$. As $\mathcal{G}_{22}^C$ has a maximal value $\max(\mathcal{G}_{22}^C) \approx 1.1$, one can find that this critical point occurs at
\begin{equation}
    \bar{\alpha}_\mathrm{crit} > \frac{6 + 3 \max(\mathcal{G}_{22}^C)}{6 + \max(\mathcal{G}_{22}^C)} \approx 1.3.
\end{equation}
Thus, for tokamaks that are horizontally elongated (beyond some $\sim14\%$) negative triangularity tends to make all particles precess against the diamagnetic drift. 
This distinction regarding the effect of triangularity is reminiscent of the effects of triangularity on MHD stability. In that case, and describing stability through the Mercier criterion \citep{rodriguez2023mhd,freidberg2014,shafranov1983,lortz_nuhrenberg_1978ballooning,Solovev1970}), one can show that for $\bar{\alpha}>\bar{\alpha}_\mathrm{MHD}=1$, negative triangularity contributes positively to stability. Thus, MHD stability seems to align with precession against the diamagnetic drift, at least for sufficiently horizontally elongated configurations. That is, there is some synergy, which is the opposite to the changes due to plasma-$\beta$.
\par
Most commonly, though, most tokamak fields are vertically elongated, and thus have $\bar{\alpha}<1<\bar{\alpha}_\mathrm{crit}$. In that usual scenario different trapped particles respond differently, some tending to precess in one direction, others in the opposite. The most deeply and barely trapped particles are a special case , though, as taking $k=0,~1$, $\mathcal{G}_\delta= 4 \bar{\alpha}/(3 + \bar{\alpha})>0$ has a maximum value (see Fig.~\ref{fig:w_a1_gen}). With a sign independent of elongation, one can conclude that positive triangularity tokamaks will always tend to make deeply and shallowly trapped particles precess in the direction of the diamagnetic drift. Thus, only through negative triangularity can this shaping be used to effect in reversing the behaviour of deeply trapped particles. In the vertically elongated regime, negative triangularity hampers MHD stability, thus opposing the tendency to improve the maximum-$\mathcal{J}$ behaviour. { As noted in the plasma $\beta$ scenario, as precession of particles is reduced, fast ion confinement can be negatively affected in an imperfect QS stellarator. The different behaviour of each trapped particle makes an assessment of the overall effect complex.} 
\par
{ It would be appropriate, as we did to illustrate the effect of plasma $\beta$, to introduce the idea of a critical triangularity: a value of triangularity such that one expects precession of deeply trapped particles to leading order to be significantly affected, i.e. $r \delta\mathcal{G}_\delta(k)/2 \sim G(k)$ for $k=0$. Precisely as in the case of pressure,
\begin{equation}
    (r\delta)_\mathrm{crit}\sim\frac{3+\bar{\alpha}}{2\bar{\alpha}}. \label{eqn:crit_delta}
\end{equation}
The interpretation of $r\delta$ as triangularity of the cross-section in the Frenet frame of the magnetic axis (see Appendix~\ref{app: relation to geometry}) is an asymptotic concept. As such, this geometric interpretation of $r\delta$ ceases to be accurate upon approaching unity (especially as a significant bean-shape is developed). This limit of large $r\delta$ is nevertheless a limit of strongly shaped flux surfaces, which could even describe surfaces that self-intersect or intersect with other flux surfaces\citep{landreman2021a, rodriguez2023mhd}. To make sense of whether $(r\delta)_\mathrm{crit}$ is feasible in practice, we should compare this measure to the maximum triangularity achievable without incurring into these geometric defects. The critical $r_c$ was defined in \cite{landreman2021a}. In any case, Eq.~(\ref{eqn:crit_delta}) indicates that a very significant triangularity (of order 1) is needed to significantly affect precession of deeply trapped particles in a tokamak. Hence we conclude that, although triangular shaping may help in achieving the maximum-$\mathcal{J}$ property together with finite $\beta$ effects, achieving it via shaping alone is more difficult in tokamaks.}
\par
Before concluding this section, let us briefly consider the full quasisymmetric case, {beyond the special case of axisymmetry}, through some examples. Unlike in the axisymmetric case, this general scenario preserves a measure of symmetry-breaking of the geometry. The measure $\bar{F}$ \citep{rodriguez2023mhd}, for which explicit expressions are presented in Appendix~\ref{app: relation to geometry}, depends on the shape of the axis  and modifies the effects of pressure and triangularity. Reminding ourselves that in such a scenario $\delta$ represents the triangularity of the up-down symmetric cross-section with the largest binormal elongation, we show in Fig.~\ref{fig:qs_config_wa_1} the values of $\mathcal{G}_\delta$ and $\mathcal{G}_{p_2}$ for a number of QS configurations.\footnote{Many of the configurations in Figure~\ref{fig:qs_config_wa_1} were analysed for their MHD behaviour in \cite{rodriguez2023mhd}, and thus make it a suitable set for discussion. We note that in Figure~5 of \cite{rodriguez2023mhd}, the configuration named `2022 QA' was represented through its less vertically elongated cross-section. This is not wrong per se, as one may represent the configuration by any of its cross-sections (and in the framework in the paper, any up-down symmetric one). However, for a discussion on the effect of bean shapes, as in said paper, it was not the appropriate choice, and we point that out here (may it serve as erratum). That this inappropriate choice of a cross-section was made may be seen from an analysis of its cross-sections (see \cite{landreman2022map}), or from the unusual value of $\bar{F}$ in Table~1 in \cite{rodriguez2023mhd}. Here we consistently consider its vertical cross-section (equivalently, the $\varphi=0$ cross-section of the rotated configuration). All other configurations are similarly represented by their most vertically elongated cross-section for consistency.}  Each segment consists of pairs $(\mathcal{G}_\delta,\mathcal{G}_{p_2})$ for different $k$ for the same configuration, taking the ideal QS limit of the configurations (which are only approximately so).
\par
From the plot it is clear that, for those quasiymmetric configurations analysed, the effect of a finite pressure gradient is the same as in the axisymmetric limit (i.e., an increase in pasma $\beta$ leads to precession in the direction opposite to the diamagnetic drift). The effect of triangularity is analogous to a tokamak that is elongated in the horizontal direction, where negative triangularity pushes particles particles against the diamagnetic drift. From the MHD stability analysis of these configurations in \cite{rodriguez2023mhd}, one recovers the synergy of horizontally elongated tokamaks for quasisymmetric stellarators. Thus, the behaviour is quite different from that of regularly shaped tokamaks. 
\par
{ An interpretation of the magnitudes of $\mathcal{G}_{\delta}$ and $\mathcal{G}_{p_2}$ may be provided by considering the critical $\beta$ and $r\delta$ once more. As in the tokamak scenario, at the edge of the configuration $\beta_\mathrm{crit}\sim 2a|\eta|/\mathcal{G}_{p_2}(0)$ and $(r\delta)_\mathrm{crit}\sim2/\mathcal{G}_\delta$. A complete table for the configurations in Fig.~\ref{fig:qs_config_wa_1} is included in Appendix~ \ref{sec:appCritEst}. We note here that in QA configurations, $\beta_\mathrm{crit}\sim5\%$, while QH stellarators generally exhibit a more resilient behaviour with $\beta_\mathrm{crit}\sim 10-15\%$. Reversing the precession of deeply trapped particles via finite $\beta$ effects thus appears to be a possibility most easily in QA configurations. Given the simple form of $(r\delta)_\mathrm{crit}$, it is straightforward to see that $O(1)$ triangularity is generally required to observe a significant effect. In many of these configurations such values are indeed achievable without incurring in forbidden flux surface shapes (see Appendix~\ref{sec:appCritEst}).}

\subsection{Verification of the expansion}
In the preceding sections we investigated the analytical behaviour of the trapped particle precession. We derived these under two important assumptions: (i) exact quasisymmetry (or symmetry) of the fields, and (ii) the near-axis expansion. It is thus natural to wonder how close trapped particle precession are to the idealised limit in realistic configurations, as these assumptions become increasingly less accurate. We check this through three numerical examples, in which we compare the bounce-averaged drift computed from a global MHD code (in this case \texttt{VMEC} \citep{hirshman1983}, using numerical methods detailed in \citet{mackenbach2023bounce}) against the analytical results. For this practical comparison, we employ the definition of $k$ in terms of the bounce point, Eq. \eqref{eqn:def-k2}, which we compute for the numerical precession calculation. Note that upon significant deviations from QS this choice ceases to be convenient, especially when there exist differently shaped wells within a flux surface.\footnote{ In the more general case it is then more convenient to group precession through the class label $\lambda$, normalised as in \citet{roach1995trapped}, namely $k^2 = (1-\lambda B_\mathrm{min})B_\mathrm{max}/(B_\mathrm{max}-B_\mathrm{min})$. This maintains the 0 and 1 values for deeply and barely trapped particles respectively. This definition is only equal to the natural near-axis definition to leading asymptotic order. Thus, in that case, a comparison would require the alternative, yet asymptotically equivalent, definition of $k$ in \eqref{eqn:def_lam_k_hat}, which defines it in terms of $\lambda$.}
\par 
\begin{figure}
    \centering
    \includegraphics[width=0.9\textwidth]{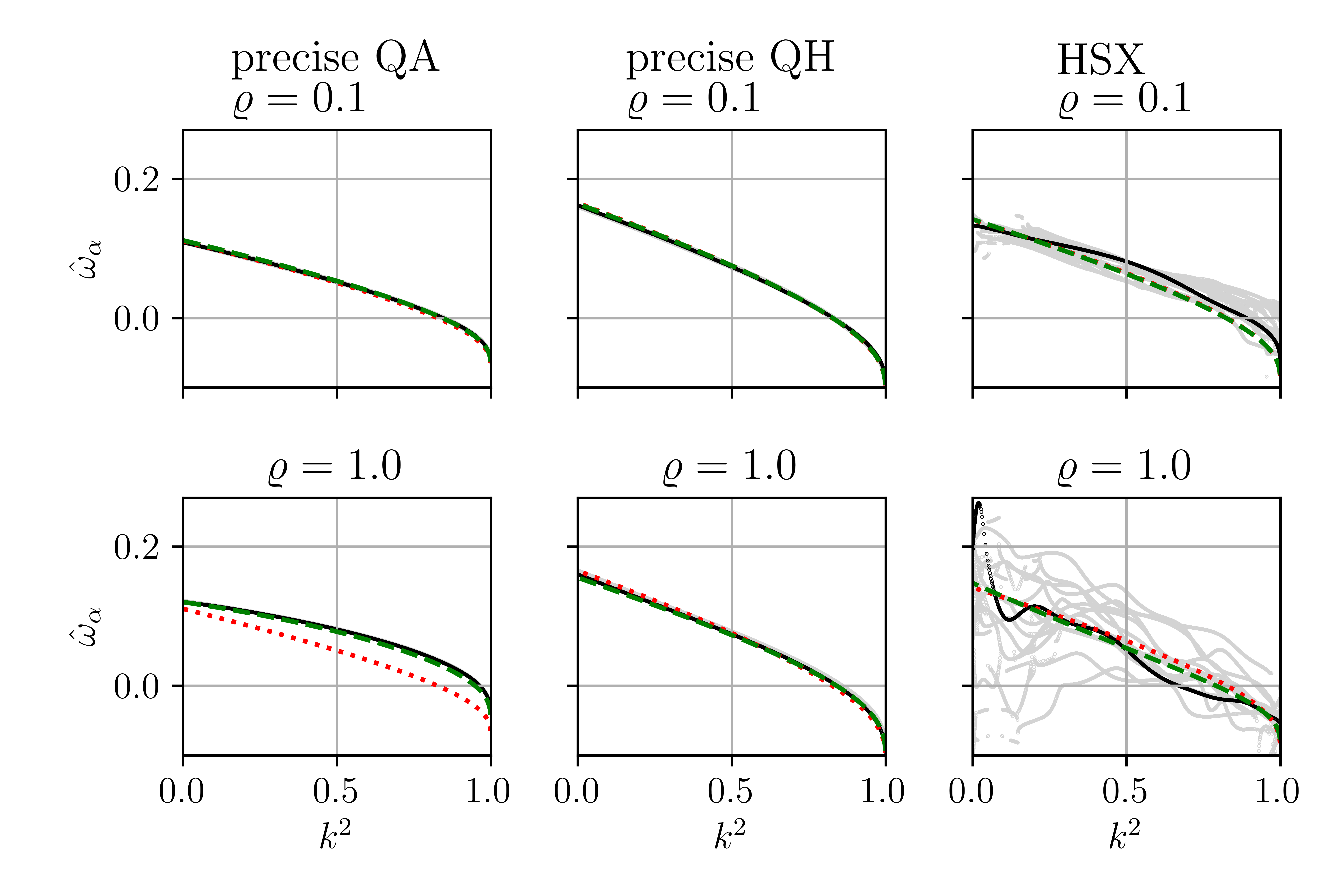}
    \caption{{\bf A comparison between the analytical and numerical precession.} The precession $\omega_\alpha$ for three QS fields and two radial positions $\varrho=\sqrt{\psi/\psi_\mathrm{edge}}$ is presented, normalized to $H / r \sqrt{2 B_0 \psi_{\mathrm{edge}}}$, where $a$ is the minor radius of the configuration, $B_0$ the $B$ on axis, and $2 \pi \psi_{\mathrm{edge}}$ the total toroidal flux. There is good correspondence for the precisely quasisymmetric configurations \citep{landreman2022magnetic}, and less so for HSX, who exhibits important deviations from QS. The black and grey lines are the numerical result of precession for different field lines (black corresponding to $\alpha=0$), whereas the dashed red and green lines are the first and second order analytic precessions respectively.}
    \label{fig:numerical-verification}
\end{figure}
% \begin{figure}
%     \centering
%     \includegraphics[width=0.9\textwidth]{figures/precession_comparison_shaded.png}
%     \caption{{\bf A comparison between the analytical and numerical precession.} The precession $\omega_\alpha$ is normalized to $H a/r B_0$, where $a$ is the minor radius of the configurations and $B_0$ the average $B$ on axis. We see good correspondence for the precisely quasisymmetric configurations \citep{landreman2022magnetic}, especially the QA, and less so for HSX, who exhibits important deviations from QS. The black lines are the numerical result at $\alpha=0$, the grey lines represent $\omega_\alpha$ at other field lines, and the dashed red and green lines are the first and second order analytic precession respectively.}
%     \label{fig:numerical-verification}
% \end{figure}
This comparison is shown in Fig.~\ref{fig:numerical-verification}, where the bounce-averaged precession is plotted as a function of $k$ and for two radial locations, $\varrho \stackrel{\cdot}{=} \sqrt{\psi/\psi_\mathrm{edge}}$. The correspondence is excellent for all displayed $\varrho$ in the precise quasisymmetric configurations, recently presented by \citet{landreman2022magnetic}, which are characterised for having an excellent degree of quasisymmetry. The theory works remarkably well even at larger radii, where one would expect the near-axis expansion to falter, although this near-axis nature of the configurations had been previously noticed \citep{rodriguez2022quasisymmetry}, and could be a feature necessary to achieve excellent global quasisymmetry. This numerical comparison evidences that the second order calculation is also appropriate. For HSX \citep{hsx1995}, we see more significant deviations from the analytical result. This is mainly a consequence of the breaking of QS (for a naive fitting of its near-axis behaviour, $B_{20}$ variations are $\sim 4$), and significant shaping beyond the near-axis behaviour (see $\varrho=1$). Although the trends and magnitude seem correct, there are clear deviations from the idealised limit.
% \par 
% It is perhaps more surprising that precise QA and QH perform well even at $\varrho=1.0$, where the correspondence is expected to be the worst as higher order effect come into play. This may partly be a consequence of the inverse aspect ratio, which we denote as $\varepsilon$, being $1/6$. Hence at $\varrho=1$ one has roughly $a \eta \sim a/R_0 \sim \varepsilon \approx 1/6$. In equation \eqref{eq: second order precession}, we realise that one can extract a factor $r \eta$, and we shall interpret this as an expansion parameter. One should then expect good correspondence up to $(1/6)^2 \approx 3\%$. Of course, the true error depends on the specifics of the higher order terms and their dependencies, but the current argument does aid in estimating the magnitude of the error.

\section{Energy available for trapped particle modes} \label{sec:AE}
The preceding study of particle precession was strongly motivated by its central role in driving trapped-particle modes (TPM). In essence, TPM turbulence arises driven by the trapped-particle precession when it resonates and destabilises the diamagnetic drift wave. Simply put, whenever the trapped particles co-drift with the diamagnetic drift wave, energy may be transferred to the drift wave, driving the instability thereof. As we have learnt, a special feature of axisymmetric and quasisymmetric fields is that $\omega_\alpha$ has, to leading order, a zero crossing. That is, there always exists a subgroup of trapped particles (which includes deeply trapped particles) that co-rotate with the drift wave, and thus potentially drive TPMs. To assess how the size of this group and the magnitude of its precession feeds TPMs, though, a more qualitative treatment is necessary. To perform such an analysis, we delve into the stability of TPMs by studying the available energy of the trapped particles \citep{helander2017available,helander2020available}.
\par
Available energy (\AE{}) is an upper bound on the free energy available to the plasma after a constrained rearrangement of the distribution function (called Gardner restacking, see \citet{kolmes2020available}), rearrangement that needs to conserve certain dynamical quantities like $\mathcal{J}_\parallel$. Restricting ourselves to the available energy contained in trapped particles, one obtains a proxy measure of non-linear turbulent activity of TPMs (and upon specialising to electrons, TEMs \cite{proll2012resilience}). This is, nevertheless, a simplified description of the turbulence, in the best of cases a bound, which does not include a discussion of accessibility. That is, although energy may be available, it might not be possible for a system to evolve to such lower energy state and access the available energy, thus the turbulence activity would be over-estimated. The \AE{} measure is nevertheless useful \citep{mackenbach2022available}, and it provides insight into TPMs.
\par
The calculation of available energy for trapped electrons in a flux-tube was recently presented by \cite{mackenbach2022available,mackenbach2023transport}. For a flux tube of length $L$ and elliptical cross section $\Delta \alpha$ and $\Delta \psi$ in $(\psi,\alpha)$-space (principal axes), the available energy in an omnigeneous ($\omega_\psi=0$) field may be written as,
\begin{equation}
\begin{aligned}
    A = \frac{1}{2 \sqrt{\upi}} \frac{\upi \Delta \psi \Delta \alpha L}{B_0} n_0 T_0 \iint & \sum_{\text{wells}(\lambda)} e^{-z} z^{5/2} \hat{\omega}_\alpha^2 \mathcal{R} \left[ \frac{\hat{\omega}_*^T}{\hat{\omega}_\alpha} - 1\right] \hat{G}^{1/2} \mathrm{d}z \mathrm{d}\lambda,   \label{eqn:AEinteg}
\end{aligned}
\end{equation}
where $\hat{G}^{1/2}=\int (1 - \lambda \hat{B})^{-1/2} \mathrm{d} \ell/L $ is the normalized bounce-time, $\mathcal{R}$ is the ramp function (indicating that only faster than the diamagnetic drift co-precessing particles contribute to $A$), and the hats denote normalisation of the frequencies $\hat{\omega}=\Delta \psi ~\omega/H$. The integral is performed over $z=H/T$ and $\lambda$, the combination of which constitute all trapped particle energies and classes (the exponential in the integrand is a consequence of the chosen distribution function of which the \AE{} is to be calculated, a Maxwellian). The sum over wells simply indicates that the available energy is the result of adding the contributions from every well along the flux tube, as trapped particles may be rearranged within each. 
% \footnote{The length of the flux tube $L$ is defined as $\int\mathrm{d}l=\int\mathcal{J}B\mathrm{d}\phi\sim B_{\alpha0}\Delta\phi/B_0$ where $\Delta\phi/2\pi$ is the number of toroidal transits considered. [{What is the appropriate length measure, is it the characteristic $1/k_\parallel$, roughly a connection length? Could change the scaling. From (7) in the PRL it seems that this is arbitrary.}]}  
\par
Of course, the amount of energy available depends on the size of the flux tube consider; the measure is extensive. To construct an intensive measure we normalise it to the total thermal energy available in the tube. The total thermal energy in the flux-tube for a plasma of temperature $T_0$ and density $n_0$ is (using $\mathbf{B}\cdot\nabla\ell=B$)
\begin{equation}
    E_t = \int \frac{n T}{B} \mathrm{d}\psi \mathrm{d}\alpha \mathrm{d}\ell \approx n_0 T_0 \frac{\upi \Delta \psi \Delta \alpha L}{B_0} \int \frac{1}{\hat{B}} \frac{ \mathrm{d} \ell}{L}.
\end{equation}
Hence the normalized \AE{} becomes
\begin{equation}
    \widehat{A} \equiv \frac{A}{E_t} = \left( \int \frac{2 \sqrt{\pi}}{\hat{B}} \frac{\mathrm{d}\ell}{L} \right)^{-1} \iint \sum_{\text{wells}(\lambda)} e^{-z} z^{5/2} \hat{\omega}_\alpha^2 \mathcal{R} \left[ \frac{\hat{\omega}_*^T}{\hat{\omega}_\alpha} - 1\right] \hat{G}^{1/2} \mathrm{d}z \mathrm{d}\lambda\label{eqn:AEnorm},
\end{equation}
where the normalising factor in front will henceforth be succinctly referred to as $\mathcal{V}=2\sqrt{\pi}\int\mathrm{d}\ell/L\hat{B}$. To make further progress we realize that the integral over $z$ is analytically tractable if one splits up $\hat{\omega}_*^T$ as
\begin{equation}
    \hat{\omega}_*^T=\hat{\omega}_{*,0}^T/z+\hat{\omega}_{*,z}^T,
\end{equation}
where $\hat{\omega}_{*,z}^T=-\Delta\psi\partial_\psi\ln T$ and $\hat{\omega}_{\star,0}^T=-\Delta\psi(\partial_\psi\ln n-\frac{3}{2}\partial_\psi \ln T)$. To ease the calculation (although it may be extended to the more general case), we shall take the temperature gradient to be zero and consider the limit of a peaked density profile (i.e. $\partial_\psi \ln n$ is negative for $\psi>0$); we are specialising to density-driven trapped-particle instabilities. This assumption makes the diamagnetic drift $\omega_\star$ particle-energy independent, leading to,
\begin{equation}
    \widehat{A}=\frac{1}{\mathcal{V}}\int\mathrm{d}\lambda \sum_\mathrm{wells(\lambda)}(\hat{\omega}^T_{*,0})^2\hat{G}^{1/2}\mathcal{F}(c_1)\Theta(\hat{\omega}_\alpha), \label{eqn:AEinteg}
\end{equation}
where,
\begin{equation}
    \mathcal{F}\left(c_1=\frac{\hat{\omega}_{*,0}^T}{\hat{\omega}_\alpha}\right)= \frac{ 2 \sqrt{c_1} \left( 15 + 4 c_1 \right)\exp(-c_1) + 3 \sqrt{\pi} (2 c_1 - 5) \mathrm{erf}(\sqrt{c_1}) }{8 c_1^2}, \label{eqn:F_ae_def}
\end{equation}
and the $\Theta$ function is a Heaviside function that vanishes for $\omega_\alpha(\lambda)<0$, which physically represents the inability of counter-rotating particles to drive the TPM.
\begin{figure}
    \centering
    \includegraphics[width=0.6\textwidth]{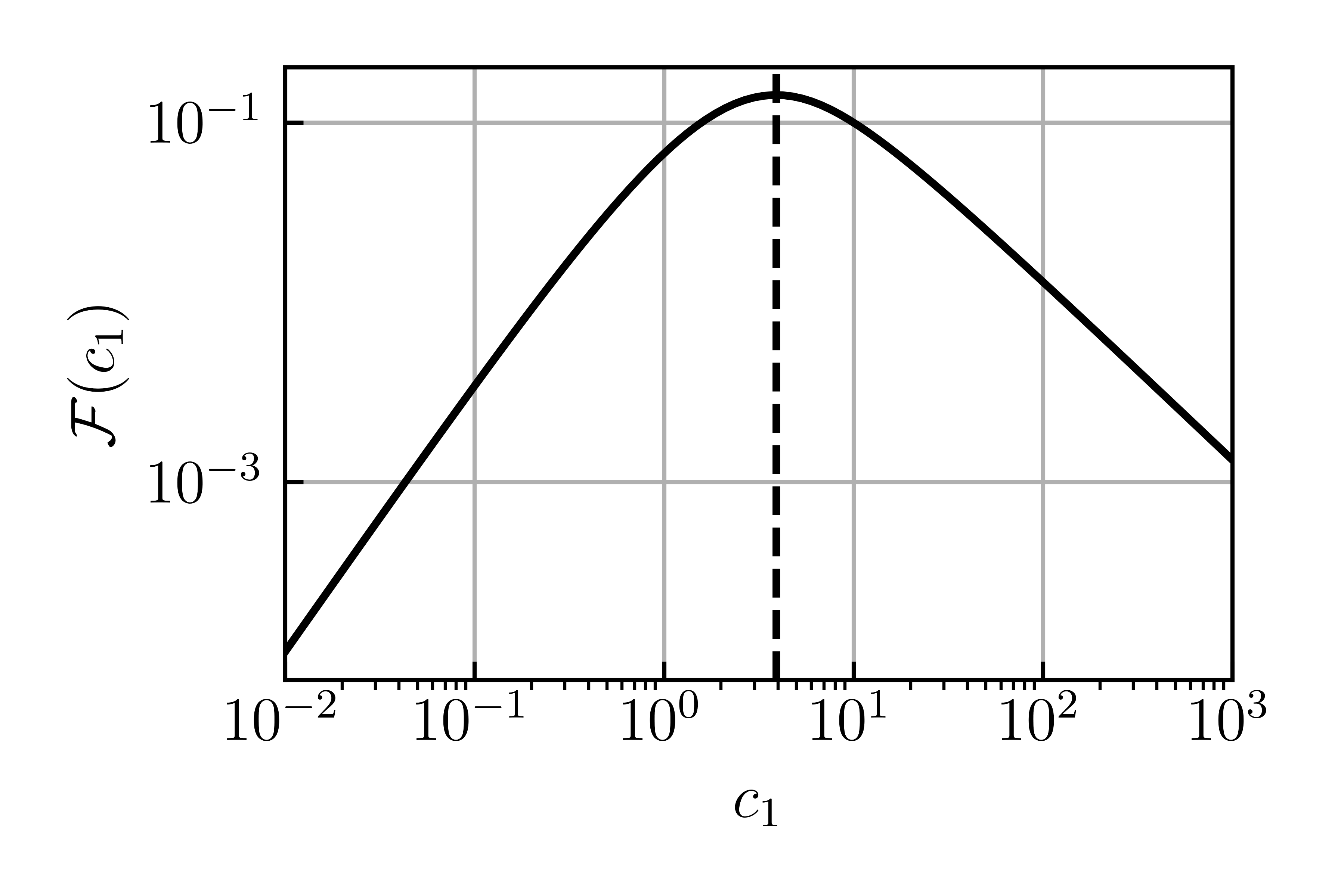}
    \caption{\textbf{Plot of the function weighing the contribution to \AE{} of various trapped particle classes.} Plot of $\mathcal{F}$ as a function of the diamagnetic to precession drifts $c_1$. The dashed line indicates the value $c_1 \approx 3.9$, which corresponds to the maximum of $\mathcal{F}$, and in a sense represents the subset of particles that most vigorously resonate and drive the drift wave.}
    \label{fig:Fc1}
\end{figure}
\par 
The function $\mathcal{F}$, see Fig.~\ref{fig:Fc1}, may be interpreted as a measure of the coupling of different particle classes to the available energy (ignoring further contributions from the normalized bounce-time). With the energy dependence averaged out after the integral over $z$, the comparison between the diamagnetic drift and precession  is captured by $c_1$. Both trapped particles that are drifting too fast (i.e. $|c_1| \ll 1$) and which are drifting too slow (i.e. $|c_1| \gg 1$) as compared to the drift wave have a vanishing contribution to the \AE{}, as $\mathcal{F}\rightarrow0$. This is a formal statement of the resonance requirement for an effective drive of the drift-wave instability, where the coupling is largest at $c_1 \approx 3.9$. 
\par
To proceed further, and since the expressions for the precession derived in the preceding section depend on the trapping parameter $k$ explicitly, it will be natural to write \AE{} in Eq.~(\ref{eqn:AEinteg}) as an integral over $k$. 

\subsection{Leading order form of \AE{}}
Let us begin the asymptotic analysis by considering the first order expression of the trapped-particle precession $\omega_\alpha\approx\omega_{\alpha,0}(k)$, defined in Eq.~\eqref{eqn: precession leading order}. To perform the integral in Eq.~(\ref{eqn:AEinteg}) we need a number of ingredients. First of all, we must consider the integral only over trapped-particle classes that co-rotate with $\hat{\omega}_{*,0}^T$ (i.e., the range allowed by the Heaviside). The domain of integration thus runs from the most deeply trapped particles to the transition point of $\omega_{\alpha,0}$, i.e. $k \in [0,k_0]$ (with the definition of $k_0|\omega_{\alpha,0}=0$ from before).
\begin{figure}
\centering
\begin{subfigure}{.5\textwidth}
  \centering
  \includegraphics[width=0.9\textwidth]{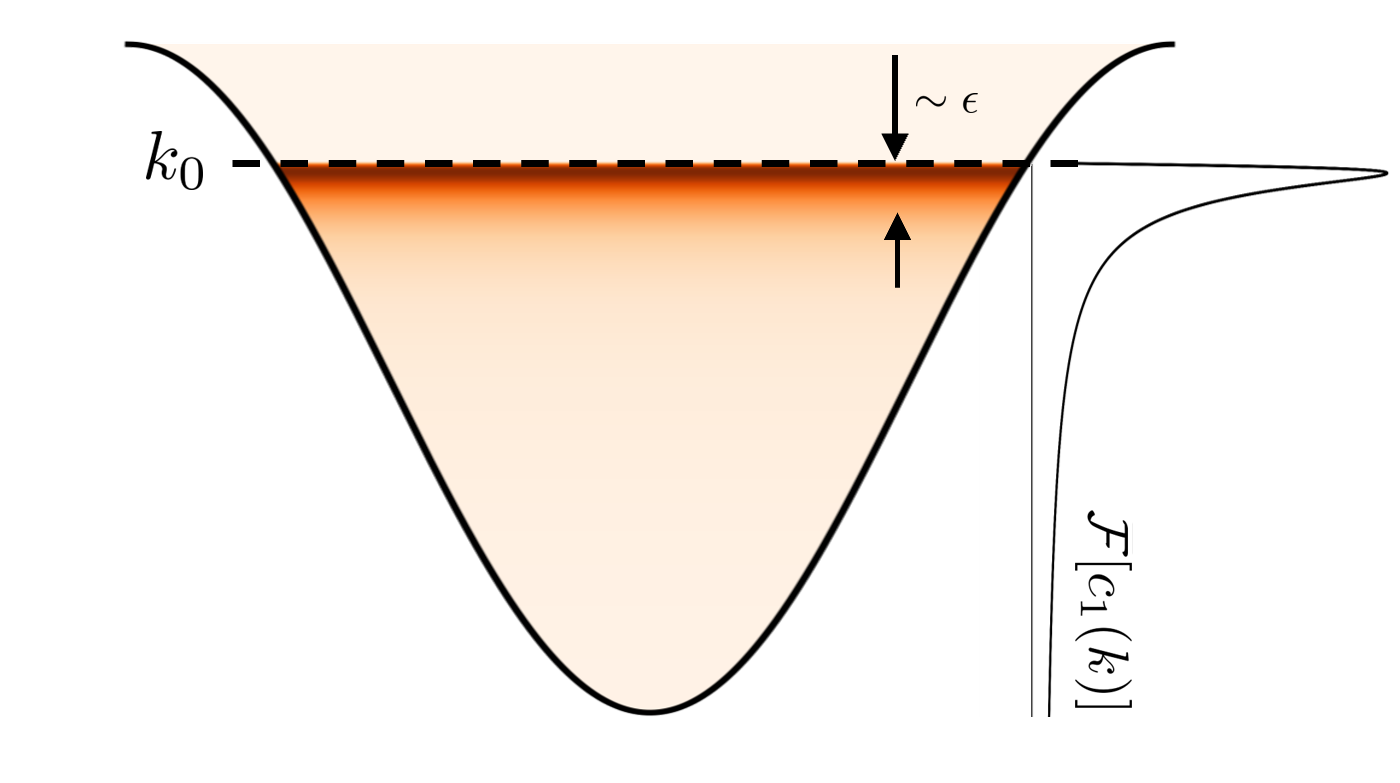}
  \caption{A schematic of the \AE{} distribution}
  \label{fig:sub1}
\end{subfigure}%
\begin{subfigure}{.5\textwidth}
  \centering
  \includegraphics[width=\textwidth]{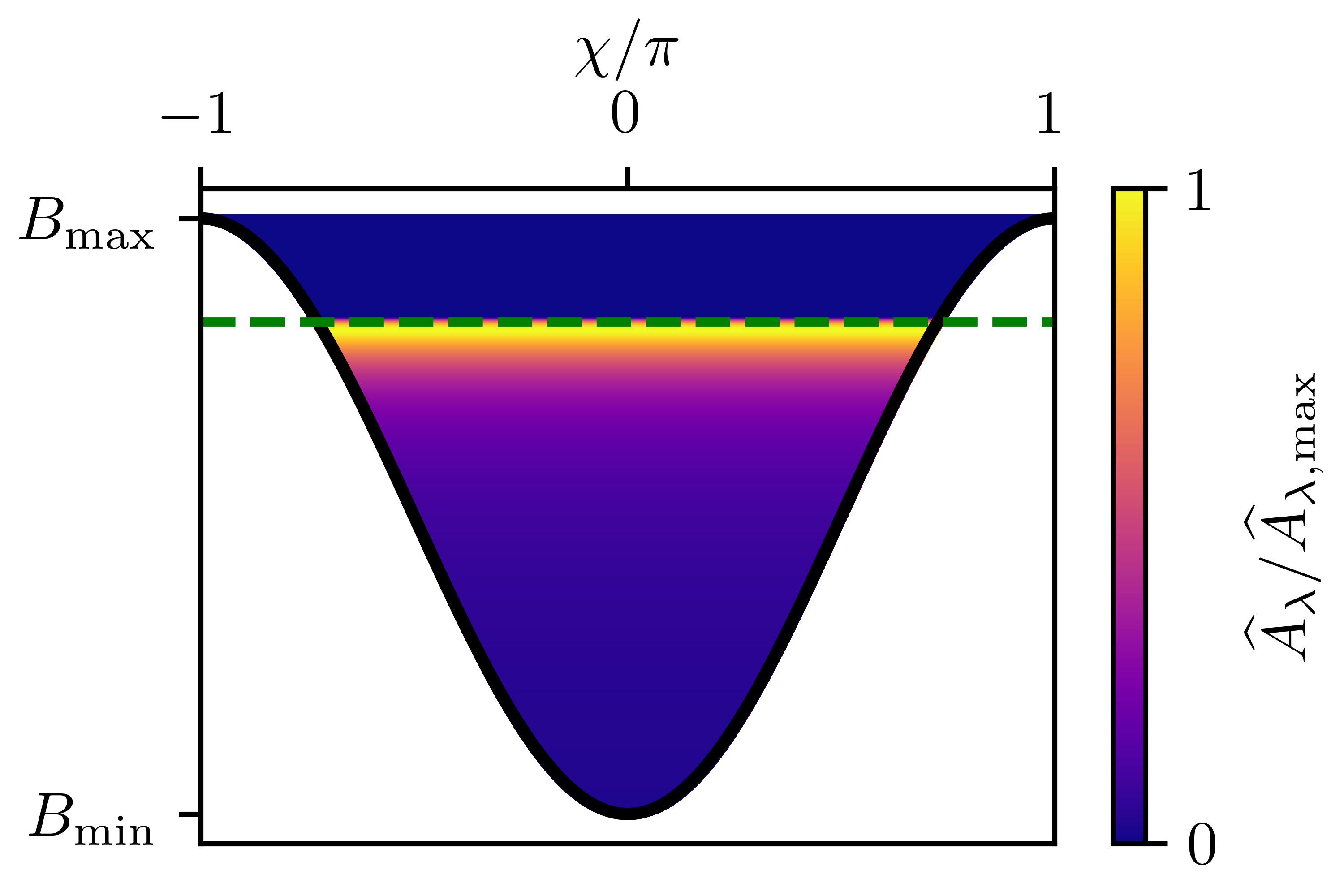}
  \caption{A numerical calculation of \AE{} distribution.}
  \label{fig:sub2}
\end{subfigure}
\caption{\textbf{Schematic of the contribution to available energy.} Left: diagram showing a narrow band of trapped particles near $k_0$ (the trapped-particle class with vanishing precession) contributing to the available energy. The broken line indicates $k_0$ in the limit of $ r\rightarrow0$, as the vertical direction denotes different classes of particles with the leading order magnetic well structure shown by the black curve. The plot of $\mathcal{F}[c_1(k)]$ is shown to the right for $ r\omega_{*,0}^T/\eta=0.1$ as an example. Right: a numerical calculation showing the distribution of \AE{} across the magnetic well, normalized by the total \AE{}. Plotted for the precise QA device at a minor radial coordinate of $r=10^{-3}$. The points where $\omega_\alpha=0$ is denoted by a green dashed line, and $\hat{\omega}_n = 0.1$.}
\label{fig: AE distribution}
\end{figure}
\par
The second ingredient that naturally arises is then the need to express the integration variable $\lambda$ in terms of $k$. The presence of $c_1=\hat{\omega}_{*,0}^T/\hat{\omega}_{\alpha,0}(k)$ as a function of $k$ inside $\mathcal{F}$ makes the integral highly nested, and thus appears to make it difficult to approach analytically. However, given the form of the precession, $c_1$ is asymptotically small in the sense that $c_1\sim r$. This appears to offer a way to proceed by expanding $\mathcal{F}$ in the small $c_1$ limit. However, that would be wrong, as it would neglect the most important contribution to \AE{}. Recall that the particle precession vanishes at $k_0$, and thus near this value of $k$ the function $c_1$ cannot be considered small. This resonant behaviour is, in the asymptotic limit, the principal contributor. The integral is significant only in a narrow region $\Delta k\sim r$, close to $k_0$, where $c_1 \sim O(1)$ (see a clarifying sketch in Fig.~\ref{fig: AE distribution}). This teaches us that in this asymptotic limit, the most important class of particles are those with relatively small precession, as in this limit $\omega_\alpha$ tends to be much larger than the diamagnetic drift. The evaluation of the integral may then be carried out analytically considering a local approximation of the integrand in this contributing narrow band (correct down to a correction $O(r^{1/2})$ on the leading contribution), the details of which are presented in Appendix~\ref{sec:appAEdet}. 
% and the resulting value of available energy is, 
% \begin{equation*}
%     \widehat{A} \approx \frac{2\sqrt{2}}{ 9 \pi} \left( \hat{\omega}_{*,0}^T \right)^3 \frac{r B_0}{\Delta \psi} \sqrt{\frac{r}{\eta}}.
% \end{equation* 
\par
Before writing the result for the \AE{} out, one last consideration is required. In this case, one needs to make an explicit assumption regarding the width of the flux tube $\Delta\psi$, which the \AE{} will depend on. This width should be interpreted as the `length' over which density gradients may be flattened by the turbulence to extract energy. Following the steps taken by \citet{mackenbach2022available,mackenbach2023transport}, we estimate such a flattening length-scale to be the correlation length, and to be on the order of the Larmor radius $\rho$. As such, we write
\begin{equation}
    \Delta \psi = B_0 r \Delta r = B_0 r C_r \rho
\end{equation}
where $C_r$ may formally be dependent on other equilibrium parameters (e.g., the rotational transform $\iota$ or the flux expansion $|\nabla\psi|$). We shall nevertheless take $C_r$ to be a constant, assuming that the flattening length-scale providing free energy to the TPMs is simply proportional to the Larmor radius. 
\par
It is customary to define a minor radius $a$ for the field configuration, so that the radial coordinate may be normalised as $\varrho = r/a$. This way, we define the density gradient $\hat{\omega}_n \equiv - a \partial_r \ln n = - \partial_\varrho \ln n$, which scales like $\varrho$ for a quadratic (in $\varrho$) density profile. In terms of these variables, the \AE{} becomes
\begin{equation}
    \widehat{A}_\mathrm{w} \approx  \frac{2\sqrt{2}}{ 9 \pi} C_r^2 \rho_*^2 \left(\frac{\hat{\omega}_n}{\varrho}\right)^3  \frac{\varrho^3\sqrt{\varrho}}{\sqrt{a \eta}}
    \label{eqn:AE_lead}
\end{equation}
where the common gyrokinetic expansion parameter is defined as $\rho_* \stackrel{\cdot}{=} \rho/a$. 
\par
The leading order expression for $\widehat{A}_\mathrm{w}$ includes important information regarding the behaviour of the available energy near the axis. We highlight various scalings here:
\begin{enumerate}
    \item One observes a strong scaling with the distance from the axis $\varrho$, whose origin may be presented in simple terms as follows (see the integral expression in Eq.~(\ref{eqn:AEinteg}) for reference). One factor of $\varrho^{1/2}$ may be accounted for due to trapping fraction of particles, which leaves one with an overall $\varrho^3$ scaling. In this scenario, the energy scale is set by the diamagnetic drift (as only precessing particles with speeds analogous to those of the diamagnetic drift contribute to the available energy), which goes like $\hat{\omega}_n\propto\varrho$ near the axis. Thus, two powers of $\varrho$ can be argued to come from the kinetic drive of the diamagnetic drift.
    % A different way of seeing this is that one is able to access the energy available in flattening the density gradient, which goes like $\Delta\psi \omega_{*,0}^T\sim \varrho^2$. 
    The final power of $\varrho$ comes from the small fraction of trapped particles that contribute to the the available energy, as the band near $k_0$ scales with $\varrho$.
    \item Another scaling of interest in Eq.~(\ref{eqn:AE_lead}) is its dependency on the stellarator shaping parameter $\eta$. Increased $\eta$ leads to a more restricted access to energy and thus, a reduced TPM activity (as measured by \AE{}). Thus, horizontally elongated shapes would seem to favour stability, and in the context of quasisymmetric stellarators, quasi-helically symmetric configurations. In tokamak terms, $a\eta \sim a\sqrt{\mathcal{E}}/R_0\sim\varepsilon\sqrt{\mathcal{E}} $, where $\varepsilon $ is the inverse aspect ratio. Thus, $\widehat{A}_\mathrm{w} \sim 1/(\mathcal{E}^{1/4}\sqrt{\varepsilon})$. Hence, the available energy increases with aspect-ratio keeping the minor radius fixed. This dependency becomes even stronger if one chooses $\rho_* \varepsilon = \rho/R_0$ to be constant.
    \item One finally observes a scaling with $(C_r \rho_*)^2$, which is the square of the assumed length-scale over which energy is available. Importantly, we note that an implicit magnetic field-strength dependency arises here (for fixed minor radius and $C_r$), as $\rho \sim 1/B_0$. Hence, in terms of \AE{}, it is beneficial to have a strong magnetic field so that the length-scale over which energy is available decreases as $\widehat{A}_\mathrm{w} \sim 1/B_0^2$.
\end{enumerate}
\par

As noted before, a full interpretation of these scalings for a comparison between different configurations would have to take into account the form of $C_r$ that most closely describes the volume that is accessible to rearrangment of energy. This may be particularly important when proceeding to a comparison between highly different configurations. The normalisation with $C_r$ being a constant appears to provide, though, a reasonable description in \AE{} as a measure of heat transport in practice (see \cite{mackenbach2022available}).\footnote{An appealing alternative to this constant value would perhaps be to consider the poloidal Larmor radius instead of the Larmor radius as the appropriate scale of the flux tube. In that case, we would have an additional factor of aspect ratio and rotational transform, which would change the comparison of behaviour between different fields. Which form is most appropriate is not a closed question.}

\subsection{A strongly-driven finite-radius asymptotic regime}
\begin{figure}
\centering
\begin{subfigure}{.5\textwidth}
  \centering
  \includegraphics[width=0.9\textwidth]{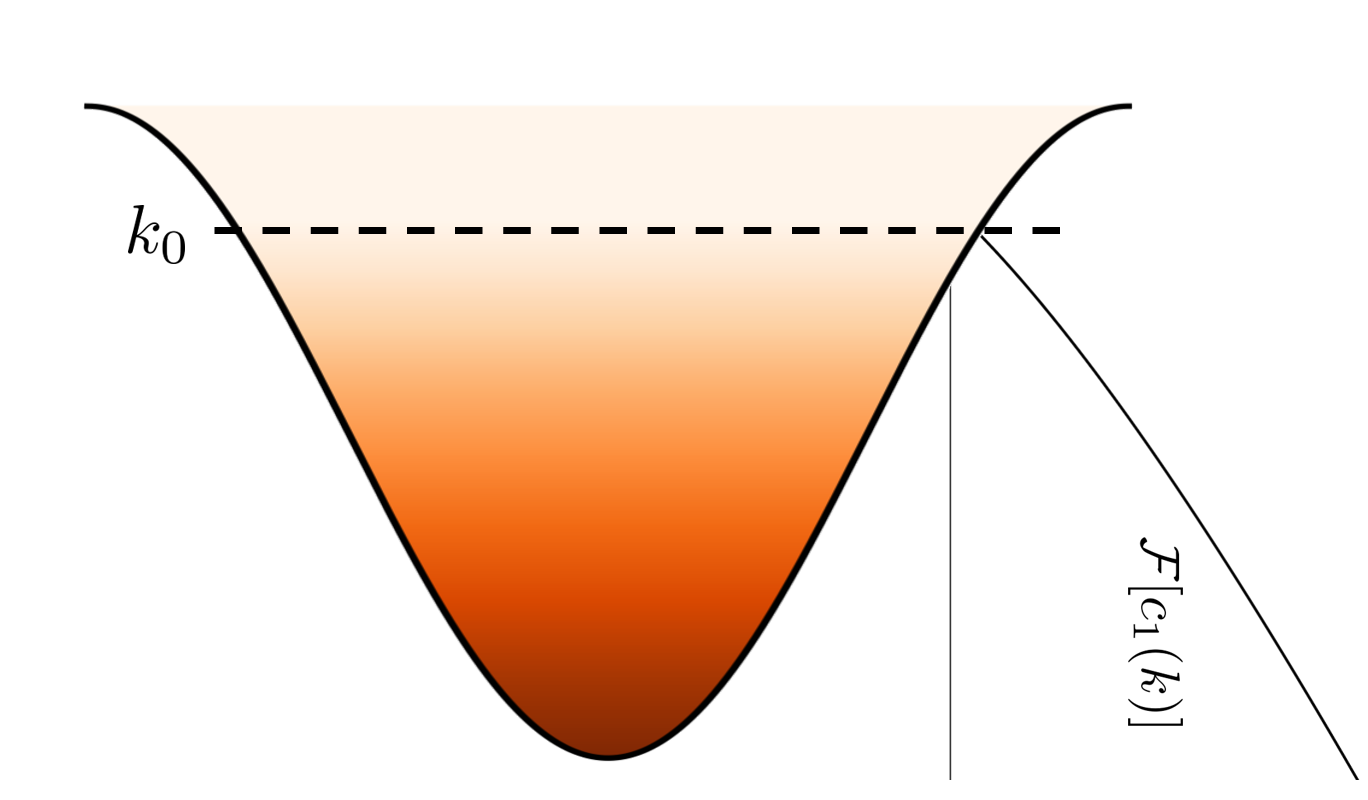}
  \caption{A schematic of the \AE{} distribution}
  \label{fig:sub1}
\end{subfigure}%
\begin{subfigure}{.5\textwidth}
  \centering
  \includegraphics[width=\textwidth]{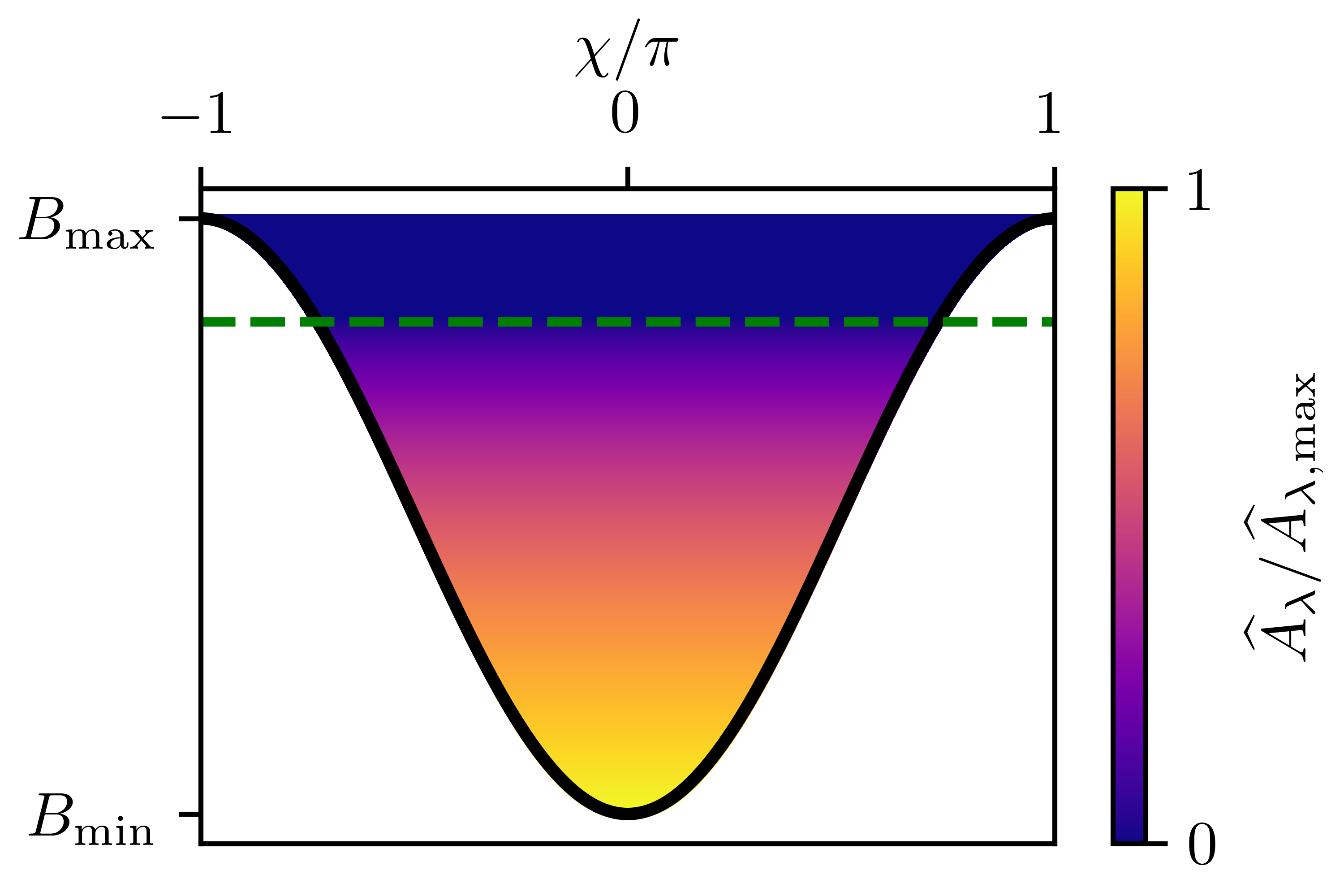}
  \caption{A numerical calculation of \AE{} distribution.}
  \label{fig:sub2}
\end{subfigure}
\caption{\textbf{Schematic of the contribution to available energy.} Left: diagram showcasing contribution of $\mathcal{F}$ to the \AE{} integrand. The broken line satifies $\omega_\alpha(k_0)=0$ to leading order. The plot of $\mathcal{F}[c_1(k)]$ is shown to the right for $ r\omega_{*,0}^T/\eta \gg 1$ as an example. Right: a numerical calculation showing the distribution of \AE{} across the magnetic well for the precise QA device at a minor radial coordinate of $r=10^{-3}$. The points where $\omega_\alpha=0$ is denoted by a green dashed line, and $\hat{\omega}_n = 10^4$.}
\label{fig: AE distribution strongly driven}
\end{figure}
In the derivation above it was key to consider the contribution to available energy from a narrow band of trapped particles with `slow' precession. As such, the particular form of the expression in Eq.~(\ref{eqn:AE_lead}) is only valid in the regime where $\omega_{\star,0}^T/ \omega_\alpha$ can be considered small. That is, when the trapped particle drifts are (as a group) much faster than the diamagnetic drift. As this roles revert, either because the turbulence becomes strongly driven (i.e., large density gradient) or the precession diminishes, the `weak' approach presented before will cease to provide us with a good approximation to the available energy. 
\par
As the precession becomes smaller relative to the diamagnetic drift, we however find another tractable limit, which we refer to as the `strongly-driven' limit. That is, we still consider a near-axis description of the magnetic field and precession, but at the same time order the diamagnetic drift to be large, i.e. vigorously driven.\footnote{Note that there might be some issues of independence here. We assume the density gradient to be independent from the field, which we know is not true, as the field must satisfy force balance, and the pressure gradient has a density gradient component to it. The large density gradient limit will thus to an extent bring second order $|\mathbf{B}|$ into the picture. However, we may formally proceed in this form.} Although this might appear inconsistent, it is not, as any ordering assumption about $\omega_n$ only affects the evaluation of the \AE{} integral, but not the particle precession itself. From the set-up, it should be clear that this ``strongly-driven'' regime gains relevance away from the magnetic axis, where the precession of trapped electrons decreases and the diamagnetic drift increases.\footnote{This is an important point. In the asymptotic sense, the weakly-driven regime will always hold sufficiently close to the axis. However, for a finite radius description, this may become quickly unimportant.} 
% Basic twiddle algebra allows us to estimate when the `strongly driven' regime is accessed. Equating the precession and diamagnetic drift $H\eta/rB_0\sim T\partial_\psi\ln n$ for thermal electrons, estimating $\eta\sim1/R$ and a quadratic density profile, for $\varrho\sim \varepsilon$ there will be a transition between the weakly and strongly driven regimes. This is relatively close to the core of the plasma, where the near-axis description may still be fitting, and thus it is an interesting limit to evaluate, and likely the dominant one in the volume.
\par
In this new scenario, the integral for available energy may be recomputed (see Appendix~\ref{sec:appAEdet}) using standard methods, as there no longer exists a narrow contributing band (see Fig. \ref{fig: AE distribution strongly driven}). All in all, one finds that the integral reduces to
\begin{equation}
    \widehat{A}_\mathrm{s} \approx 1.1605~\frac{C_r^2\rho_*^2}{\pi\sqrt{2}}   \frac{\hat{\omega}_n}{\varrho} (a \eta)^{3/2}\varrho^{3/2}.
\end{equation}
A different regime brings a different scaling with $\varrho$ and $\hat{\omega}_n$, in both cases with weaker dependencies than in the narrow-band regime. These changes are a result of, (i) the particle precession that serves as energy source contributing directly to \AE{}, and thus introducing a $1/\varrho\hat{\omega}_n$ factor compared to the weak regime (simply because on average particles do not quite reach the diamagnetic drift), and (ii) the contributing trapped particle fraction corresponding to the whole population with a positive precession, which no longer is a narrow band, and thus does not contribute with an additional $\hat{\omega}_n\varrho$ factor. Importantly, the scaling with $\eta$ \emph{inverts} compared to the weak regime, $\widehat{A}_\mathrm{w}$. Using the same tokamak-estimation for $\eta$ as there, one finds $\widehat{A}_\mathrm{s} \sim \varepsilon^{3/2} \mathcal{E}^{3/4}$, 
suggesting that a large-aspect-ratio tokamak which is vertically elongated reduces \AE{}. Once again, this is under the assumption of keeping the minor radius $a$ fixed. The behaviour will change depending on which features of the equilibrium are kept constant.
\par
The existence of these two different regimes of available energy naturally defines a transition. One can calculate this transition point by equating $\widehat{A}_\mathrm{w}\approx \widehat{A}_\mathrm{s}$, denoting the `critical' transition gradient as $-a  \partial_r \ln n|_\mathrm{crit} = a / L_{n}|_\mathrm{crit} $. We find, 
\begin{equation}
     \left. \frac{a}{L_{n}}\right|_\mathrm{crit} \approx 1.61591~ a \eta.
\end{equation}
For a quadratic density profile ($\hat{\omega}_n/\varrho\approx2$), the radial position $\varrho$ at which this critical transition occurs is
\begin{equation}
    \varrho_\mathrm{crit} \approx 0.80795~  a \eta.
\end{equation}
Using some typical tokamak values, we estimate $a \eta \sim \varepsilon \sqrt{\mathcal{E}} \sim 0.3$, and thus $\varrho_\mathrm{crit} \sim 0.2$. Hence, in a standard tokamak one can transition to the strongly driven regime fairly close to the axis, and we expect the strongly-driven regime description to be most suitable for most of its volume. 
\par
\begin{figure}
\centering
\begin{subfigure}{.5\textwidth}
  \centering
  \includegraphics[width=\textwidth]{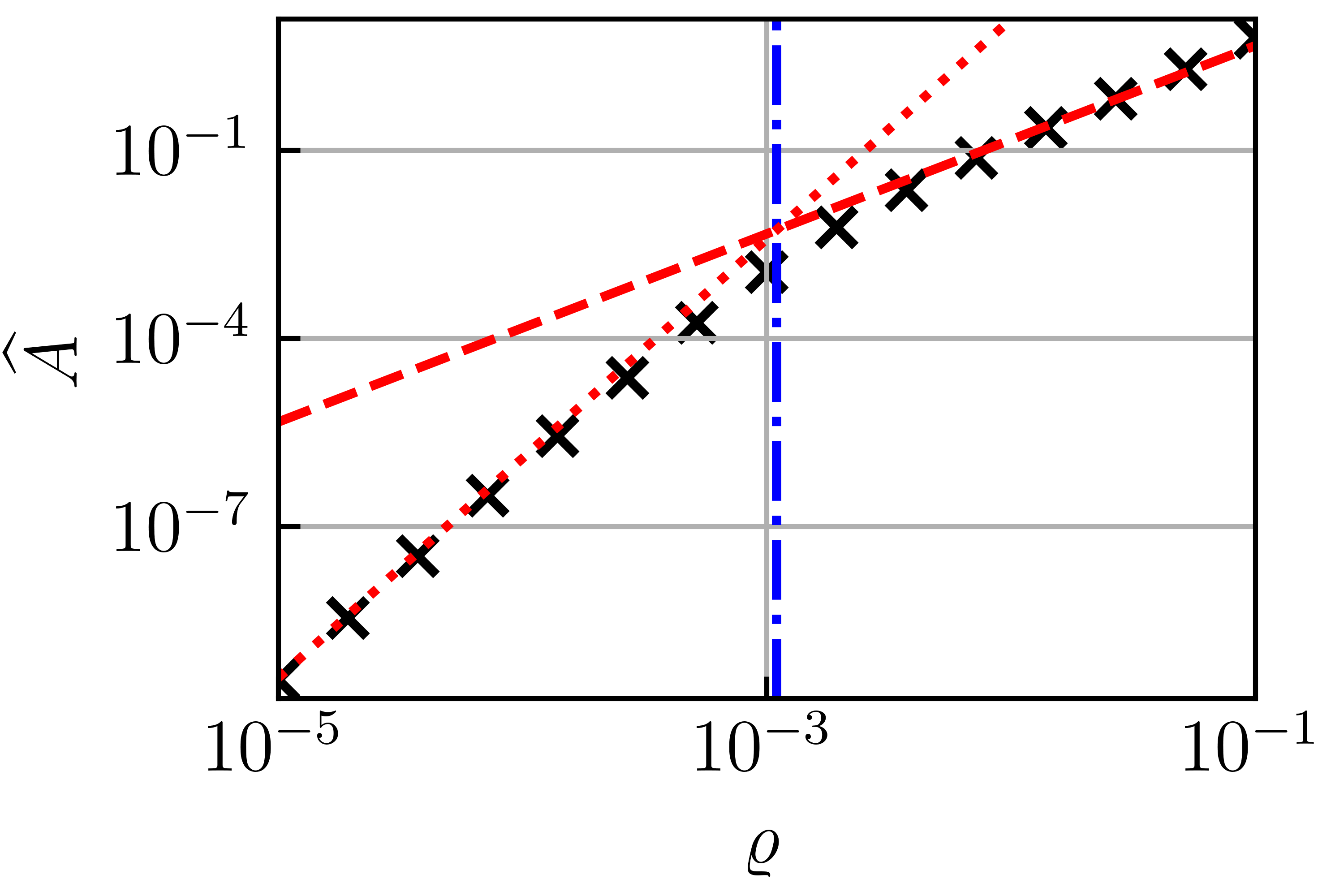}
  \caption{Precise QA}
  \label{fig:sub1}
\end{subfigure}%
\begin{subfigure}{.5\textwidth}
  \centering
  \includegraphics[width=\textwidth]{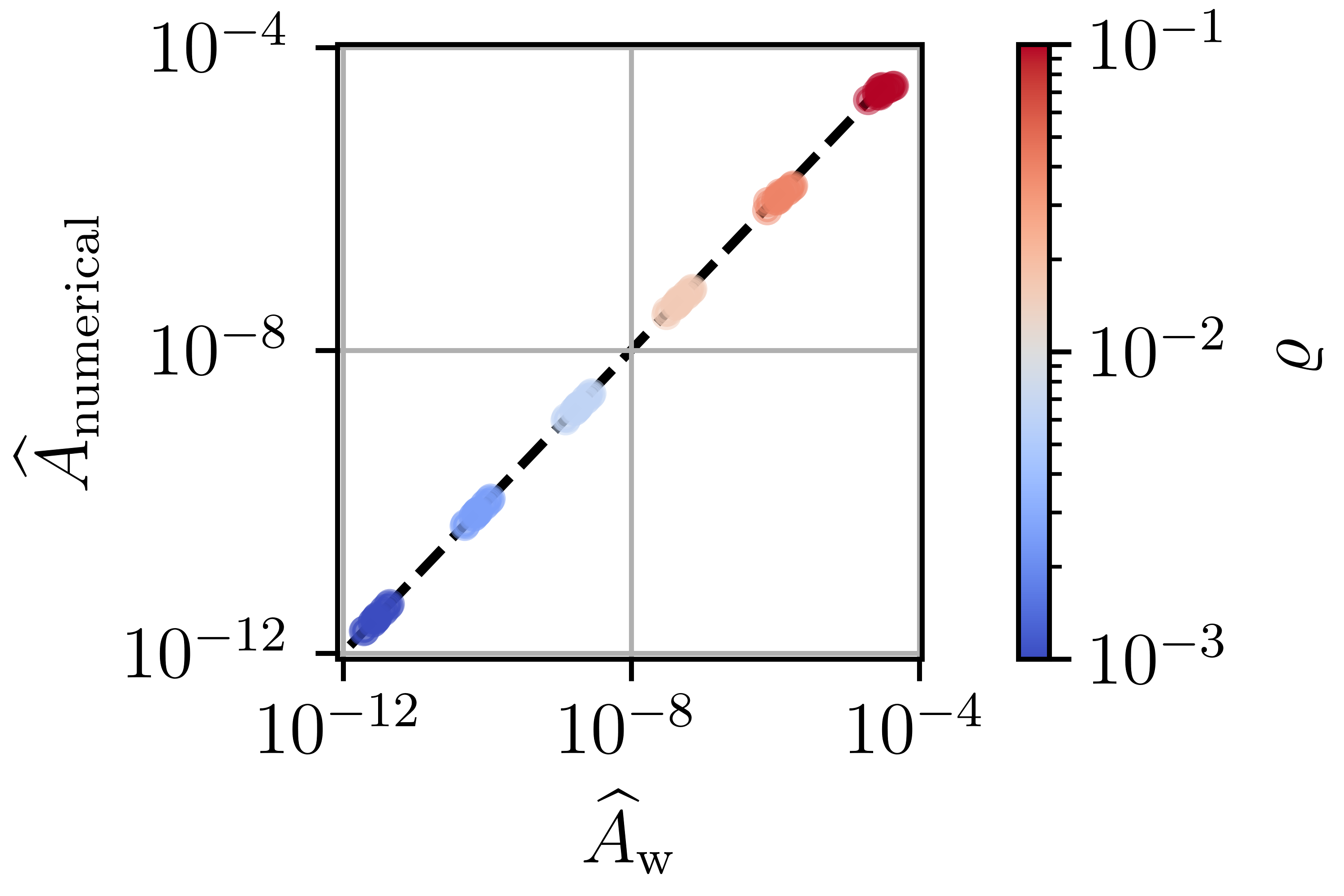}
  \caption{Multiple configurations}
  \label{fig:sub2}
\end{subfigure}
\caption{\textbf{A comparison of the numerical calculation of \AE{} against the analytical result}. Left: comparison of  the \AE{} in the two asymptotic regimes in the precise QA configuration as a function of $\varrho$. The red dashed and dotted lines denote the analytic asymptotic results in the strongly and weakly driven regime, respectively. The critical radial coordinate $\varrho_\mathrm{crit}$ is shown by a blue dash-dotted line. The crosses are numerical evaluations of the \AE{} using the near-axis equilibrium of the precise QA configuration in \cite{landreman2020magnetic}. The plot has been generated with a gradient value of $\hat{\omega}_n/\varrho=10^3$ for visualisation purposes. Right: correlation of the numerical and analytic result in the weakly driven regime for a wide number of quasisymmetric devices \citep{landreman2022map}. The ordinates correspond to the numerically evaluated $\widehat{A}$, whereas the abscissa corresponds to the asymptotic result of Eq. \eqref{eqn:AE_lead}, $\widehat{A}_\mathrm{w}$. For this plot, $\hat{\omega}_n/\varrho=1$. A close correspondence can be seen across many orders of magnitude. Both plots have been generated using the \texttt{pyQSc} code, which does not have a notion of minor radius, and as such they have $\varrho=r$}
%     \label{fig:avail_ener_comparison}
\label{fig: numerical verification AE}
\end{figure}

 
We conclude this leading order \AE{} analysis by presenting a numerical verification in Fig.~\ref{fig: numerical verification AE}, where we compare both asymptotic regimes in one device, and the weakly asymptotic regime across multiple devices\footnote{The code is freely available on \url{https://github.com/RalfMackenbach/AEpy}}. We observe excellent agreement in the asymptotic behaviour between the found results and the numerical calculation.

\subsection*{Additional dependence of \AE{}}
To learn anything about how triangularity of flux surfaces and pressure may affect this available energy, and thus how TPMs may be affected by them, we need to proceed to higher order in the calculation of $\widehat{A}$. We show how to do this to obtain the dependence of $\widehat{A}$ on $p_2$ and $\delta$ to leading order at the end of Appendix~\ref{sec:appAEdet}.\footnote{Note that to do so it is not necessary to compute the whole next order correction to $\widehat{A}$, but only the pieces concerned with the second order parameters directly.} After such considerations, we may write $\widehat{A}\approx\widehat{A}_0+\widehat{A}_1+\dots$, where $\widehat{A}_0$ is the leading order expression and $\widehat{A}_1$ the pressure and triangularity dependent piece. As before, for the discussion in the text we specialise to the generally shaped up-down symmetric tokamak. Full expressions that apply to the quasisymmetric case may be found in Appendix~\ref{sec:appAEdet}.
\par
In the weakly driven regime one finds
 \begin{equation}
     \left. \frac{\widehat{A}_1}{\widehat{A}_0} \right|_\mathrm{weak} \approx \mathcal{R}_{20} \left(  - \frac{r}{\eta} \frac{\mu_0 p_2}{B_0^2} \left[1+\frac{4\sqrt{\bar{\alpha}}}{(\bar{\alpha}+3)}\left(\frac{\eta B_{\alpha0}}{B_0\bar{\iota}_0}\right)^2\right] +  \frac{3}{2} \frac{1-\bar{\alpha}}{3+\bar{\alpha}}r\delta\right), \label{eqn:A1_weak}
 \end{equation}
 where $\mathcal{R}_{20}\approx 1.37$.
\par
It follows from this that, regardless of the choice of parameters, increasing the pressure gradient always leads to an increase in the available energy. Note that this is the case even if the density gradient, here controlled by the diamagnetic frequency $\hat{\omega}_n$, is fixed.\footnote{The change in pressure without a change in the density gradient may appear impossible, but it may be achieved straightforwardly by keeping the density gradient fixed and increasing the ion temperature.} To picture what is happening, we resort to the discussion on precession presented before. As we increase the pressure gradient, we learnt that the trapped-electron precession goes against the diamagnetic drift, which means that the trapped population is brought further away from resonance. But from this behaviour, one would expect the drive of the instability to decrease and with it \AE{} to do so. So, how is getting further away from the diamagnetic resonance making things worse? 
\par
To figure this out it is illuminating to assess, formally, the origin of the sign of $\mathcal{R}_{20}$. The sign is, to a large extent, a result of the correction to the integral measure $\mathrm{d}k/\mathrm{d}c_1$ needed when writing the \AE{} integral in $c_1$ (as it is necessary for the weak regime calculation, see Appendix~\ref{sec:appAEdet}). This piece of the integral measures the number of trapped particles that exist with a precession that is similar to the resonant one. The question is then how this population fraction changes as the precession slows down. And the answer is that the population that has a near-vanishing precession grows, as most directly seen in the smaller gradient of $\omega_\alpha$ with $k$ (see Fig.~\ref{fig:circ_tok_wa_1_e}). And because this fraction of the population is the only one that may contribute to the total available energy, the result is the increase of \AE{} with plasma $\beta$. This is an important feature of available energy, which not only assigns value to the magnitude of $\omega_\alpha(\omega_{*,0}^T-\omega_\alpha)$, but also to the number of particles with a particular value for its precession. As a consequence, we expect this behaviour to change in the strongly-driven regime, which we will visit later.
\par
The effect of triangularity, $\delta$, in Eq.~(\ref{eqn:A1_weak}) depends critically on whether cross-sections are elongated vertically or horizontally, as we saw MHD stability to do in the preceding discussion. In the most common case of vertically elongated cross-sections, negative triangularity is seen to reduce the \AE{} (which increases the precession of the trapped-particle class at $k_0$). Thus, the effect is not synergistic with MHD stability, as triangularity has precisely the opposite effect there. This anti-correlation holds also in the more general case of quasisymmetric configurations, which is readily seen by comparing Eq.~(\ref{eqn:R_qs}) for $\mathcal{R}_\delta$ directly to Eq.~(4.2) in \cite{rodriguez2023mhd} for $\mathcal{T}_\delta$. This intimate relation between MHD stability and what may be interpreted as TPM activity has been observed in many occasions (in fact, could be interepreted as the driver for many reverse triangularity studies in advance tokamak scenarios). Here we have formally proven in the weak asymptotic regime that a compromise between the two properties is needed in this regime. This opposed behaviour is not shared by plasma $\beta$, which acts in a detrimental form on both MHD and TPM activity. 
\par
Performing a similar analysis in the strongly driven regime, we find $\widehat{A}_1/\widehat{A}_0|_{\mathrm{strong}}\approx-2.85\widehat{A}_1/\widehat{A}_0|_{\mathrm{weak}}$, which presents the opposite sign to the weak regime. That is, an increased plasma $\beta$ (in the form of pressure gradient) always decreases the \AE{}, and in a standard tokamak with $\bar{\alpha}<1$ positive triangularity becomes TEM-stabilising. In the strongly-driven regime, reducing precession brings the zero-crossing point closer to $k=0$, thus reducing the total $k$-space available to drive TEMs. In that limit, then, with both precession and accessible population decreasing with increased pressure gradient and positive triangularity, we expect available energy to decrease, { and regarding fast particle confinement due to non-QS behaviour, to momentarily grow before eventually decreasing as precession grows in the direction of maximum-$\mathcal{J}$. The details will depend on how different trapped particles are affected, and how important their contribution to confinement is.} Unlike in the weak regime, the whole trapped population becomes important, and not just a narrow-band, Fig.~\ref{fig: AE distribution}. In the strongly-driven regime then there is a synergy between MHD-stability and TEM activity with respect to the triangular shaping of cross-sections, but opposed in the effect of plasma $\beta$. The expected behaviour of a stellarator will thus depend critically on the particular regime considered. 
\par
{ Besides the sign, there is also a difference in magnitude of roughly a factor $3$ between the relative effects of triangularity and plasma $\beta$ in the strong regime compared to the weak regime. For the discussion following, we focus on the strongly driven regime. This effect can be quantified as we did in the discussion of precession, which we do as follows. When the first-order correction significantly affects the available energy, i.e. $\widehat{A}_1/\widehat{A}_0\sim1$, we state that we have a critical scenario. At the edge, the critical $\beta$ becomes $\beta_\mathrm{crit}^{\AE{}}\sim 2a|\eta|/\mathcal{R}_{20}(1+f)$, with $f$ as defined before (with its QS generalisation, which may be found in Eq. \eqref{eqn:Gp2_qs}). This shows that plasma $\beta$ becomes effective in significantly changing \AE{} in the strong regime for QAs in the regime of $\beta_\mathrm{crit}^{\AE{}}\sim2-3\%$, while QH $\beta_\mathrm{crit}^{\AE{}}\sim4-7\%$. Finite $\beta$ QA equilibria appear, therefore, to significantly affect the behaviour of TPM-like behaviour, while QHs remain more resilient, as expected from the behaviour of precession. As far as triangularity is concerned, the expression in Eq.~(\ref{eqn:crit_delta}) may be used for \AE{} with $\mathcal{G}_\delta=3[(3+\bar{\alpha})-(\bar{\alpha}+1)\bar{F}]/[(1-\bar{\alpha})+(1+\bar{\alpha})\bar{F}]$ there. The values for QS configurations may be found in Appendix~\ref{sec:appCritEst}, and range $(r\delta)_\mathrm{crit}\sim 0.4-1.5$, which is a significant triangularity, nevertheless consistently achievable in many configurations (see Appendix~\ref{sec:appCritEst}). Note that in the circular tokamak limit, triangularity has no effect on \AE{} (only a very small one in the weak regime from $\mathcal{R}_{22}^C$).}

\par 
Given the changes in the weak and the strong regime, the critical transition gradient also changes and may be computed,
\begin{equation}
    \begin{aligned}
        \left. \frac{a}{L_{n}}\right|_\mathrm{crit} \approx \left. \frac{a}{L_{n}}\right|_{\mathrm{crit},0}\Bigg[ & 1-2.6389 \times  \\
        & \left. \left(- \frac{r}{\eta} \frac{\mu_0 p_2}{B_0^2} \left[1+\frac{4\sqrt{\bar{\alpha}}}{(\bar{\alpha}+3)}\left(\frac{\eta B_{\bar{\alpha}0}}{B_0\bar{\iota}_0}\right)^2\right] +  \frac{3}{2} \frac{1-\bar{\alpha}}{3+\bar{\alpha}} r \delta\right)\right].
        \label{eq: crit grad}
    \end{aligned}
\end{equation}
This means that the critical gradient decreases for an increased pressure gradient (as the factor multiplying the pressure is always positive), and increases for negative triangularity tokamak which are vertically elongated. 
\par 
To close the paper, we make a comparative study of the insight and results presented in this paper with the literature. The comparison to a recent numerical analysis of the \AE{} for tokamak equilibria described by a Miller \citep{miller1998noncircular} geometry \citep{mackenbach2023Miller} is most straightforward. One can see that many of the trends found there are recovered in the current paper on an analytical basis; in particular, negative triangularity decreases the \AE{} for vertically elongated tokamaks only if the gradient is sufficiently small, and the gradient-threshold has the same dependencies as derived here. Of course, our work sheds light on the origin of such behaviour, and can be applied beyond axisymmetric configurations. 
\par 
The comparison to other turbulent study results and experiments requires us to carefully discern between the two distinct regimes that we have shown the \AE{} exhibits. Depending on which regime a given scenario is in, the beneficial or detrimental nature of various equilibrium shaping parameters may change. In general, though, and bearing this important caveat in mind, the strongly driven regime is most often entered fairly close to the magnetic axis (as argued before). It is also, in magnitude, the principal contributor to \AE{}, and the very character of  the weak regime may make it numerically elusive (as the narrow \AE{} band would have to be resolved in simulations). As such, it is natural to take the \AE{} in the strongly driven regime as indicative of overall behaviour of a configuration to TPM mediated transport. In terms of leading order effects, then, the prediction that increasing the vertical elongation in tokamaks improves transport agrees with existing knowledge \citep{chu1978effects,qi2019characteristics}. Concerning higher order effects, the beneficial nature of a pressure gradient on the trapped electron mode has been noted by many authors \citep{rosenbluth1968,connor1983effect,li2002role}. The effect of triangularity on \AE{}, however, is more paradoxical. In experiments, it has been shown that negative triangularity exhibits improved confinement whilst remaining in L-mode \citep{marinoni2019h}. The current model, however, would predict an \emph{increase} in the \AE{} in such a scenario and hence more unfavourable transport. Part of this discrepancy may be explained by the role of magnetic shear, which is positive and significant near the edge of the tokamak, but we have not explicitly considered it here. The trapped particle precession increases with increasing magnetic shear, which may push one to a more weakly driven regime in which negative triangularity is beneficial. 
\par  
However, the discrepancy may also come from the consideration that behaviour \emph{within} the strongly driven regime may not be the most adequate to describe the turbulent performance of a configuration. To illustrate this, take a scenario in which \AE{} is large. Turbulence is expected to be virulent in such a scenario, which will enhance transport and ultimately limit the attainable density gradient (as a maximum transport may be supported). This limiting factor to the gradient naturally leads on to consider profile stiffness \citep{garbet2004profile}; profiles are stuck to the maximal sustainable gradient values, which are fixed by the point at which a regime of increased turbulence is accessed. Under such prism, it is not that important what occurs within the strong and weak turbulent regimes, but rather what happens to the transition point. In such context, support of larger gradients is seen as beneficial, as higher central densities and higher confinement times can be achieved. The key is then to understand the behaviour of this threshold. In practice this transition point is found through gyrokinetic simulations \citep{dimits2000comparisons} to find when a steep decrease in the non-linear heat-flux is seen when the gradient value is decreased below some threshold value. Recognizing an analogous behaviour in \AE{}, where $\widehat{A}_\mathrm{w} \sim \hat{\omega}_n^3$ below criticality and $\widehat{A}_\mathrm{s} \sim \hat{\omega}_n$ above it, one may postulate that the behaviour in Eq.~\eqref{eq: crit grad} is similar to that one would observe for the common conception of critical gradient. As a consequence of this threshold behaviour, one would then expect that the attained gradient in experiments is the one which we have calculated in \eqref{eq: crit grad}: the system would reside on the weak-to-strong \AE{} boundary. We do not attempt to verify this relationship here, which for a consistent consideration would require consistent plasma profiles based on (estimated) heat-fluxes, which would feedback onto the geometry (e.g. Shafranov shifts). We make no attempt of solving this problem here, and leave it to a future investigation, but merely note similarities in the behaviour of our transition threshold and \citet{merlo2015investigating,merlo2023interplay} in the increase of gradient-thesholds with negative triangularity tokamaks.

\section{Conclusions and outlook}
In this paper, we explored the trapped-particle precession and its effects on the available energy to trapped particle modes in axisymmetric and quasisymmetric fields. We do so by considering a near-axis description of the fields, in which the magnitude of the magnetic field is directly prescribed and interlinked to other geometric and equilibrium features. As such, this may be regarded as an extension or alternative to previous local considerations \citep{connor1983effect,roach1995trapped,connor1983effect,hegna2015effect}. The precession of trapped particles is constructed explicitly and analytic expressions are given to leading order, and the first order correction. This allows us to prove the impossibility of the maximum-$\mathcal{J}$ property in such fields to leading order, as follows from \cite{boozer1983transport}. The role of a pressure gradient in helping to attain this property at a finite radius is presented. Investigating the effect of triangularity in axisymmetric devices, we find that negative triangularity may aid in attaining the maximum-$\mathcal{J}$ property, but the influence on different classes of trapped particles is generally different. In the full quasisymmetric case, closed forms are also provided and some practical examples discussed.
\par
The influence of such precession on density-gradient driven trapped particle modes in the system is then analysed by assessing its effect on the available energy \citep{helander2017available,helander2020available,mackenbach2023transport}. Two different asymptotic regimes naturally arise in the form of a ``weakly'' and ``strongly'' driven regimes, each with a different behaviour and physical mechanism, which furthermore allows one to define a critical transition density gradient. In the weakly driven regime, a narrow band (in $\lambda$-space) of the trapped particle population is responsible for the available energy, whilst in the strongly driven regime all resonating trapped particles contribute. This physical difference between the two asymptotic regimes leads to a difference in the dependencies of \AE{} on the field.
\par  
This is certainly true for the effects of pressure gradient and triangularity: its beneficial nature depends on the asymptotic regime considered. Interestingly, we find that the dependence of \AE{} on triangularity is of the same form of that in Mercier's criterion for MHD-stability, up to a sign that changes depending on the driving regime. In the strongly driven regime a synergy is found between improving MHD-stability and lowering energy available to the trapped particles through triangularity, meaning that improving one will improve the other (the opposite being true of plasma $\beta$). { The reduction in precession of deeply trapped particles behind this behaviour will, whenever deviations from ideal QS exist, lead to increased fast particle losses, at least until a regime of significant precession in the direction of maximum-$\mathcal{J}$ is reached.} The synergy between MHD and turbulent activity through triangularity inverts in the weakly driven regime, where it is under plasma $\beta$ that this synergy is observed. This difference in behaviour affects the critical-gradient estimate for the transition between the two regimes. This gradient grows with negative triangularity, which could align with some of the existing observations in advanced tokamak scenarios. 
\par 
All in all, one finds that the near-axis framework is a convenient model to assess properties of trapped particles in quasi-symmetric magnetic fields. The notions of maximum-$\mathcal{J}$, omnigeneity  (through the bounce-averaged radial drift) or \AE{}, for which analytical expression may be found allows one to readily evaluate several aspects of performance of different stellarator configurations at negligible computational cost (as measured by \AE{}). This may be helpful as a proxy for turbulence optimisation. Finally, though we have specialised to quasi-symmetric configurations, many of the techniques presented may be applied in other contexts (such as quasi-isodynamic fields or Miller geometry models) allowing one to make similar statements. For the quasi-isodynamic case such an investigation is currently being undertaken, and an even more distinct case in which the bounce-averaged radial drift may play a significant role could also benefit from the current framework.

\section*{Data availability}
The data that support the findings of this study are openly available at the Zenodo repositories with DOI/URL 10.5281/zenodo.8185571 and 10.5281/zenodo.8199904.

\section*{Acknowledgements}
The authors would like to acknowledge fruitful discussion with Per Helander, Jaime Caballero and Iv\'{a}n Calvo. E. R. was supported by a grant by Alexander-von-Humboldt-Stiftung, Bonn, Germany, through a postdoctoral research fellowship. R.M. is partly supported by a grant from the Simons Foundation (560651, PH), and through the project “Shaping turbulence—building a framework for turbulence optimisation of fusion reactors,” with Project No. \texttt{OCENW.KLEIN.013} of the research program “NWO Open Competition Domain Science” which is financed by the Dutch Research Council (NWO).

\appendix

\section{Calculation of the second adiabatic invariant}
\label{app: second adiabatic invariant details}
In this appendix we write out the full derivation of the second adiabatic invariant $\mathcal{J}_\parallel$. Our starting point is the leading order integral given in Eq. \eqref{eq: expression for Jparr(1)} after a straightforward expansion in the ordering parameter $\epsilon$, defined in the main text. 
\par
To make progress with this integral, start by defining a so-called trapping parameter $k$, which relates to the pitch-angle parameter $\lambda$ via
\begin{equation} 
    \lambda = \frac{1}{1 + r \eta (2 k ^2 -1) + \delta B_b }. \label{eqn:k_def_lambda}
\end{equation}
Such a definition gives $k\in[0,1]$, with the limits corresponding to deeply and barely trapped particles respectively. It must be noted that despite its simple appearence, this definition hides complexity in the trapped particle class dependence of $\delta B_b$, Eq.~(\ref{eqn:lam_chi_b}, the field perturbation at the bouncing point. With this definition, we rewrite the integral
\begin{equation}
   \mathcal{J}_\parallel^{(0)} = \underbrace{ \sqrt{2H}\frac{B_\alpha}{\bar{\iota}B_0} \sqrt{r \eta\lambda}}_{\stackrel{\cdot}{=}\bar{\mathcal{J}}} \int\frac{\sqrt{2 k^2 - 1 + \cos \chi}}{1 - r \eta \cos \chi}\mathrm{d}\chi.
\end{equation}
This integral can be cast into a form close to the one required for elliptic integrals by employing the double-angle identity $\cos \chi = 1 - 2\sin^2 \left(\chi/2 \right)$, with which the integral reduces to
\begin{equation}
    \mathcal{J}_\parallel^{(0)} = 2 \sqrt{2} \bar{\mathcal{J}} \int \frac{\sqrt{k^2 - \sin^2 \overline{\chi}}}{1 -  r \eta (1 - 2 \sin^2 \overline{\chi})} \mathrm{d} \overline{\chi},
\end{equation}
where the integration variable has been set to $\overline{\chi}=\chi/2$. The final subtlety that one needs to take into account is that the limits of integration are currently set by the zeros of the argument of the numerator in the integrand (namely, the bouncing points), which also gives an intuitive (and equivalent) definition of the trapping parameter $k$,
\begin{equation}
    k^2 = \sin^2 \left( \frac{\chi_b}{2} \right),
\end{equation}
where we remind the reader that $\chi_b$ denotes the bounce angle. This shows, as did Eq.~(\ref{eqn:k_def_lambda}), that $k$ includes some of the higher order corrections to $B$. This arises naturally from the integration, and importantly, preserves the meaning of deeply and barely trapped particles at $k=0,~1$, regardless of the perturbation. 
\par
\begin{figure}
    \centering
    \includegraphics[width=0.6\textwidth]{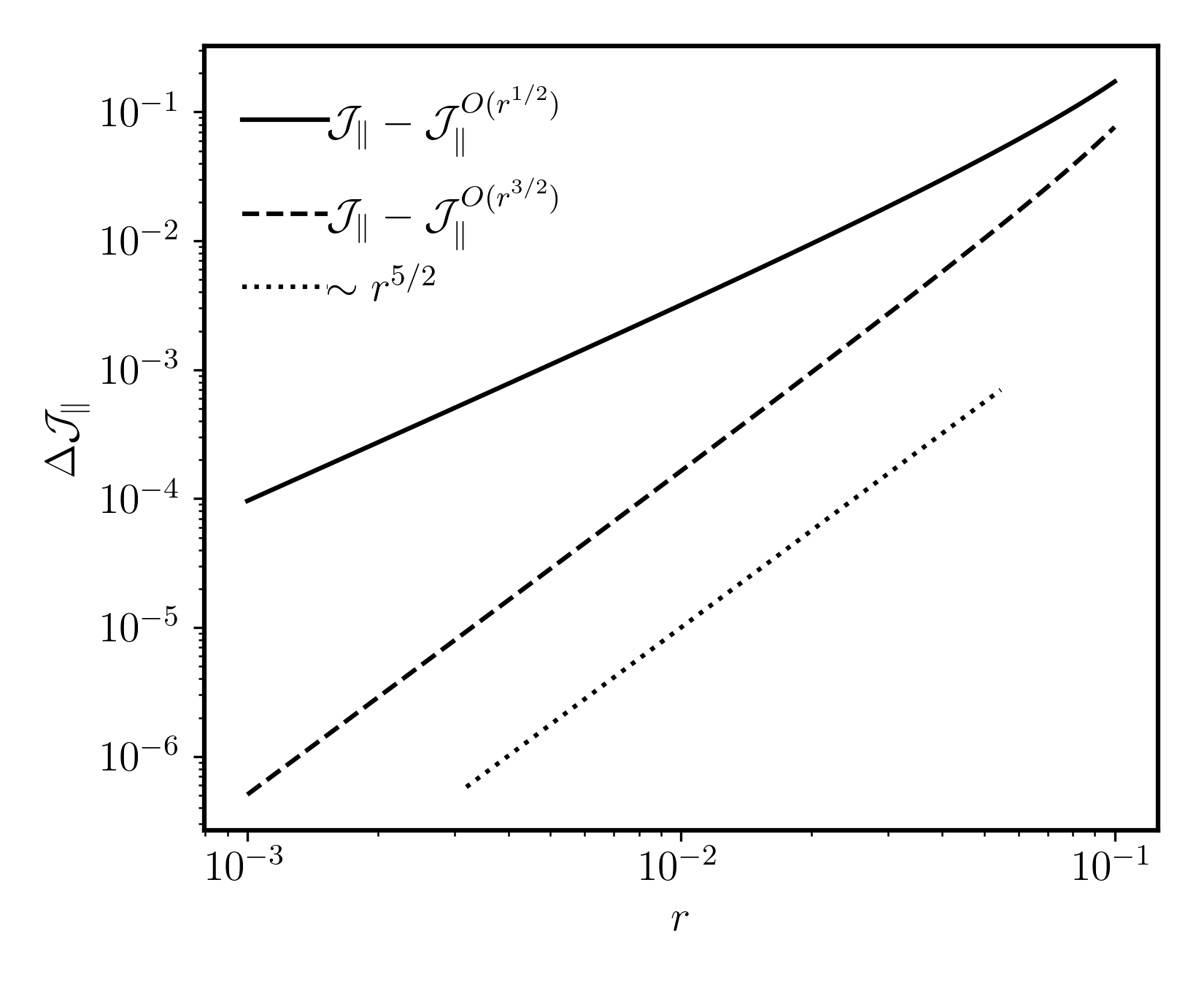}
    \caption{\textbf{Residual between numerical $\mathcal{J}_\parallel$ and analytic approximation.} The plot shows, in log scale, the difference between the numerically computed $\mathcal{J}_\parallel$ and the analytic expression, Eq.~(\ref{eqn:2nd_ad_inv_full}), to $O(r^{1/2})$ (solid line) and $O(r^{3/2})$ (broken line). The dotted line shows a reference $\sim r^{5/2}$ scaling, which agrees with the broken line as predicted by the theory. This particular case was run for an artificial ideal second-order near-axis $|\mathbf{B}|$ profile with $\eta=1$, $B_{20}=1$, $B_{22}^C=3$ and $k=0.5$. }
    \label{fig:my_label}
\end{figure}

A final substitution puts the integral into the standard form required for elliptic integrals. Define,
\begin{equation}
    k \sin \zeta = \sin \overline{\chi} \implies \mathrm{d} \overline{\chi} = \frac{\sqrt{k^2 - \sin^2 \overline{\chi}}}{\sqrt{1 - \sin^2 \overline{\chi}}} \mathrm{d} \zeta,
\end{equation}
in which case the bounce points simply become $\zeta=\pm\pi/2$, independent of $k$. This transformation is well-defined because $k\in[0,1]$. The leading order contribution to the second adiabatic invariant is now equal to
\begin{equation}
    \mathcal{J}_\parallel^{(0)} = 2 \sqrt{2} \bar{\mathcal{J}} \int_{-\pi/2}^{\pi/2} \frac{1}{1 -  r \eta (1 - 2 k^2 \sin^2 \zeta)} \frac{k^2\cos^2\zeta}{\sqrt{1 -  k^2\sin^2 \zeta}}\mathrm{d} \zeta.
\end{equation}
As part of the asymptotic near-axis treatment, $r$ is to be considered small, and we may thus expand the non-singular denominator of the integrand in powers of $r$. Including terms up to the first order,\footnote{If one wants to include higher order effects into the presented calculation, the Boozer Jacobian must be treated with some care. The reason is that, after all, the Jacobian represents the geometry of flux surfaces, and thus, the form of the Jacobian itself depends on how the near-axis expansion is interpreted. If the near-axis expansion is treated as a model that is truncated at second order to construct flux surfaces, then the Jacobian is actually not equal to $B_{\alpha}/B^2$ beyond the first order. The integral at second order would need to be reconsidered. A comment in this regard may be found in \cite{landreman2021a}.}, we find
\begin{equation}    \mathcal{J}_\parallel^{(0)} = 2 \sqrt{2} \bar{\mathcal{J}} \left[ I_1(k) +  I_2(k) r \eta + O(r^2)\right],
\end{equation}
where we have introduced,
\begin{equation}
    \begin{aligned}
        I_1 &= \int_{-\pi/2}^{\pi/2} \left(\frac{k^2 -1  }{\sqrt{1 -  k^2\sin^2 \zeta}} + \sqrt{ 1 - k^2 \sin^2 \zeta} \right) \mathrm{d} \zeta \\
        &= 2\left[ (k^2 - 1) K(k) + E(k) \right] \\
        \: \\
        I_2 &= \int_{-\pi/2}^{\pi/2} \frac{k^2\cos ^2\zeta\left(1-2 k^2 \sin ^2\zeta\right)}{\sqrt{1-k^2 \sin ^2\zeta }}  \mathrm{d} \zeta \\
   &= \frac{2}{3} \left[(2 k^2-1) E(k)-(k^2-1)
   K(k)\right]\\
   \: \\
   %      I_2 &= \int_0^{\pi/2} \frac{k^2\cos ^2 \zeta\left(4 k^4 \sin ^4\zeta -4 k^2
   % \sin ^2 \zeta +1\right)}{\sqrt{1-k^2 \sin ^2\zeta }} \\
   % &= \frac{1}{15} \left[(-4 k^4-3 k^2+7) K(k)+(8 k^4-8
   % k^2-7) E(k)\right].
    \end{aligned}
\end{equation}
and define complete elliptic integrals of the first and second kind \citep[Sec.~19]{DLMF},
\begin{align}
    K(k) & \stackrel{\cdot}{=} \int_0^{\pi/2} \left( \sqrt{1 - k^2 \sin^2 \zeta} \right)^{-1} \mathrm{d}\zeta,  \tag{\ref{eq:expansion-new-coordinates}} \\
    E(k) & \stackrel{\cdot}{=} \int_0^{\pi/2} \sqrt{1 - k^2 \sin^2 \zeta} \: \mathrm{d}\zeta.
    \tag{\ref{eq:expansion-new-coordinates-b}}
\end{align}
Our final step is to expand $\bar{\mathcal{J}}$ to the required order in $r$. This expansion is readily done and one can show that, neglecting terms of order $O(r^2)$, this reduces to
\begin{equation}
    \begin{aligned}
        \bar{\mathcal{J}} \approx\sqrt{2H r\eta}\frac{B_{\alpha0}}{\bar{\iota}_0B_0} \left[1+ r\eta\left(\frac{1}{2} - k^2\right)  + O(r^2)\right]. 
    \end{aligned} \label{eqn:J_bar_expansion}
\end{equation}
Collecting all the terms of order $O(r)$ in $\mathcal{J}_\parallel^{(0)}$,
\begin{equation}
    \mathcal{J}_\parallel^{(0)} = 4\sqrt{H r\eta }\frac{B_{\alpha0}}{\bar{\iota}_0B_0} \left[ I_1(k) +r\eta\left(I_1(k)\left(\frac{1}{2}-k^2\right) + I_2(k)\right) + O(r^2)\right].
\end{equation}
We now turn to the next order correction in $\epsilon$ following Eq.~(\ref{eqn:I_J_par}), 
\begin{equation}
        \mathcal{J}_\parallel^{(1)}=
        -\sqrt{2H}\frac{B_\alpha}{\bar{\iota}B_0}\int_{-\chi_b}^{\chi_b} \left( \frac{\lambda}{2} \frac{\delta B-\delta B_b}{\sqrt{1-\lambda(\bar{B}+\delta B_b)}} + \frac{\delta B \sqrt{1 - \lambda (\bar{B} + \delta B_b)}}{\bar{B}^2}  \right)\mathrm{d}\chi.
    \label{eq:J-epsilon-expansion}
\end{equation}
The second term in the integrand is a factor $r$ smaller than the first, and hence, for our current expansion, only the first term needs to be taken into account. To perform that remaining integral, we ought to express the integrand explicitly as a function of the integration variable $\chi$. Using the near-axis expansion form of $B$ for a quasi- and stellarator symmetric fields, we know that
\begin{subequations}
    \begin{gather}
        \delta B-\delta B_b= r^2 B_{22}^C \left( \cos 2\chi - \cos 2 \chi_b  \right), \\
        1-\lambda(\bar{B}+\delta B_b)=\frac{r \eta (2 k^2 - 1 + \cos \chi)}{1 + r \eta (2 k^2 - 1) + \delta B_b}.
    \end{gather}
\end{subequations}
With these, introducing the trapping parameter, using double angle identities, and employing $\zeta$ as the integration parameter (like in the previous order), the integral reduces to
\begin{equation}
\begin{aligned}
    \mathcal{J}_\parallel^{(1)} &\approx \bar{\mathcal{J}} \frac{r B_{22}^C}{\eta} \times \sqrt{2} \int_0^{\pi/2} \frac{\cos 4 \bar{\chi}_b - \cos 4 \bar{\chi}}{\sqrt{1 - k^2 \sin^2 \zeta}} \mathrm{d} \zeta \\ 
    &\approx - 4 \sqrt{H r \eta} \frac{B_\alpha}{\bar{\iota}B_0 } \frac{r B_{22}^C}{\eta} \mathcal{I}_{22}^C(k) ,
\end{aligned}
\end{equation}
where we have kept leading order terms in $r$. The function describing the behaviour of different trapped particle classes is $\mathcal{I}_{22}^C(k)$, which we have defined as
\begin{equation}
    \mathcal{I}_{22}^C(k)\stackrel{\cdot}{=} I_2(k)-(2k^2-1)I_1(k).
\end{equation}
Expanding to order $r^2$ and combining the first and second order result, Eq.~(\ref{eqn:I_J_par}), gives us our final expression for the second adiabatic invariant,
\begin{equation}
    \mathcal{J}_\parallel\approx4\sqrt{H r\eta }\frac{B_{\alpha 0}}{\Bar{\iota}_0 B_0}\left\{I_1(k) +r\eta\left[\left(\frac{1}{2}-k^2\right)I_1(k)+I_2(k)-\frac{B_{22}^C}{\eta^2}\mathcal{I}_{22}^C(k)\right]\right\}. \tag{\ref{eqn:2nd_ad_inv_full}}
\end{equation}
Figure~\ref{fig:my_label} shows the comparison of this analytic expression to the numerical calculation of $\mathcal{J}_\parallel$.

\section{Integral expressions for the radial drift} \label{sec:app-radial-drifts}
In this section, we use the near-axis framework to find expressions for the bounce-averaged radial drifts to leading order in the near-axis expansion. Of course, only if the system is not exactly quasisymmetric will the radial drift be non-vanishing. We will assume that quasisymmetry is broken at second order in the near-axis expansion. 
\par
As is well-known, it is generally not possible to guarantee the symmetry of $|\mathbf{B}|$ to second-order in the expansion. Thus, generally, to form a consistent description to second-order one formally allows $B_{20}$ to be a function of $\varphi$, rather than a constant. This is the conventional choice, and keeps the other pieces of $|\mathbf{B}|$ constant. Let us then consider here a field that is quasisymmetric to first order, but, at second order, has $B_2=B_{20}(\varphi)+B_{22}^C \cos 2\chi$, which retains stellarator symmetry for an even $B_{20}(\varphi)$. Including the contributions from the other terms if their $\varphi$-independence was relaxed would also be straightforward.
\par
Let us then consider the $\alpha$ derivative of the second adiabatic invariant needed for assessing the averaged radial drift explicitly, 
\begin{equation}
    \partial_\alpha\mathcal{J}_\parallel=\sqrt{2H}\frac{B_\alpha}{B_0\Bar{\iota}}\partial_\alpha\left[\int \frac{\sqrt{1-\lambda\hat{B}}}{\hat{B}}\mathrm{d}\chi\right]. \label{eqn:dJdalphaorig}
\end{equation}
The integrand vanishes at the bouncing points by construction, and therefore, by Leibniz's rule, the boundary terms may be dropped when enacting the derivative of the integral. This is not completely true for barely and deeply trapped particles, as their bounce points may change in a non-smooth way as we move from field line to field line. To picture this, think of a top of a magnetic well coming down on one side of the well as we move to a different field line. The original barely-trapped particle `leaks', undergoing a sudden change in behaviour, leading to a new class of trapped particle. Such a behaviour cannot be appropriately captured in a perturbative sense, but the importance of this particle `leak' may be assessed by constructing a measure of the leaked particle fraction $f_\mathrm{leak}=\max_\alpha[|B_{20}((\pi-\alpha)/\Bar{\iota})-B_{20}((\pi+\alpha)/\Bar{\iota})|]/(B_\mathrm{max}-B_\mathrm{min})$, where we assumed stellarator symmetry (i.e., $B_{20}(-\varphi)=B_{20}(\varphi)$ at second order).
\par
With this in mind, and ignoring this fringing case, we may pull the derivative through inside the integral. The integral is taken along field lines (i.e., at constant $\alpha$), which meand that the Boozer toroidal angle $\varphi=-(\alpha-\chi)/\bar{\iota}_0$ becomes a function of both $\alpha$ and $\chi$. That means that $\partial_\alpha f(\varphi)=-\partial_\varphi f/\bar{\iota}_0$, which keeping the leading order near-axis term and using the same notation as in Appendix~\ref{app: second adiabatic invariant details} gives,
\begin{align}
          \partial_\alpha\mathcal{J}_\parallel=&-\sqrt{2H}\frac{B_\alpha}{B_0\Bar{\iota}}\frac{\lambda}{2}\int \frac{\partial_\alpha \hat{B}}{\sqrt{1-\lambda\hat{B}}}\mathrm{d}\chi\\
          =&\sqrt{2H}\frac{B_\alpha}{B_0\Bar{\iota}_0^2}\frac{r}{\eta}\sqrt{\frac{\eta r}{2}} \overbrace{\int_{-\pi/2}^{\pi/2} \frac{ B_{20}'[\varphi(\bar{\chi},\alpha)]}{\sqrt{1-k^2\sin^2\zeta}}\mathrm{d}\zeta}^{\stackrel{\cdot}{=}\mathcal{I}_{20,\alpha}} 
          % &+ \overbrace{\int_{-\pi/2}^{\pi/2} \frac{B_{22}^C{}'[\varphi(\bar{\chi},\alpha)] \cos 4\bar{\chi}}{\sqrt{1-k^2\sin^2\zeta}}\mathrm{d}\zeta}^{\stackrel{\cdot}{=}\mathcal{I}_{22,\alpha}^C} \nonumber \\
    % &+ \underbrace{\int_{-\pi/2}^{\pi/2}\frac{B_{22}^S{}'[\varphi(\bar{\chi},\alpha)] \sin 4\bar{\chi}}{\sqrt{1-k^2\sin^2\zeta}}\mathrm{d}\zeta}_{\stackrel{\cdot}{=}\mathcal{I}_{22,\alpha}^S}  \Bigg)
          \label{eqn:dJdalphaorig}
\end{align}
where the prime indicates derivative respect to the only argument of $B_{20}$, $\varphi$. The integration variable is $\zeta$, and  one should interpret $\overline{\chi}(\zeta,k) = \arcsin \left( k \sin \zeta \right)$.
\par
% In a form closer to that of elliptic integrals, 
% \begin{equation}
% \begin{aligned}
%     \mathcal{I}_{22,\alpha}^C &= \int_{-\pi/2}^{\pi/2} \frac{B_{22}^C{}'[\varphi(\bar{\chi},\alpha)] \left( 8k^4\sin^4\zeta-8k^2\sin^2\zeta+1 \right)}{\sqrt{1-k^2\sin^2\zeta}}\mathrm{d}\zeta.
% \end{aligned}
% \end{equation}
% and, using the multiple angle identity involving sines,
% \begin{equation}
%     \begin{aligned}
%         \sin 4 \overline{\chi} &= \cos \overline{\chi} \left( 4 \sin \overline{\chi} - 8 \sin^3 \overline{\chi}  \right) \\
%         &= \left( 4 k \sin \zeta - 8 k^3 \sin^3 \zeta \right) \sqrt{1 - k^2 \sin^2 \zeta},
%     \end{aligned}
% \end{equation}
% the term,
% \begin{equation}
%     \begin{aligned}
%         \mathcal{I}_{22,\alpha}^S &= \int_{-\pi/2}^{\pi/2} B_{22}^S{}'[\varphi(\bar{\chi},\alpha)]\left( 4 k \sin \zeta - 8 k^3 \sin^3 \zeta \right)\mathrm{d}\zeta.
%     \end{aligned}
% \end{equation}
The integral may be readily evaluated using standard numerical methods. To find $\omega_\psi$, the bounce-averaged radial drift, though, we need to evaluate the leading order bounce time, $\tau_b$. Using the expression for $\mathcal{J}_\parallel^{(1)}$, the bounce time is equal to
\begin{equation}
    \tau_b = \partial_H \mathcal{J}_\parallel\approx\frac{B_{\alpha0}}{B_0\bar{\iota}_0}\frac{2 K(k)}{\sqrt{Hr\eta}}
\end{equation}
Hence, the bounce-averaged radial drift to leading order can readily be found to be
\begin{equation}
    \omega_\psi = - \frac{H r^2}{\bar{\iota}_0} \frac{ \mathcal{I}_{20,\alpha}(k,\alpha)}{2K(k)}.
\end{equation}
% \begin{equation}
%     \omega_\psi = - \frac{H r^2}{\bar{\iota}_0} \frac{ \mathcal{I}_{20,\alpha}(k,\alpha) + \mathcal{I}_{22,\alpha}^{C}(k,\alpha) + \mathcal{I}_{22,\alpha}^{S}(k,\alpha)}{2K(k)}.
% \end{equation}
% \par 
% In the scenario of constructing quasisymmetric configurations through near-axis expansion, as is well-known, it is generally not possible to guarantee the symmetry of $|\mathbf{B}|$ to second-order. The way in which this is generally dealt with formally is by allowing $B_{20}$ to be a function of $\varphi$, but none of the other pieces of $|\mathbf{B}|$. Thus, the bounce-average radial drift becomes simply driven by the $\mathcal{I}_{20,\alpha}$ term, 
% \begin{equation}
%     \omega_\psi=\frac{r^2H}{\bar{\iota}_0}\frac{ \mathcal{I}_{20,\alpha}(k,\alpha)}{2  K(k)}.
% \end{equation}
% where we remind ourselves that
% \begin{equation}
%     \varphi(\zeta,\alpha) = - \frac{\alpha - 2 \arcsin (k \sin \zeta)}{\bar{\iota}_0}.
% \end{equation}

\par
% To gain insight into the radial drift, we expand $B_{20}'[\varphi]$ as a power series in $k \sin \zeta$, thus
% \begin{equation}
% \begin{aligned}
%     B_{20}' \approx & B_{20}'(-\alpha/\bar{\iota}_0) + \frac{\partial B_{20}'}{\partial \varphi} \frac{\partial \varphi}{\partial k \sin \zeta} k \sin \zeta + ... \\ 
%     & \approx B_{20}' + \frac{2 B_{20}''}{\bar{\iota}_0} k \sin \zeta + \frac{2 B_{20}'''}{\bar{\iota}_0^2} (k \sin \zeta)^2 + ...
% \end{aligned}
% \end{equation}
% where in the final line we have omitted that $B_{20}$ is evaluated at $-\alpha/\bar{\iota}_0$. Inserting this expansion into the expression for $\mathcal{I}_{20,\alpha}$, one finds
% \begin{equation}
% \begin{aligned}
%     \mathcal{I}_{20,\alpha} \approx & \int_{-\pi/2}^{\pi/2} \frac{B_{20}'(-\alpha/\bar{\iota}_0)}{\sqrt{1 - k^2 \sin^2 \zeta}} \mathrm{d} \zeta + \int_{-\pi/2}^{\pi/2} \frac{2B_{20}'''(-\alpha/\bar{\iota}_0) (k \sin \zeta)^2}{\bar{\iota}_0^2\sqrt{1 - k^2 \sin^2 \zeta}} \mathrm{d} \zeta + ...  \\ 
%     \approx & 2 B_{20}' K(k) + \frac{4 B_{20}'''}{\bar{\iota}_0^2} \left[ K(k) - E(k) \right] + \\ 
%     & \frac{2}{9} \frac{\bar{\iota}_0^2 B_{20}''' + B^{(5)}_{20}}{\bar{\iota}_0^4} \left[ 2 ( 2 + k^2) K(k) - 4 (1 + k^2) E(k) \right] + ...
% \end{aligned}
% \end{equation}
% where we introduced the notation $B'''=B^{(3)}$. It should be evident that, all higher order terms in the expansion go to zero as $k \rightarrow 0$, and only the first term of the expansion survives. 
\par
The averaged radial-drift $\omega_\psi$ serves as a physically meaningful measure of the quasisymmetry quality of a configuration in this near-axis construction, vanishing when it is omnigeneous. The expression for $\omega_\psi$ is however a function of both $\alpha$ and $k$, and thus, for a single scalar measure that characterises the radial drift performance of a field at a given flux surface we must reduce it. Note that given the periodicity of $B_{20}$, the average of $\omega_\psi$ over all field lines (i.e., $\alpha$) vanishes. This might suggest that there is no net detrimental effect to having this radial drift, as on average, there is the `same' amount of particles going one way or the other. However, the neoclassical transport in the low collisionality limit as measured by $\epsilon_\mathrm{eff}$ is insensitive to the sign of $\omega_\psi$. To find a single measure then of its magnitude, then, we attempt to find an upper bound for $\omega_\psi$.
\par
\begin{figure}
    \centering
    \includegraphics[width=\textwidth]{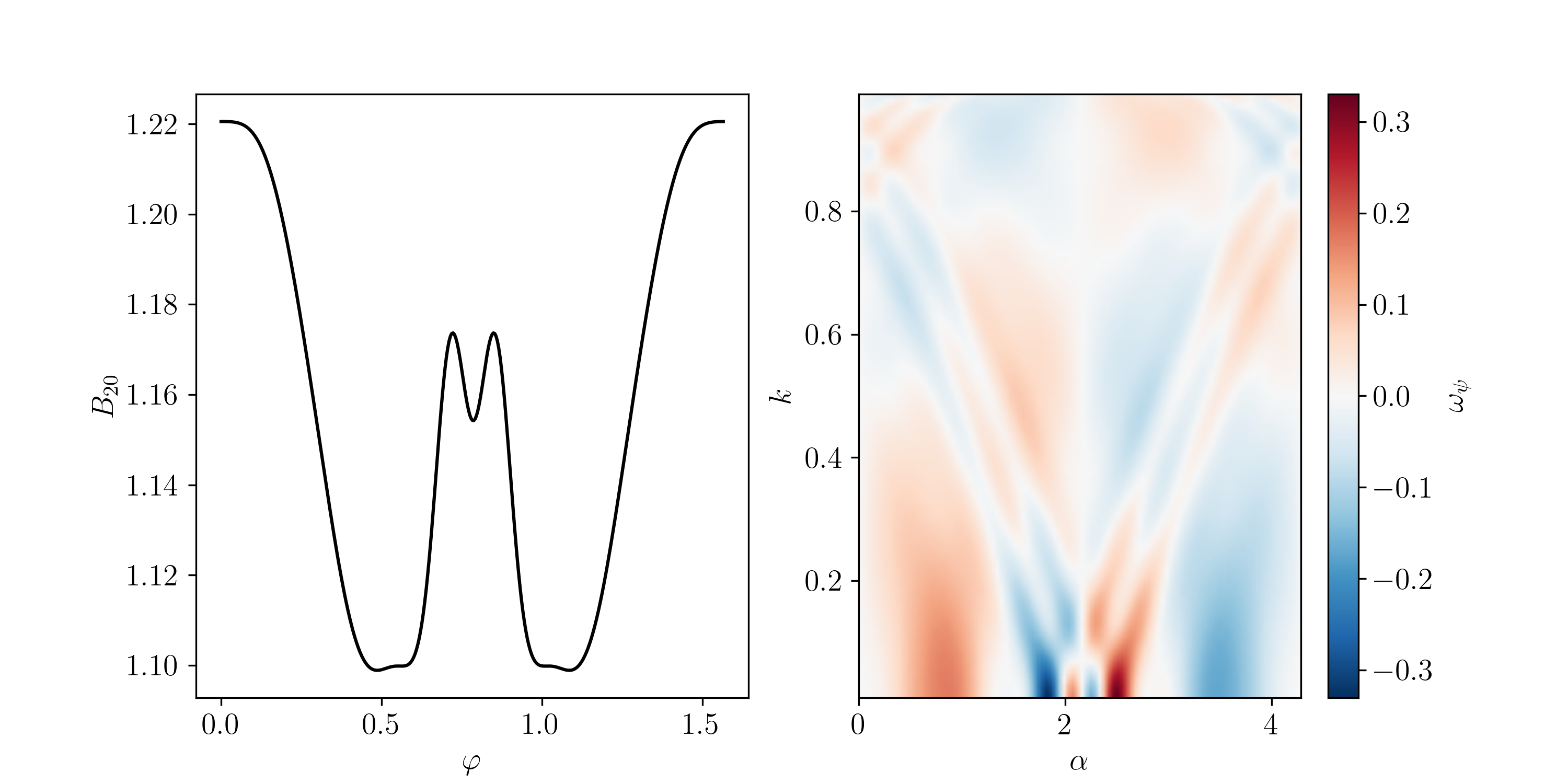}
    \caption{\textbf{Radial drift of precise QH in the near-axis description.} The plot shows the $B_{20}$ function of the precise QH near-axis construction as a function of $\varphi$, and (right) the bounce-average radial drift $\omega_\psi$ as a function of $\alpha$ and $k$. The maximum bounce-average radial drift $\omega_\psi$ occurs for the deeply-trapped population at the field line where the largest gradient of $B_{20}$ aligns with the minimum of the well.}
    \label{fig:B20_w_psi_precise_QH}
\end{figure}

Noting that the denominator in the integrand of $\mathcal{I}_{20,\alpha}$ is an always positive function within the integration domain, 
\begin{equation}
    \mathcal{I}_{20,\alpha} \leq 2 B_{20,\mathrm{max}}' K(k).
\end{equation}
Thus,
\begin{equation}
    \omega_\psi\leq \frac{r^2H}{\bar{\iota}_0}B_{20,\mathrm{max}}', \label{eqn:nae_measure_qs}
\end{equation}
with the equality only holding when the field line label $\alpha$ makes the largest gradient $B_{20,\mathrm{max}}'$ match the bottom of the well, and deeply trapped particles are considered (see Fig.~\ref{fig:B20_w_psi_precise_QH}). Only in this limit the particle samples the largest non-QS value of $B_{20}'$. Any other value will necessarily be smaller. 
\par
This bound provides a simple relation between the derivative of $B_{20}$ and the radial drift of particles. The derivative of $B_{20}$ is thus a more physical form of measuring quasisymmetry breaking within the near-axis framework compared to simply using the peak-to-peak $B_{20}$ variation as it is customary \citep{landreman2019constructing,landreman2022map,rodriguez2022phases}. 
\par
\section{Full expressions for the particle precession} \label{sec:appPrecess}  \label{sec:correspondence-to-connor}
In this appendix we present some key elements and expression for the calculation of the precession of trapped particles, $\omega_\alpha=\partial_\psi\mathcal{J}_\parallel/\partial_H \mathcal{J}_\parallel$. Obtaining such expressions from the asymptotic form of $\mathcal{J}_\parallel$ is straightforward, but it requires taking care of partial derivatives appropriately. 
\par
To that end, let us remind ourselves of the set of independent variables: $\psi$, $\alpha$, $H$ and $\mu$. Because we are using the toroidal flux over $2\pi$ as our flux surface label,
\begin{subequations}
\begin{gather}
    \frac{\partial r}{\partial\psi}=\partial_\psi\left(\sqrt{\frac{2\psi}{B_0}}\right)=\frac{1}{B_0 r},.
    \end{gather}
Furthermore, as shown in Appendix~\ref{app: second adiabatic invariant details}, the pitch angle is related to the trapping parameter via
\begin{equation}
    \lambda = \frac{1}{1 + r \eta (2 k ^2 -1) + \delta B_b }. 
    \label{eqn:def_lam_k_hat}
\end{equation}
We thus require expressions for $\delta B_b$, which may readily be computed as
\begin{equation}
\begin{aligned}
    \delta B_b  &= r^2 \left( B_{20} + B_{22}^C \cos 2 \chi_b \right) \\ 
                &= r^2 \left[ B_{20} + B_{22}^C \left( 1 - 8 k^2 + 8 k^4 \right)  \right].
\end{aligned}
\end{equation}
With this expression at hand, it is straightforward to compute the radial derivative of $k$ (at constant $\mu$ and $H$, and thus constant $\lambda$),
\begin{gather}
    \frac{\partial k}{\partial  r}\approx \frac{1 - 2k^2}{4kr} + \frac{B_{22}^C- B_{20}}{2 k \eta}.
\end{gather}
Similarly the derivative with respect $H$ can be found as
\begin{gather}
    \frac{\partial k}{\partial H}\approx \frac{1}{4kHr\eta} + \frac{(2k^2-1)(\eta^2 - 4 B_{22}^C)}{4 k H \eta^2}.
\end{gather}
\end{subequations}
The particle class label $k$ changes both with radius and particle energy, as it follows from the change in $|\mathbf{B}|$. 
\par 
We first use the above expressions to find the total bounce-averaged excursion in $\alpha$, $\Delta \alpha = \partial_\psi \mathcal{J}_\parallel$. Using the chain rule and keeping the leading two non-zero orders,
\begin{equation}
\begin{aligned}
    \Delta \alpha = & \sqrt{\frac{H \eta}{r}} \frac{B_{\alpha 0}}{\bar{\iota}_0 B_0} \frac{1}{B_0 r} \Bigg\{ 2\left[2 E(k) - K(k)\right] + \\
    & +r \eta \left[ (2 - 4k^2) E(k) + (1+2k^2) K(k) \right] - \\ 
    & -\frac{4r B_{20}  }{\eta}K(k)  + \frac{4r B_{22}^C}{\eta} \left[ (2k^2-1) E(k) + (1-k^2)K(k) \right] \Bigg\}.
    \label{eq: delta alpha}
\end{aligned}
\end{equation}
Next, the bounce time $\tau_b=\partial_H \mathcal{J}_\parallel$ is
\begin{equation}
\begin{aligned}
    \tau_b=&\frac{B_{\alpha0}}{B_0\bar{\iota}_0}\frac{1}{\sqrt{Hr\eta}}\Bigg\{ 2K(k) + \\
    & +r\eta \left[ 4 E(k) + (2k^2 - 3) K(k) \right] +\frac{4r B_{22}^C}{\eta} \left[ E(k) -k^2 K(k) \right] \Bigg\}.
\end{aligned} \label{eqn:t_b}
\end{equation}
The trapped-particle precession is now readily found by taking the ratio of $\Delta \alpha/\tau_b$ and expanding to include the leading-order and the first correction terms. Doing the expansion, one finds this to be equal to
\begin{equation}
\begin{aligned}
    \omega_\alpha =& \frac{H\eta}{B_0 r} \Bigg\{ 2 \frac{E(k)}{K(k)} - 1 +  \\
    & +r\eta \left[ - 4\left( \frac{E(k)}{K(k)} \right)^2 + 2(3 - 2k^2)\frac{E(k)}{K(k)} + (2k^2 - 1)  \right] - \\ 
    & -\frac{2rB_{20}}{\eta} +\frac{2r B_{22}^C}{\eta}\Bigg[ -2\left( \frac{E(k)}{K(k)} \right)^2 +4k^2 \frac{E(k)}{K(k)}+ (1 - 2 k^2) \Bigg] + O(r^2) \Bigg\},
    \label{eq: omega alpha appendix}
\end{aligned}
\end{equation}
which is equivalent to Eq. \eqref{eq:trapped-particle-precession-full}.

\section{Precession in terms of triangularity and pressure gradient} \label{app: relation to geometry}
In the main text and Appendix~\ref{sec:appPrecess} we showed how to get an expression for the precession of trapped particles at second order in the near-axis expansion. This involved the parameters $B_{22}^C$ and $B_{20}$, natural parameters in the near-axis expansion which directly describe the shape of $|\mathbf{B}|$. Although most natural in the calculations of trapped-particle precession, they do however not offer a clear connection to the physical nor geometric features of the field.
\par
Within the near-axis framework, though, such a connection can be made. At second order in the near-axis expansion of an exactly quasisymmetric, stellarator-symmetric configuration, $B_{22}^C$ and $B_{20}$ can be re-expressed as linear combinations of pressure gradient, $p_2$, and triangularity of an up-down cross-section, $\delta$. The latter two can then be used as the independent parameter choices at second order in the expansion of the field.
\par
The definition of pressure gradient is straightforward as the flux derivative of the pressure on axis, $p_2=(B_0/2)\mathrm{d}p/\mathrm{d}\psi$. The notion of triangularity, $\delta$, requires additional care. We will define $\delta$ as the relative displacement of the vertical turning points of the cross-section to the width of the cross section along the up-down symmetry line (normalised by $r$), and do so in the $(\kappa,~\tau)$ Frenet frame. That is, in the plane orthogonal to the magnetic axis where the cross-section is up-down symmetric. Asymptotically, and in terms of the near-axis expansion quantities \citep{landreman2019constructing,rodriguez2023mhd},
\begin{equation}
    \delta= 2~\mathrm{sign}(\eta)\left(\frac{Y_{22}^S}{Y_{11}^S}-\frac{X_{22}^C}{X_{11}^C}\right). \label{eqn:triang-def}
\end{equation}
Here the $X$ and $Y$ expansion parameters describe the flux surface shape in the Frenet-Serret frame, and details on them may be found in \cite{landreman2019constructing}. Note that dimensionally, the definition in Eq.~(\ref{eqn:triang-def}) has units of inverse length. This is so because $\delta$ is defined not as triangularity, but rather as $1/r$ times triangularity, which accounts for the fact that close to the magnetic axis, where cross-sections are elliptical, the triangularity vanishes. This notion of normalised triangularity in the plane normal to the axis, as explained in \cite{rodriguez2023mhd}, is generally different to the common definition of triangularity in the lab-frame, $\delta_\mathrm{lab}$. That is, it is not equivalent to the triangularity of the cross-section that results from making a cut of the configuration at a constant cylindrical angle. If the magnetic axis has a relative tilt respect to the cylindrical coordinate system, then $\delta_\mathrm{lab}=\delta+\Lambda$, where $\Lambda$ is a term that depends only on the axis shape and first order near-axis shaping.\footnote{We note that the expression in Eq.~(C2) in \cite{rodriguez2023mhd} for $\Lambda$ is incorrectly simplified, as it assumes certain symmetry that does generally not exist. The correct expression will be presented in a future publication, but makes no difference to the discussion in the present paper.} In the special case of an axisymmetric field, this geometric transformation term $\Lambda$ vanishes. In general, though, varying $\delta_\mathrm{lab}$ or $\delta$ is equivalent other near-axis features kept constant. 
\par
With the notions of triangularity and pressure gradient in place, the equilibrium field of a quasisymmetric configuration can be uniquely defined at second order \cite{rodriguez2023mhd}. In the axisymmetric limit this is evident following the behaviour of the Grad-Shafranov equation and the set-up of its solutions. In the general quasisymmetric configuration this is more subtle, as the cross-section shapes change around the torus driven by the asymmetry of the magnetic axis. Specifying the triangularity of a single cross-section might then appear insufficient to describe uniquely the whole field, but the conditions of quasisymmetry and equilibrium are sufficient to grant this uniqueness. Schematically, one may picture the situation as a kind of initial value problem in which the single cross-section is the initial value, and the axis-shape describes the flow of evolution. There are therefore different ways of describing the very same field, as different cross-sections may be chosen.
\par
It should appear clear that the shape of the axis thus plays a crucial role in the problem, especially through its asymmetry. Described by its curvature, $\kappa$, and torsion, $\tau$, it is natural to construct a measure of said asymmetry like $\bar{F}$, introduced in \cite{rodriguez2023mhd}, and defined to be,
\begin{equation}
    \bar{F}=2\left[\frac{(I_2-B_0\tau)/\kappa^2}{\int_0^{2\pi}\mathrm{d}\varphi(I_2-B_0\tau)/\kappa^2}\frac{\int_0^{2\pi}\mathrm{d}\varphi(1+\sigma^2+\eta^4/\kappa^4)}{1+\eta^4/\kappa^4}-1\right],
    \label{eqn:Fbar}
\end{equation}
where $I_2$ is the toroidal current and $\sigma$ is a measure of up-down asymmetry, solution to the Riccati $\sigma$ equation (see \cite{landreman2019constructing}), and thus not a degree of freedom. The measure $\bar{F}$ must be evaluated at the location of an up-down symmetric cross-section. As we are assuming stellarator symmetry, there are at least two such distinct poisitions (in the axisymmetric limit, an infinite number of them, but $\bar{F}=0$ in that case). This emphasises the freedom mentioned before about the description of stellarator fields; there are two naturally simple ways of identifying the very same field, depending on which up-down symmetric cross-section is chosen to identify the configuration with. Depending on this choice, the meaning of the `effects of changing the cross-section' (and in particular triangularity) will of course change, and with it the conclusions derived. One must interpret it as the effects of changing the shape of that very cross-section, modifying the remaining of the configuration in a consistent way, keeping the axis and profiles fixed. For consistency and analogy with the typical axisymmetric case, we shall choose to identify configurations with their most vertically elongated, up-down symmetric cross-section, which often exhibit a bean-shape.
\par
With the definitions above, the magnetic parameters $B_{20}$ and $B_{22}^C$ may be written explicitly in terms of the pressure gradient $p_2$ and the triangularity $\delta$ (defined to be positive in the direction of the normal curvature) of the cross-section at $\varphi=0$, \begin{subequations}
    \begin{equation}
        B_{20}=-\frac{\mu_0 p_2}{B_0^2}\left[1+\frac{4\sqrt{\bar{\alpha}}}{(\bar{\alpha}+3)-(\bar{\alpha}+1)\Bar{F}}\left(\frac{\eta B_{\alpha0}}{B_0\bar{\iota}_0}\right)^2\right]-\frac{3}{2}|\eta|\frac{(1-\bar{\alpha})+(\bar{\alpha}+1)\Bar{F}}{(\bar{\alpha}+1)\Bar{F}-(3+\bar{\alpha})}\delta +\dots,
    \end{equation}
    \begin{equation}
        B_{22}^C=-\frac{2\mu_0 p_2}{B_0^2}\frac{\sqrt{\bar{\alpha}}}{(\bar{\alpha}+1)\Bar{F}-(\bar{\alpha}+3)}\left(\frac{\eta B_{\bar{\alpha}0}}{B_0\bar{\iota}_0}\right)^2-\frac{|\eta|}{2}\frac{(3-\bar{\alpha})+(\bar{\alpha}+1)\Bar{F}}{(3+\bar{\alpha})-(\bar{\alpha}+1)\Bar{F}}\delta +\dots,
    \end{equation}
\end{subequations}
where $\bar{\alpha}=(\eta/\kappa(0))^4$ and the dots denote terms independent of second order choices that we shall not focus on. 
\par
With these expressions in place, we may then rewrite the effects of second-order on the trapped-particle precession (noting that we assumed $\eta>0$ in the adiabatic invariant calculation in this paper),
\begin{equation}
    \omega_{\alpha,1}=\frac{H \eta}{B_0}\left[\eta\tilde{\mathcal{G}} - \frac{\mu_0p_2}{\eta B_0^2}\mathcal{G}_{p2} + \frac{\delta}{2}\mathcal{G}_{\delta}\right],
\end{equation}
and write,
\begin{subequations}
    \begin{equation}
        \mathcal{G}_{p_2}= \mathcal{G}_{20}(k)+\left(\frac{\eta B_{\alpha0}}{B_0\bar{\iota}_0}\right)^2\frac{2\sqrt{\bar{\alpha}}}{(\bar{\alpha}+3)-(\bar{\alpha}+1)\Bar{F}}[2\mathcal{G}_{20}(k) -\mathcal{G}_{22}^C(k)] \label{eqn:Gp2_qs}
    \end{equation}
    \begin{equation}
        \mathcal{G}_{\delta}= \frac{3(1-\bar{\alpha})\mathcal{G}_{20}(k)-(3-\bar{\alpha})\mathcal{G}_{22}^C(k)}{(3+\bar{\alpha})-(\bar{\alpha}+1)\Bar{F}}+\frac{(\bar{\alpha}+1)[3\mathcal{G}_{20}(k)-\mathcal{G}_{22}^C(k)]\Bar{F}}{(3+\bar{\alpha})-(\bar{\alpha}+1)\Bar{F}}.
    \end{equation}
\end{subequations}
These two expressions describe the effect of the pressure gradient and the cross-section triangularity on the trapped-particle precession as a function of the class of trapped particle. Note that $\tilde{\mathcal{G}}$ is independent of the pressure gradient and triangularity both. As such, when investigating dependencies on these parameters (at fixed $\eta$ and axis), we may safely ignore contributions of $\tilde{\mathcal{G}}$. The derivation may be checked with computer algebra, as provided in the repository associated to this paper.

\subsection*{Relation to the large-aspect ratio circular tokamak limit}
Here we relate the found results to the large-aspect ratio tokamak limit investigated by \citet{connor1983effect}. Let us start comparing the pressure dependence of precession, and as such let us define $\alpha_p= - 4 \mu_0 r p_2/ |\eta | B_0^2 \bar{\iota}_0^2 $. One may then write
\begin{equation}
\begin{aligned}
    -\frac{\mu_0 r p_2}{|\eta|B_0^2}\mathcal{G}_{p2} &= \frac{\alpha_p \bar{\iota}_0^2}{4} \mathcal{G}_{p2} \\
    &= - \frac{ \alpha_p \bar{\iota}_0^2}{2}+ \frac{ \alpha_p}{2} \left(\frac{\eta B_{\alpha0}}{B_0}\right)^2 \frac{4\sqrt{\bar{\alpha}}}{(\bar{\alpha}+3)-(\bar{\alpha}+1)\Bar{F}} \left(   -1 -  \frac{\mathcal{G}_{22}^C}{4}\right).
\end{aligned}
\end{equation}
Specialising to the circular tokamak case (i.e. $\bar{\alpha}=1$ and $\bar{F}=0$), we find
\begin{equation}
    -\frac{\mu_0 r p_2}{|\eta|B_0^2}\mathcal{G}_{p2} = - \frac{ \alpha_p \bar{\iota}_0^2}{2} - 2\alpha_p \cdot \underbrace{ \frac{1}{4} \left[  \frac{E(k)}{K(k)} \left( 2 k^2 - \frac{E(k)}{K(k)}\right)  +\frac{3}{2} - k^2 \right] }_{\stackrel{\cdot}{=} G_{\alpha,p}(k)}.
\end{equation}
% \begin{equation}
%     \mathcal{G}_{22}^C(k)\stackrel{\cdot}{=}-4\Bigg[\left(\frac{E(k)}{K(k)}\right)^2 - 2 k^2 \frac{E(k)}{K(k)} + \left(k^2-\frac{1}{2}\right) \Bigg].
% \end{equation}
In the circular limit then, we may write,
\begin{equation}
    \omega_{\alpha,\mathrm{circle}} = \frac{H \eta}{B_0 r} \left( G(k)- \frac{\alpha_p \bar{\iota}_0^2}{2} - 2 \alpha_p G_{\alpha,p}(k) \right),
\end{equation}
acknowledging that we are mixing terms of different order in $r$, and not keeping all relevant terms consistently to the right order. 
% \begin{equation}
%     \omega_{\alpha,\mathrm{circle}} = \frac{H \eta}{B_0 r} \left( G(k) + 4s G_s(k) - \frac{\alpha_p \bar{\iota}_0^2}{2} - 2 \alpha_p G_{\alpha,p}(k) \right)
% \end{equation}
\par
This expression has a similar form to Eq.~(9) in \citet{connor1983effect}. The first two terms are exactly identical, while the latter is different from, 
\begin{equation}
    G_{\alpha,\mathrm{Connor}}(k) = \frac{2}{3} \left( \frac{E(k)}{K(k)} \left( 2 k^2 - 1\right)  +1 - k^2 \right).
\end{equation}
We thus see that, though the functional form is similar, it is not the same, especially near deeply trapped particles.
% , and a more stark difference may be found in the magnitude which is different by a factor of $8/3$. 
These differences correspond to the difference between the models considered and the meaning of changing a certain parameter thereof. In the approach of Connor, the field is being constructed 
% "artifice of introducing an enhanced pressure gradient" \citep{connor1983effect} is made, and most importantly, 
in a way that locally, at finite radius, equilibrium is satisfied (as explicitly shown in \citet{roach1995trapped}). Such a method requires the shape of the flux surface and radial variations of various geometric and equilibrium quantities to fully specify the local equilibrium conditions. Furthermore, many of these parameters are treated independently, and one requires subsidiary assumptions in order to set these values (in particular, the `artifice' \citep{connor1983effect} of ordering the pressure in a particular form). Though this does allow one to investigate the effect of such parameters independently, it is not guaranteed that there exists a global solution adhering to the set of local conditions. The near-axis approach presented, although asymptotic in nature, is valid in some region near the axis, and as such can perhaps be thought of a `more global' solution. This translates into a larger degree of coupling between the geometry and the magnitude of the field (due to the singular nature of the axis) compared to the radially local approach, which ends up reducing the number of free parameters while it retains realism. 
% { This discussion, have a look at that. Maybe reference Hegna and local considerations (complication as input become full shapes). I guess we must make a point about why the near-axis approach is approopriate, and what it provides beyond the large aspect ratio case. I guess an important point is that it minimises the requirement of subsudary assumptions, guarantees consistency, and allows for generalisations and connections to other physics straightforwardly. Simplicity. Note that we could repeat the considerations here of AE for the large aspect ratio tokamak as well. }
\par
The difference is especially noticeable in the involvement of the magnetic shear, which appears explicitly in \cite{connor1983effect}, but becomes a higher order effect in the current treatment. In fact, we may formally obtain a sense of the involvement of the magnetic shear by considering the next order in the expansion for the precession, and focusing on the variation of the field line length due to the change in $
\iota$ when taking the derivative of $\mathcal{J}_\parallel$ with respect to $\psi$, it can straightforwardly be shown to give,
\begin{equation}
\begin{aligned}
    \omega_{\alpha,\mathrm{shear}} =&- 4 \frac{H \eta}{B_0 r}  \frac{r \partial_r \bar{\iota}}{\bar{\iota}_0}  \left( \frac{E(k)}{K(k)} + (k^2 - 1) \right)  \\ 
    \stackrel{\cdot}{=}& 4 s  \frac{H \eta}{B_0 r} \underbrace{ \left( \frac{E(k)}{K(k)} + (k^2 - 1) \right)}_{\stackrel{\cdot}{=} G_s(k)},
\end{aligned}
\end{equation}
where we have defined the magnetic shear as $s = - r \partial_r \bar{\iota}/\bar{\iota}_0$, and the term is clearly a second order effect. This term has precisely the same form as in \citep{connor1983effect}. To arrive at such term we considered the magnetic shear separate from other elements in the model (an independence that is partially correct given the freedom in the toroidal current profile), but hides connections to higher order considerations within the near-axis framework \citep{rodriguez2021weak}, especially its ties to the third order harmonics of $|\mathbf{B}|$, which we would expect to modify this dependence at least partially.  

\section{Critical value estimates for plasma $\beta$ and triangularity}
\label{sec:appCritEst}
{
In the main text we presented some estimates for the magnitude of plasma $\beta$ and cross-section triangularity that made their effect on precession of deeply trapped particles and available energy significant. In this Appendix we present the values for these estimates for the sample near-axis QS configurations in Fig.~\ref{fig:qs_config_wa_1}. 
\par
Table~\ref{tab:criticalValues} includes the following parameters. First, the critical plasma $\beta$ for reversal of the deep-particle precession at the edge of the configuration, Eq.~(\ref{eqn:crit_beta}),
\begin{equation}
    \beta_\mathrm{crit}=a|\eta|\frac{2}{\mathcal{G}_{p_2}(0)},
\end{equation}
with $a$ the minor axis, and taking $r=a$. The second is the critical beta but for the available energy (see Appendix~\ref{sec:appAEdet} and Sec.~\ref{sec:AE}), 
\begin{equation}
    \beta_\mathrm{crit}^\mathrm{AE}=-a|\eta|\frac{2}{\mathcal{R}_{20}(\mathcal{G}_{p_2}(0)+1)},
\end{equation}
where the denominator is given in Eq.~(\ref{eqn:A1_weak}).
\par
For the critical triangularity, the value of $(r\delta)$ to revert the precession of deeply trapped particles, we consider the definition in Eq.~(\ref{eqn:crit_delta}),
\begin{equation}
    (r\delta)_\mathrm{crit}=\frac{2}{\mathcal{G}_\delta(0)}.
\end{equation}
Similarly, for the available energy, we shall use the same expression but for the reinterpretation of $\mathcal{G}_\delta$ with
\begin{equation}
    \mathcal{G}_\delta^\mathrm{AE}=3\mathcal{R}_{20}\frac{(1-\bar{\alpha})+(1+\bar{\alpha})\bar{F}}{(3+\bar{\alpha})-(\bar{\alpha}+1)\bar{F}},
\end{equation}
which follows directly from Eq.~(\ref{eqn:A1_weak}). Finally, for a reference, we shall compare this critical triangularity to $r_c\delta$, the maximum attainable triangularity in a given near-axis configuration without incurring in unphysical flux surface shapes. 
\begin{table}
    \centering
    \begin{tabular}{c|c|c|c|c|c|c|c|c|c|c|}
      & PQA & PQAw & PQH & PQHw & 22QA & 22QH2 & 22QH3v & 22QH3b & 22QH4l & 22QH4w  \\\hline
      $\beta_\mathrm{crit}$ & 0.06 & 0.05 & 0.14 & 0.12 & 0.05 & 0.06
 & 0.10 & 0.11 & 0.11 & 0.11  \\
     $\beta_\mathrm{crit}^\mathrm{AE}$ & 0.02 & 0.02 & 0.07 & 0.06 & 0.02 & 0.02
 & 0.04 & 0.05 & 0.05 & 0.06 \\
 $(r\delta)_\mathrm{crit}$ & 0.96 & 1.09 & 1.47 & 1.42 & 1.05 & 0.99 & 1.22 & 1.32 & 1.40 & 1.36  \\ 
  $(r\delta)_\mathrm{crit}^\mathrm{AE}$ & 0.47 & 0.61 & 1.43 & 1.25 & 0.57 & 0.50
 & 0.81 & 1.00 & 1.19 & 1.09 \\ 
 $r_c\delta$ &  2.02 & 3.46 & 1.60 & 2.23 & 3.64 & 0.09 & 1.08 & -0.29 & 1.18 & 5.13 \\
 $\Delta_\mathrm{QS}$ & 0.06 & 0.20 &  0.34 & 0.99 & 0.18 & 0.51 & 0.001 & 0.16 & 0.09 & 0.0003
 \\\hline

 \end{tabular}
    \begin{tabular}{c|c|c|c|c|c|c|}
         & 22QH4m & 22QH7 & ARIESCS & GAR & HSX* & QHS48 \\\hline
       $\beta_\mathrm{crit}$  & 0.16 & 0.14 & 0.06 & 0.05 & 0.15 & 0.11 \\
       $\beta_\mathrm{crit}^\mathrm{AE}$ & 0.06 & 0.05 & 0.03 & 0.02 & 0.07 & 0.06 \\
       $(r\delta)_\mathrm{crit}$ & 0.78 & 1.44 & 1.02 & 0.87 & 1.40 & 1.51 \\
       $(r\delta)_\mathrm{crit}^\mathrm{AE}$ & 0.33 & 1.30 & 0.53 & 0.40 & 1.19 & 1.57 \\ 
       $r_c\delta$ & -0.91 & 0.13* & 0.47 & 1.90 & 0.04 & 0.48 \\ 
       $\Delta_\mathrm{QS}$ & 0.02 & 0.0005 & 0.70 & 0.53 & 8.05 & 0.05 \\\hline
    \end{tabular}
    \caption{\textbf{Critical plasma $\beta$ and triangularity in QS configurations.} The table gathers critical plasma $\beta$, triangularity $r\delta$ and QS measure $\Delta_\mathrm{QS}$ values for a number of near-axis QS configurations, configurations used in Fig.~\ref{fig:qs_config_wa_1}. The short names for the configurations follow straightforwardly from their full labels in Fig.~\ref{fig:qs_config_wa_1}. The starred triangularity corresponds to the triangularity for an aspect ratio 10 consideration (as there is difficulty in computing $r_c$). An aspect ratio of 10 was assumed to evaluate the $\beta_\mathrm{crit}$ values. One should take the case of HSX sceptically, as the near-axis description is far from being quasisymmetric (a clearly indicated by $\Delta_\mathrm{QS}$; in fact, it also has a very small $r_c$, hence the small value of attainable triangularity). One could attempt modifying the near-axis configuration to make it more quasisymmetric and a better description, but we shall not do this here. }
    \label{tab:criticalValues}
\end{table}
\par
Finally, to include a sense of the QS quality, we introduce the QS measure motivated in Appendix~\ref{sec:app-radial-drifts}, Eq.~(\ref{eqn:nae_measure_qs}),
\begin{equation}
    \Delta_\mathrm{QS}=\left|\frac{B_{20,\mathrm{max}}'}{\bar{\iota}_0}\right|,
\end{equation}
where a vanishing value indicates an exactly QS configuration to 2nd order. This measure shows that some near-axis configurations (especially that of HSX) lie far from ideal QS. In the special case of HSX, the near-axis model is constructed to reproduce the leading harmonic content of $|\mathbf{B}|$; small variations can however lead to significant effects especially at second order in the near-axis expansion (that is, where we asssess QS or compute triangularity). Because we are using these as illustrating examples, we do however not consider a further refinement of these configurations.
\par
The values in Table \ref{tab:criticalValues} show that physically relevant finite beta effects can have significant effect on the leading order behaviour of both deeply trapped particle precession and \AE{}. This effect is stronger in quasi-axisymmetric configurations, and we should remind ourselves, on the outermost surface. As the magnetic axis is approached, the critical $\beta$ needed will grow. Interestingly, \AE{} is more susceptible than the precession itself. The case of triangularity shows that a significant degree of shaping is required to affect either precession or \AE{}. This level of shaping is in many configurations present or exceeded, as the relative comparison of the critical $r\delta$ to $r_c\delta$ shows. Once again, these effects become strongest as the outer surface is approached.

}
\section{Details of asymptotic available energy integral} \label{sec:appAEdet}
In this appendix we provide mathematical and algebraic details concerning the asymptotic evaluation of \AE{} in the two considered regimes. Note that these considerations may be extended simply to other approximation schemes, such as a large aspect ratio model.
\subsection{The weakly driven regime}
To perform the available energy integral in this regime, we showed how it is the popuation of trapped particles that have an almost vanishing precession that contribute. Thus, we first re-define the zero crossing of the trapped particle precession, $k_0$,
\begin{equation}
    G(k_0) = 0.
\end{equation}
To make further progress with the integral, we ought to transform the integration variable $\lambda$ into $k$, and thus require $\mathrm{d} \lambda / \mathrm{d} k$. Details of this derivation are included in Appendix~\ref{sec:appPrecess}, and to leading order it reduces to
\begin{equation}
    \frac{\mathrm{d}\lambda}{\mathrm{d}k} \approx -4 r\eta k.
\end{equation}
The final component needed for the integral is the normalized bounce-time $\hat{G}^{1/2} \approx \int (1 - \lambda \hat{B})^{-1/2} \mathrm{d}\ell / L$, for which the leading order expression reduces to
\begin{equation}
    \hat{G}^{1/2}(k) = \frac{2 \sqrt{2} K(k) }{\sqrt{r \eta}}\frac{ B_{\alpha0}}{B_0\bar{\iota}_0L},
\end{equation}
which may be calculated analytically or verified via the expressions for the bounce-time given in Appendix \ref{sec:appPrecess}.
\par
To explit the existence of a narrow contributing band, we will then make a local approximation of the integrand about $k_0$. The region can be shown to be small,
\begin{equation}
    \hat{\omega}_\alpha \approx \frac{\Delta \psi \eta}{r B_0} G'(k_0)(k-k_0) \sim O(1) \implies (k-k_0) \sim r,
    \label{eq:app-linearisation}
\end{equation}
so that conveniently using $c_1$ as integration variable, 
\begin{equation}
    k = \frac{\hat{\omega}_{*,0}^T}{\frac{\Delta \psi \eta}{r B_0} G'(k_0) c_1} + k_0 \longrightarrow  \frac{\mathrm{d} k}{\mathrm{d} c_1} = -\frac{1}{c_1^2} \frac{r B_0 \hat{\omega}_{*,0}^T}{\Delta \psi \eta G'(k_0)}.
\end{equation}
the remaining integral becomes,
\begin{equation}
    \int_0^\infty \frac{\mathcal{F}(c_1)}{c_1^2} \mathrm{d} c_1 = \frac{\sqrt{\pi}}{6}.
\end{equation}
In doing so, we approximated the integration domain $ c_1 \in \left( \hat{\omega}_{*,0}^T B_0 r/\Delta \psi \eta | G'(k_0) | k_0, \infty \right]$ as $ (0,\infty ]$. Doing so only introduces an error of order $\sqrt{r}$ on the integral, which is the asymptotic value of $\mathcal{F}$ at small argument.
\par
Collecting all the factors involved, the leading order expression of the \AE{} simply becomes
\begin{equation}
    \widehat{A} \approx \frac{4 \sqrt{2 \pi}}{3 \mathcal{V}} \frac{B_{\alpha 0}N_\mathrm{wells}}{B_0 \bar{\iota} L}  \frac{k_0 K(k_0)}{|G'(k_0)|} \left( \hat{\omega}_{*,0}^T \right)^3 \frac{r B_0}{\Delta \psi}\sqrt{\frac{r}{\eta}}  .
\end{equation}
Computing the dimensionless factor $\mathcal{V}$ to leading order, 
\begin{equation}
    \mathcal{V}=\int \frac{2 \sqrt{\pi}}{\hat{B}} \frac{\mathrm{d}\ell}{L}\approx 4\pi^{3/2}\frac{B_{\alpha0}N_\mathrm{wells}}{B_0 \bar{\iota}_0 L}.
\end{equation}
and easing notation by approximating
\begin{equation*}
    \frac{k_0 K(k_0)}{|G'(k_0)|} = 0.666834\dots\approx\frac{2}{3} \pm 0.03 \%,
\end{equation*}
we find the result
\begin{equation*}
    \widehat{A} \approx \frac{2\sqrt{2}}{ 9 \pi} \left( \hat{\omega}_{*,0}^T \right)^3 \frac{r B_0}{\Delta \psi} \sqrt{\frac{r}{\eta}}.
\end{equation*}
\par 
One may now retrieve the result stated in the main text by imposing that $\Delta \psi = C_r r B_0 \rho$, $\varrho = r/a$, $\rho_* = \rho/a$, and $-\partial_\varrho \ln n = \hat{\omega}_n$, resulting in
\begin{equation}
    \widehat{A} \approx  \frac{2\sqrt{2}}{ 9 \pi} C_r^2 \rho_*^2 \left(\frac{\hat{\omega}_n}{\varrho}\right)^3  \frac{\varrho^3\sqrt{\varrho}}{\sqrt{a \eta}}.
\end{equation}
This concludes the analysis of the weakly driven asymptotic regime.

\subsection{The strongly driven regime}
The integral in the strongly driven regime is straightforward. Our starting point is the full integral given in
\eqref{eqn:AEinteg},
\begin{equation}
    \widehat{A}=\frac{(\hat{\omega}^T_{*,0})^2}{\mathcal{V}}\int\mathrm{d}\lambda \sum_\mathrm{wells(\lambda)}\hat{G}^{1/2}\mathcal{F}(c_1)\Theta(\omega_\alpha). \tag{\ref{eqn:AEinteg}}
\end{equation}
We again employ the fact that $\mathrm{d} \lambda / \mathrm{d} k \approx-4 k r \eta$ to write
\begin{equation}
     \widehat{A}= \frac{2 \sqrt{2}}{\pi^{3/2}} \left( \hat{\omega}_{*,0}^T \right)^2 \sqrt{r \eta} \int k K(k) \mathcal{F}\left(c_1\right) \Theta(\hat{\omega}_\alpha) \mathrm{d} k
\end{equation}
In the strongly driven limit, the main assumption is that $c_1\gg1$ for all $k$, and thus we may replace $\mathcal{F}$ with its asymptotic form for large $c_1$. Using \cite[Secs.~7.2,~7.12]{DLMF}, $\mathcal{F}(c_1) \approx 3 \sqrt{\pi}/4c_1$, which yields 
\begin{equation}
     \widehat{A}= \frac{3\Delta r}{\pi\sqrt{2}}  (\hat{\omega}_{*,0}^T) \eta \sqrt{r\eta} \int_0^{k_0} k K(k) G(k)\mathrm{d} k.
\end{equation}
One can numerically evaluate the integral to find
\begin{equation}
    \int_0^{k_0} k K(k) G(k)\mathrm{d} k \approx 0.386842.
\end{equation}
Introducing the dimensionless variables as we did in the weak regime, we find
\begin{equation}
    \widehat{A}= 1.1605~\frac{(C_r\rho_*)^2}{\pi\sqrt{2}}   \hat{\omega}_n (a \eta)^{3/2} \sqrt{\varrho},
\end{equation}
which is our final result.
\subsection*{Dependence of \AE{} on $B_{20}$ and $B_{22}^C$}
Before concluding this appendix, we show how to proceed to assess the effect of $B_{22}^C$ and $B_{20}$ on the expression of available energy. The main idea here is that we would like to evaluate these contributions without having to evaluate all the consistent asymptotic dependence of \AE{}. For instance, we are not interested in the order $r^{1/2}$ correction to the weak $\widehat{A}$ in Eq.~(\ref{eqn:AE_lead}) due to the finite extent of $\mathcal{F}$. To see how to proceed let us write the relevant parts of the available energy integral (setting overall factors aside for simplicity of notation),
\begin{equation}
    \widehat{A}\sim\sum_\mathrm{wells}\int\mathrm{d}c_1\frac{\mathrm{d}\lambda}{\mathrm{d}k}\frac{\mathrm{d}k}{\mathrm{d}c_1}\tau_b\mathcal{F}(c_1).
\end{equation}
To evaluate higher order corrections to the integral in the weak regime, we remind the reader first that the calculation involves the local expansion of the integrand. In that spirit, the changes due to second order on the first factor in the integral is straightforward, and may be simply read off Eq.~(\ref{eqn:def_lam_k_hat}). It gives as a correction on the leading order integral 
\begin{equation}
    \mathcal{R}_1=\frac{4rB_{22}^C}{\eta}(2k_0^2-1)
\end{equation}
where we have evaluated it at the original $k_0$ to leading order. A similarly simple correction arises from the corrections to the bounce time $\tau_b$, which may simply be read off Eq.~(\ref{eqn:t_b}),
\begin{equation}
    \mathcal{R}_2=\frac{2rB_{22}^C}{\eta}\left(\frac{E(k_0)}{K(k_0)}-k_0^2\right).
\end{equation}
The contributions left come from the $c_1$ dependent terms, $\mathcal{F}(c_1)\mathrm{d}k/\mathrm{d}c_1$. Here the change in the zero-crossing of precession due to second order corrections on $|\mathbf{B}|$ contribute to the available energy. We must thus assess where the new zero is. With the precession defined as $\omega_\alpha=\omega_{\alpha,0}+\omega_{\alpha,1}$, Eq.~(\ref{eq:trapped-particle-precession-full}), we rewrite it as
\begin{equation}
    \omega_{\alpha}=(\eta/rB_0)[G(k)+(r/\eta)\hat{\mathcal{G}}(k)],
\end{equation}
ignoring every term that does not depend on $B_{20}$ or $B_{22}^C$, and recalling that $\hat{\mathcal{G}}=B_{20}\mathcal{G}_{20}+B_{22}^C\mathcal{G}_{2c}$, Eq.~(\ref{eqn:G_funcs_2}). Then, define the new $\omega_\alpha(k^\star) = 0$ point with $k^\star=k_0+r\delta k$, so that 
\begin{equation}
    \delta k\approx-\frac{\hat{\mathcal{G}}(k_0)}{\eta G'(k_0)}.
\end{equation}
With this ``displaced'' $k^\star$, we may assess the change in the available energy by simply considering the perturbation of the leading order $k_0$ dependence, namely $k_0K(k_0)/G'(k_0)$, noting that the denominator comes from $\omega_{\alpha}'$ and thus it has an additional contribution. Thus, the third correction may be written as,
\begin{equation}
    \mathcal{R}_3=-\frac{r}{\eta}\frac{1}{G'(k_0)}\left[\left(1+\frac{K'(k_0)}{K(k_0)}-\frac{G''(k_0)}{G'(k_0)}\right)\hat{\mathcal{G}}(k_0)+\hat{\mathcal{G}}'(k_0)\right].
\end{equation}
Evaluating all these terms numerically,\footnote{There is a closed form for the numerical factor in this correction. However, this is a complicated function of $k_0$ and elliptic integrals, and thus we do not present the expression which does not add much to the picture.} the total correction due to $B_{22}^C$ and $B_{20}$ is,
\begin{equation}
    \mathcal{R}=\sum_i\mathcal{R}_i\approx \frac{r}{\eta}\left(1.36782 B_{20} - 0.0316789 B_{2c}\right),
\end{equation}
where this factor should be understood to be a relative modification of the leading order available energy $\widehat{A}=\widehat{A}_0(1+\mathcal{R})$. That is, the correction to available energy due to $B_{20}$ and $B_{22}^C$, which we denote as $\widehat{A}_1$, is equal to
\begin{equation}
    \widehat{A}_1 = \widehat{A}_0\frac{r}{\eta}\left(\mathcal{R}_{20}B_{20} - \mathcal{R}_{22}^CB_{22}^C\right),
\end{equation}
where $\mathcal{R}_{20}\approx1.37$ and $\mathcal{R}_{22}^C\approx0.03$. \par
In the case of the strongly driven scenario, the integrals are over the whole $k$-space, but a term by term analysis can be performed in an analogous way to that above and the various terms numerically evaluated. The result of the calculation is,
\begin{equation}
   \widehat{A}_1= -\widehat{A}_0 \mathcal{R}_{20}^\mathrm{s} \frac{rB_{20}}{\eta}
\end{equation}
where $\mathcal{R}_{20}^\mathrm{s} =3.91$. Astonishingly, the $B_{22}^C$ contribution completely drops out. In terms of dependence on $B_{22}^C$, the two limits of the \AE{} are not that different, considering that $\mathcal{R}_{20}\gg\mathcal{R}_{22}^C$ for the former case (dropping it makes approximately $\sim 2\%$ error). For the remaining analysis we shall thus only consider dependence on $B_{20}$.
\par
As was the case in the discussion of the trapped electron precession, it is most physical to express these effects at second order in terms of the triangularity of a cross-section and the pressure gradient. This is explained in more detail in Appendix~\ref{app: relation to geometry}. Splitting $B_{20}$ up as
\begin{equation}
    B_{20}\approx-\frac{\mu_0 p_2}{B_0^2}  \mathcal{P}+\frac{\delta |\eta| }{2}  \mathcal{T},
\end{equation}
where we define
 \begin{subequations}
     \begin{gather}
         \mathcal{P} = \left[1+\frac{4\sqrt{\bar{\alpha}}}{(\bar{\alpha}+3)-(\bar{\alpha}+1)\Bar{F}}\left(\frac{\eta B_{\alpha0}}{B_0\bar{\iota}_0}\right)^2\right], \\
         \mathcal{T} = 3 \frac{(1-\bar{\alpha})+(\bar{\alpha}+1)\Bar{F}}{(3+\bar{\alpha}) - (\bar{\alpha}+1)\Bar{F}},
     \end{gather} \label{eqn:R_qs}
 \end{subequations}
% where $\mathcal{R}\approx-(\mu_0p_2/B_0^2)\mathcal{R}_{p_2}+(\delta/2)\mathcal{R}_\delta$. 
the correction in the weakly driven regime reduces to\begin{equation}
    \left. \frac{\widehat{A}_1}{\widehat{A}_0} \right|_{\mathrm{weak}} = \mathcal{R}_{20} \left(  - \frac{r}{\eta} \frac{\mu_0 p_2}{B_0^2} \mathcal{P} +  \frac{r\delta}{2}\mathcal{T} \right),
\end{equation}
and in the strongly driven regime,
\begin{equation}
    \left. \frac{\widehat{A}_1}{\widehat{A}_0}\right|_{\mathrm{strong}} = - \mathcal{R}_{20}^{\mathrm{s}} \left(  - \frac{r}{\eta} \frac{\mu_0 p_2}{B_0^2} \mathcal{P} +  \frac{r\delta}{2}\mathcal{T} \right),
\end{equation}
which are in the form presented in the main text

\bibliographystyle{jpp}
% Note the spaces between the initials

\bibliography{jpp-instructions}

\end{document}